\documentclass[pre,amsmath,amssymb,aps,10pt,one-column,showpacs,showkeys,floatfix,nofootinbib,longbibliography]{revtex4-1}

\usepackage{graphicx}
\usepackage{dcolumn}
\usepackage{bm}
\usepackage[caption=false]{subfig}
\usepackage{nicefrac}

\newcommand{\jfm}{{J. Fluids Mech.}}
\newcommand{\pof}{{Phys. Fluids}}

\newcommand{\njp}{{New J. Phys.}}

\begin{document}

\preprint{APS/123-QED}

\title{Instability of precession driven Kelvin modes: Evidence of a detuning effect}

\author{Johann Herault}
\email{johann.herault@imt-atlantique.fr}
\affiliation{IMT Atlantique, LS2N, UBL, 44307 Nantes, France} 
\author{Andr{\'e} Giesecke, Thomas Gundrum, and Frank Stefani}
\affiliation{
Helmholtz-Zentrum Dresden-Rossendorf, Bautzner Landstrasse 400,
D-01328 Dresden, Germany.} 

\date{\today}

\begin{abstract}
We report an experimental study of the instability of a
nearly-resonant Kelvin mode forced by precession in a cylindrical
vessel.  The instability is detected above a critical precession ratio
via the appearance of peaks in the temporal power spectrum of pressure
fluctuations measured at the end-walls of the cylinder. The
corresponding frequencies can be grouped into frequency sets
satisfying resonance conditions with the forced Kelvin mode.  

We show
that one set forms a triad that is associated with a parametric
resonance of Kelvin modes.  We observe a significant frequency
variation of the unstable modes with the precession ratio, which can
be explained by a detuning mechanism due to the slowdown of the
background flow. By introducing a semi-analytical model, we show that
the departure of the flow from the solid body rotation leads to a
modification of the dispersion relation of Kelvin modes and to a
detuning of the resonance condition.

The second frequency set includes a very low frequency and does not
exhibit the properties of a parametric resonance between Kelvin modes.
Interestingly this frequency set always emerges before the occurence
of the triadic resonances, i.e. at a lower precession ratio, which
implies that it may correspond to a new type of instability. We
discuss the relevance of an instability of a geostrophic mode
described by Kerswell [J. Fluid Mech. 1999, 382, 283--306] although
other mechanisms cannot be completely ruled out.
\end{abstract}

\maketitle


\section{\label{sec:intro} Introduction}

Since the original work of Kelvin \citep{Kelvin1880} it is known that
rotating columns of fluid support helical modes composed of two
inertial waves, which nowadays are called Kelvin modes.  Experiments
have shown that Kelvin modes can be directly forced in precessing or
librating flows \cite{mcewan1970inertial} or by emulating tidal
forcing \cite{le2014flows}. Kelvin modes may also appear via an
instability, as observed in the Lamb-Oseen vortex
\cite{fabre2006kelvin}, or in a column of fluid that is elliptically
deformed \cite{kerswell2002elliptical}.  Global stability analysis
shows that a single directly forced Kelvin mode may become unstable
via a parametric (triadic) resonance with two further free Kelvin
modes \cite{kerswell1999secondary}, i.e. two modes that are not
present before the onset of the instability.  Whereas the linear
mechanisms have been extensively studied, the experimental
investigation of the non-linear evolution of the instability, which
involves the saturation process, is still an open issue. Two
scenarios, which may coexist, have been suggested to explain the final
stage of the instability. First, saturation can be achieved by
non-linear interactions between the unstable modes and small scale
structures. This leads to an energy transfer from the linearly
unstable modes to small scale dissipative structures
\cite{kerswell1999secondary}. This mechanism can be intermittent, as
observed by McEwan \citep{mcewan1970inertial} during the "resonant
collapse" of Kelvin modes. Saturation can also be achieved by a
secondary mean flow generated directly by the unstable modes.  The
resulting modification of the background flow\footnote{ 
Here, the notion {\it{background flow}} refers to this axisymmetric
azimuthal flow component and does not mean the time-averaged flow
which is determined by the forcing.  For a precessing fluid the
time-averaged flow essentially consists of the directly forced
standing inertial wave (and its harmonics from nonlinear
self-interaction when the forcing is sufficiently strong) superimposed
to the azimuthal circulation which in turn consists of the solid body
rotation and a shear flow which effectively causes a braking of the
solid body rotation \cite{Giesecke2018}.
}
leads to a saturation of the amplitude of the unstable mode, via a
frequency detuning or a phase shifting between the unstable modes and
the forced mode \cite{WaleffePhD}.  This problem can be decomposed
into two parts: the modification of the background flow produced by
Kelvin modes, and the feedback of this flow on the Kelvin modes. The
first point is a complex problem because the direct non-linear
interaction between a single Kelvin mode and the background flow is
not permitted at first order for the steady case in the long time
scale \citep{greenspan1969}. However, the emergence of strong mean
flows via non-linear interactions between inertial modes is allowed
for finite Rossby number (the ratio between non-linear advection
$|\textbf u \nabla \textbf u|$ over the Coriolis term
$|2{\boldsymbol{\Omega}}\times\textbf u|$) or finite Ekmann number
(the ratio between viscous forces and the Coriolis force)
\cite{LeReun2017}.

In the present study we deal with the second point by examining
experimentally and theoretically the detuning of resonant Kelvin modes
caused by the induced modification of the background flow in a
precessing fluid.  The experiments utilize a cylinder filled with
water which rotates around its symmetry axis at the rotation rate
$\Omega_{\rm{c}}$. This cylinder is mounted on a turntable rotating at
the precession rate $\Omega_{\rm{p}}$.  When the angle between the
axes of rotation and precession (the nutation angle $\alpha$) is
non-zero, the flow deviates from the solid body rotation, due to the
acceleration of the rotation axis.  As originally observed by McEwan
\citep{mcewan1970inertial}, the primary response of the fluid is a
laminar flow that is composed of Kelvin modes superimposed on a solid
body rotation
\cite{manasseh1992breakdown,kobine1995inertial,meunier2008rotating,liao2012flow}.
The gyroscopic force due to precession directly forces a Kelvin mode
with azimuthal wave number $m=1$ and angular frequency
$-\Omega_{\rm{c}}$ in the frame of reference of the cylinder. This
mode can be resonant if its eigenfrequency $\omega$ is close to the
angular frequency of the forcing $\Omega_{\rm{c}}$.  Out of the
resonance, the amplitude of the mode scales like $\vert \omega-
\Omega_{\rm{c }} \vert^{-1}$ and increases with the strength of the
precession, quantified by the precession ratio $\epsilon=\Omega_{\rm{p
}} /\Omega_{\rm{c}}$.  At resonance, viscous and non-linear effects
must be taken into account for explaining the saturation of the forced
Kelvin mode
\cite{gans1970,meunier2008rotating,kobine1996azimuthal,liao2012flow}.
It is worth noting that non-linear and viscous effects may lead to the
generation of a geostrophic mode
\cite{meunier2008rotating,kobine1996azimuthal} in the laminar regime
before the occurrence of the instability of the forced Kelvin modes
\cite{kerswell1999secondary}.  The instability of the forced Kelvin
mode sets in above a critical precession ratio
\cite{mcewan1970inertial,manasseh1992breakdown,kobine1995inertial},
and the linear analyses \cite{kerswell1999secondary} predicts a
triggering of a parametric (triadic) resonance which has been
confirmed in experimental and numerical studies
\cite{lagrange2008instability,lin2014experimental,albrecht2015triadic,herault2015subcritical,giesecke2015triadic,albrecht2018}.
The parametric instability can either saturate to a periodic state
\cite{lagrange2011precessional} or trigger a non-stationary sequence
of bifurcations which may account for a chaotic state called resonant
collapse as it has been observed experimentally
\cite{mcewan1970inertial,manasseh1992breakdown,lin2014experimental}. The
weakly non-linear analysis of the instability
\cite{lagrange2011precessional} shows that saturation can also be
achieved by a detuning of the forced Kelvin mode due to a modification
of the solid body rotation.

This scenario is supported by measurements of the modification of the
background flow in the laminar
\cite{kobine1996azimuthal,meunier2008rotating} as well as in the
nonlinear regime \cite{lagrange2011precessional,mouhali2012evidence}.
In this regard, simulations and experiments indicate a fundamental
difference between the emerging flow at small and at large nutation
angle $\alpha$, which can be seen in the transition from laminar
to turbulent flow.  At small $\alpha$ different types of the
transition into the turbulent regime can be observed, which are
usually referred to as {\it{resonant collapse}}
\cite{manasseh1992breakdown}.  Recent simulations by
\citet{albrecht2018} give evidence that a cascadic emergence of
triadic resonances determines the transient evolution during the
resonant collapse.  In contrast, at large $\alpha$ only one type of
definite transition is found at a critical precession ratio
\cite{herault2015subcritical} which presumably is not related to the
first instabilities because it only occurs for a precession ratio
approximately two times larger than the threshold of the parametric
instability.  Moreover, measurements from our previous study
\cite{herault2015subcritical} do not support the scenario of a
resonant collapse of Kelvin modes but rather a subcritical transition
with sudden and definitive relaminarizations, a phenomenon frequently
observed in shear-flows \cite{hof2006finite}.  A strong shear flow is
indeed observed in our experiments and simulations which is reflected
in a significant braking of the initial solid body rotation
\cite{Giesecke2018}.  The associated strong velocity gradients near
the outer walls tend to become unstable which may trigger the
transition in the turbulent regime at large $\alpha$ by eruptions that
originate in the lateral boundary layers and penetrate into the bulk
flow \cite{lopezmarques2016}.

The generation of mean flows from precessional forcing is a rather
complex process strongly impacted by nonlinearities, and the
analytical calculation of the feedback of the instability, i.e. how
the unstable modes backreact on the background flow, is hardly
tractable, although particular studies have been applied to the
laminar case for forced Kelvin modes \cite{meunier2008rotating} and to
the unstable regime \cite{lagrange2011precessional}.  In this study,
we do not focus on this part, but we assume a priori the functional
form of the modified background flow which results from some
unspecified mechanism. In particular, we do not assume that the
instability is responsible for the modification of the background
flow, but that a modified background flow impacts the instabilities by
detuning, which finally causes the saturation of the instability.

Up to present, the experimental evidence of the saturation via a
modification of the background flow that goes along with the detuning
of Kelvin modes is still lacking.  The detuning of the frequency of
the forced mode could only be observed indirectly by measuring a
decrease of its amplitude, because its frequency always remains the
one of the forcing.  In contrast, the detuning of free Kelvin modes
can be measured directly because only the condition of resonance must
be maintained, i.e. the difference or the sum of their
frequencies. The present paper aims at demonstrating the detuning of
free Kelvin modes with experimental measurements and complementary
analytic calculations that are based on a perturbation approach. After
introducing the experimental set-up and the theoretical background in
sections \ref{sec_dispersion_relation} and \ref{sec:experiment}, we
report our experimental observations in section
\ref{sec_application_experiment}.  We briefly discuss the qualitative
behavior of the pattern of the unstable Kelvin mode and study the
amplitudes and the frequencies of the contributing modes. The
remainder of the paper aims at an analytical explanation of the
detuning of the frequency as a function of the modification of the
background flow (Section \ref{sec_dispersion_relation_shear}) followed
by a brief sketch of the theoretical background that describes the
impact of shear on the eigenmodes in a rotating fluid (Section
\ref{sub_sec_motivation_background}).  Finally, we discuss the effect
of the detuning on the parametric instability (section
\ref{sec_parametric}).


\section{Mathematical preliminaries}\label{sec_dispersion_relation}

\subsection{Basic equations}

We consider a viscous and incompressible fluid with kinematic
viscosity $\nu$, which is contained in a rigid cylindrical cavity of
radius $R$ and height $H$. We apply cylindrical coordinates with the
origin of the reference frame located at the center of the cylinder
(Fig.~\ref{fig_schema}b) so that the fluid domain is defined by $(r,
\varphi,z) \in \left( [0,R] \times [0,2 \pi] \times [-H/2,H/2]
\right)$. The cylinder rotates at an angular frequency
$\Omega_{\rm{c}}=2 \pi f_{\rm{c}}$ around the axial direction $\textbf
e_z$.  This axis $\textbf e_z$ in turn rotates around a second axis,
the axis of precession $\textbf e_{\rm{p}}$, at the angular frequency
$\Omega_{\rm{p}}=2 \pi f_{\rm{p}}$.  In the present study the angle
between $\textbf e_z$ and $\textbf e_{\rm{p}}$ is fixed at
$\alpha=90^{\circ}$.  In the following, all computations are performed
in the reference frame of the cylinder, where the walls are at rest.
In this system, it is the axis of precession, $\textbf e_{\rm{p}}$,
that rotates around the axis $\textbf e_z$. The equation describing
the velocity field $\textbf v$ in the cylinder reference frame
rotating at ${\boldsymbol{\Omega}}= \Omega_{\rm{c}} \textbf e_z
+\Omega_{\rm{p}} \textbf e_{\rm{p}}$ is
\begin{equation}
 \partial_t \textbf v+ \textbf v \cdot \nabla \textbf v+2 
\left[\Omega_{\rm{c}} \textbf e_z+\Omega_{\rm{p}} \textbf e_{\rm{p}} (t)\right] 
\times \textbf v = -\frac{1}{\rho}\nabla P +\textbf F + \nu  \Delta \textbf v  
\label{NSeq} 
\end{equation}
where $\textbf{v}$ additionally  satisfies the mass conservation 
\begin{equation}
\nabla \cdot \textbf v=0 \label{incomp}
\end{equation}
and $\textbf{e}_{\rm{p}}(t)$ is given by
\begin{equation}
\textbf e_{\rm{p}}(t)=
\sin\alpha\cos(\Omega_{\rm{c}}t+\varphi)\hat{\textbf r}
-\sin\alpha\sin(\Omega_{\rm{c}}t+\varphi)\hat{\boldsymbol{\varphi}}
+\cos\alpha\hat{\textbf z}.
\end{equation}
In equation~(\ref{NSeq}) $P$ denotes the modified pressure that
includes the centrifugal contributions. The forcing term $\textbf F$,
also called Poincar{\'e} acceleration, corresponds to the gyroscopic
force which is responsible for the departure of the flow from the
solid body rotation and is given by
\begin{equation}
\textbf F  =
-\Omega_{\rm{c}} \Omega_{\rm{p}}  
r \cos(\Omega_{\rm{c}} t + \varphi) 
\textbf e_z.
\end{equation}  
The velocity field $\textbf v$ satisfies no-slip
conditions at the wall ($r=R$ and  $z=\pm H/2$) 
\begin{equation}
\textbf v=\textbf{0}. \label{eq_BC}
\end{equation}
The variables are made dimensionless using the time scale
$\Omega_{\rm{c}}^{-1}$ and the length scale $R$.  The rescaled
coordinates become $(r',z')=(r/R,z/R)$ and the time $t'=t
\Omega_{\rm{c}}$. The equation governing the dynamics of the
dimensionless velocity field $\textbf U=\textbf v /(R\Omega_{\rm{c}})$
and pressure field $P'=P/ (\rho R^2 \Omega_{\rm{c}}^2)$ is then
\begin{equation}
\label{eq_NSeqAd}
 \partial_{t'} \textbf U+ \textbf U \cdot \nabla \textbf U+2 
\left(  \textbf e_z+ \epsilon  \textbf  e_{\rm{p}}   \right) 
\times \textbf U  =
- \nabla P' +\textbf F'  + {\rm{Ek}}  \Delta \textbf U  
\end{equation}
with $\epsilon=\Omega_{\rm{p}}/\Omega_{\rm{c}}$ the precession ratio
and ${\rm{Ek}}=\nu/(\Omega_{\rm{c}} R^2)$ the Ekman number. The
forcing term is then given by 
\begin{equation}
\label{eq_dimensionless_force}
\textbf F'  =-\epsilon r   \cos(  t+\varphi) \textbf e_z. 
\end{equation}  
For the sake of simplicity, we remove the prime index in the remainder
of the paper. In the limit ${\rm{Ek}} \rightarrow 0$, the viscous
effects are essentially localized in the Ekman boundary layers,
varying with $\delta_{{\rm{Ek}}} \sim {\rm{Ek}}^{1/2}$. In this limit,
the bulk flow satisfies only a no-outward flow condition close to the
wall, by omitting the pumping at first order \cite{greenspan1968},
\begin{equation}
\label{BC2}
\textbf U \cdot \textbf n= {0}  \quad  \hbox{at the wall} 
\end{equation}
with $\textbf n$ the unitary vector normal to the wall.

\subsection{Linearized equations}
\subsubsection{General case} 

In order to establish the dispersion relation of the eigenmodes of a
rotating fluid in a cylinder (the Kelvin modes), we linearize the
Navier--Stokes equation (\ref{eq_NSeqAd}) by introducing the
infinitesimal velocity perturbation $\textbf u= \textbf U -\textbf
U_{\beta}$, where $\textbf U_{\beta}(r)$ is a mean azimuthal
circulation that is given by $\textbf U_{\beta} = r \beta
\Omega_\beta(r)\textbf e_{\varphi}$ with $\Omega_\beta(r)$ a
continuous function of $r$ characterizing the departure from the solid
body rotation and $\beta$ its real amplitude.  Note that
${\Omega}_{\beta}$ only depends on the radius and is independent of
$z$ (see section \ref{sub_sec_motivation_background}).  Note also that
all calculations are performed in the cylinder reference frame, i.e.,
the frame co-rotating with the cylinder wall so that $\textbf
U_{\beta}$ does not include the solid body rotation.

We linearize around $\textbf U_{\beta}(r)$ and neglect the non-linear
term $\textbf u \cdot \nabla \textbf u$, the viscous term, the forcing
term $\textbf F$ and the non-stationary component of the Coriolis term
$\epsilon \textbf e_{\rm{p}}{ (t) } \times \textbf u$.  The equation
thus simplified describes the linear eigenmodes of a rotating fluid
with a modified rotation profile in the limit of vanishing precession
$\epsilon\rightarrow 0$ and reads
\begin{equation}
\left( \partial_t +\beta \Omega_\beta \partial_\varphi \right) 
\textbf u-2  \beta \Omega_\beta  u_\varphi    
\textbf{e}_r+\beta \Omega_\beta  u_r   
\textbf{e}_\varphi+u_r \partial_r \textbf U_{\beta}+\textbf{e}_z \times
\textbf u = -\nabla P.\label{eq::lininviscidNS}
\end{equation}
We look for infinitesimal perturbations in the form of eigenmodes,
characterized by an axial wave number $k$, an azimuthal wave number
$m$ and a complex frequency $\omega$,  
\begin{equation}
 \label{perturb_form}
 (\textbf u,p) =  \exp(i \omega t+i m \varphi+ik z)
\left(
\begin{array}{rl}
\tilde u_r(r)   \\ 
\tilde u_\varphi(r)  \\ 
\tilde u_z(r)  \\ 
\tilde p(r)   \\
\end{array}
\right)+c.c 
\end{equation} 
where $(\tilde u_r,\tilde u_\theta,\tilde u_z,\tilde p)$ are complex
functions of $r$, and $c.c.$ refers to the complex conjugate in order
to obtain a real velocity field. The eigenmodes satisfy the boundary
condition (\ref{BC2}) at the top and the bottom of the cylinder, 
$z=\pm \Gamma^{-1}/2$, if $k=\pi n \Gamma$ with $n$ an integer. The
linearized equation with the given perturbations~(\ref{perturb_form})
leads to the following set of equations
\begin{equation}
\label{sys_euler}
\left\{
\begin{array}{ll}
\displaystyle{ i \left( \omega    
+ m \beta \Omega_\beta \right) 
\tilde  u_r -2 \left(1+\beta \Omega_\beta  \right) 
\tilde u_\varphi+\partial_r \tilde p=0  }\\ \\
\displaystyle { i \left( \omega    
+ m \beta\Omega_\beta  \right) 
\tilde u_\varphi + \left(2+ \beta \zeta_\beta  \right)  
\tilde  u_r+\frac{i m}{r} \tilde p=0 }\\ \\
\displaystyle{   i \left( \omega    
+  m \beta \Omega_\beta  \right) 
\tilde u_z +  i k \tilde p=0 } \\ \\
\displaystyle{  \frac{1}{r} \partial_r ( r \tilde u_r) 
+\frac{i m}{r} \tilde u_\varphi + ik \tilde u_z=0 }  
\end{array} \right.
\end{equation}  
with $\zeta_\beta={ {2}} \Omega_\beta+r \partial_r \Omega_\beta $
denoting the $z$ component of the vorticity associated to
$\Omega_\beta$.  The system of equations~(\ref{sys_euler}) can be
rewritten using a four-component formulation for the complex
velocity-pressure field $(\tilde{ \textbf u}, \tilde p)$ which leads
to the compact form
\begin{equation}
\label{lin_eq}
\left( i \omega {\mathcal{I}}+ {\mathcal{L}}\right) 
(\tilde{ \textbf  u}, \tilde P) =\textbf{0}, 
\end{equation}
with ${\mathcal{L}}={\mathcal{L}}_0+ \beta {\mathcal{L}}_\beta$, and
the operators $\mathcal{I}$, ${\mathcal{L}}_0$ and
${\mathcal{L}}_\beta $ given in appendix~\ref{app::a1}.  The operator
${\mathcal{L}}_0$ represents the action of the Coriolis force with the
uniform rotation rate $\Omega_{\rm{c}}$ and the incompressibility
condition, and ${\mathcal{L}}_\beta$ characterizes the effect of the
modification of the background flow due to $\Omega_\beta(r)$.


\subsubsection{Kelvin modes with uniform rotation}\label{subsec_kelvin_betaz}

First, we consider the classical problem of a solid body rotation,
i.e. $\beta=0$, which gives
\begin{equation}
\label{eq_uniform}
\left( i \omega_{0l} {\mathcal{I}}+{\mathcal{L}}_0 \right) (\tilde{
  \textbf u}, \tilde p)_{0l} =\textbf{0}
\end{equation}
where the index $0$ is used to denote the eigenmodes and frequencies
for $\beta=0$. The index $l$ denotes a radial eigenmode 
$(\tilde{\textbf u},\tilde p)_{0l}$ with azimuthal and axial wave 
numbers $m$ and $k$ and characterizes the series of solutions of the
dispersion relation (see next paragraph).  

The corresponding dispersion relation
provides the radial wave numbers $\delta_{0l}$ and
frequencies $\omega_{0l}$ and consists of two equations given by 
\begin{equation}
\label{eq_disp1}
\omega_{0l}^2=\frac{4 k^2}{
\delta_{0l}^2+k^2}  
\end{equation}
and 
\begin{equation}
\label{eq_disp11} 
(2+\omega_{0l}) m J_m(\delta_{0l})- 
\omega_{0l} \delta_{0l} J_{m+1}(\delta_{0l})=0 
\end{equation}
with $m$ and $k$ given and $J_m$ the Bessel function of order $m$. The
first equation is the well known dispersion relation of inertial
waves. The second equation results from the enforcement of the
no-outflow condition $u_r=0$ at the lateral wall.  For $m$ and $k$
fixed, there exists an infinitely countable number of couples
$(\omega_{0l},\delta_{0l})$ satisfying equations (\ref{eq_disp1})
and~(\ref{eq_disp11}).  We distinguish three families of waves: Kelvin
modes are either prograde with $\omega_{0l}/m<0$, rotating in the
azimuthal direction of the background flow, or retrograde with
$\omega_{0l}/m>0$, rotating in the opposite direction. For the
retrograde (resp. prograde) Kelvin modes, the radial wave numbers
$\delta_{0l}$ are numbered in ascending (resp. descending) order with
$l$ a positive (negative) integer.  The absolute value $\vert l\vert$
corresponds to the number of zeroes in the radial direction plus one
(associated with the origin).  The third family, called geostrophic
modes, is characterized by $k=0$ and $\omega_{0l}=0$, and the radial
wave number arises as a solution of $J_m(\delta_{0l})=0$.  

Introducing the solution of Eq.~(\ref{eq_uniform}) in
(\ref{perturb_form}) and imposing free-slip boundary conditions at the
wall, the velocity field satisfying Eq.~(\ref{eq::lininviscidNS}) is
given by a sum of Kelvin modes
\begin{equation}\label{perturb_form2}
\textbf u (\textbf r,t)= \sum_m \sum_n \sum_l   
\exp(i \omega_{0l} t+i m \varphi ) ~ a_{mnl}   
\textbf u_{mnl} (r,z)+c.c.
\end{equation} 
with $(r, \varphi,z) \in \left( [0,1] \times [0,2 \pi] \times
[-\Gamma^{-1}/2,\Gamma^{-1}/2] \right)$.  The complex amplitude of the
Kelvin mode, $a_{[m,n,l]}$ with the wave numbers $m, n$ and $l$ is
associated to the velocity components $\textbf u_{mnl} (r,z)$
given by 
\begin{equation}
\label{perturb_form3}
\textbf u_{mnl} (r,z)=
\left(
\begin{array}{rl}
\tilde u^r_{mnl}(r) \sin(\pi n \Gamma z )   \\[0.2cm]
\tilde u^\varphi_{mnl}(r) \sin(\pi n \Gamma z )  \\[0.2cm] 
\tilde u^z_{mnl}(r) \cos(\pi n \Gamma z )    
\end{array}
\right) 
\end{equation} 
and the radial structure of the Kelvin mode reads 
\begin{eqnarray}  
\begin{array}{rcl}
\displaystyle \tilde{u}^r_{mnl}(r) & = & 
\displaystyle { \frac{-i}{4-\omega_{0l}^2}  
\left(\frac{(2+\omega_{0l}) m}{r}
J_m(\delta_{0l} r)- \omega_{0l}\delta_{0l} J_{m+1}(\delta_{0l} r)
\right) },
\\ \\ 
\displaystyle \tilde{u}^{\varphi}_{mnl}(r) &=& 
\displaystyle {\frac{1}{4-\omega_{0l}^2}  
\left(\frac{ (2+\omega_{0l})m}{r}
J_m(\delta_{0l} r)-2 \delta_{0l} 
J_{m+1}(\delta_{0l} r) \right)},
\\ \\ 
\displaystyle \tilde{u}^z_{mnl}(r) &=& 
\displaystyle {-\frac{k}{\omega_{0l} } J_m(\delta_{0l} r) }.     
\end{array}  
\label{kelvin_mode0}
\end{eqnarray} 
The velocity fields of the Kelvin modes form an orthogonal basis
\cite{greenspan1968} for the scalar product    
\begin{equation}
\int\limits_{r=0}^1 
\int\limits_{\varphi=0\vphantom{\Gamma}}^{2\pi}
\int\limits_{z=-\nicefrac{1}{2}\Gamma}^{1/2\Gamma} 
\left( \textbf u_{mnl} (r,z)  e^{ i m \varphi }\right)^{*}
\cdot 
\left( \textbf u_{m'n'l'} (r,z)  
e^{ i m'\varphi }\right)  r dr d \varphi d z
=\frac{\pi}{\Gamma}  e_{l'l}  \delta_{n'n} \delta_{m'm}
\label{eq::sp}
\end{equation} 
if they satisfy the boundary condition (\ref{BC2}).  The elements
$e_{l'l}$ (where we omit the indices $m$ and $n$ for the sake of
brevity) on the right hand side of definition~(\ref{eq::sp}) are given
by $e_{l'l}= \langle \tilde{ \textbf u}_{0l'}, \tilde{ \textbf u}_{0l}
\rangle \delta_{l'l}$ $e_{l'l}= \langle \tilde{ \textbf u}_{l'm'n'},
\tilde{ \textbf u}_{lmn} \rangle$ with
\begin{equation}
\langle \tilde{ \textbf  u}_{m'n'l'},
\tilde{ \textbf u}_{mnl}  \rangle
=  
\int\limits_{r=0}^1
\left[\tilde{u}^r_{m'n'l'}  {}^{*}
\tilde{u}^r_{mnl}+
\tilde{u}^{\varphi}_{m'n'l'} {}^{*}
\tilde{u}^{\varphi}_{mnl}+
\tilde{u}^{z}_{m'n'l'}{}^{*}
\tilde{u}^{z}_{mnl} 
\right] r dr.   
\label{def:ell}
\end{equation}

 
\subsubsection{Laminar flow driven by precession}

We briefly recall the response of a rotating fluid to precessional
forcing. Two new terms appear due to precession: the Coriolis term
associated with the axis $\textbf e_{\rm{p}}$, and the Poincar\'e
acceleration caused by the temporal variation of $\textbf e_z$.  Only
the latter is usually considered at leading order which leads to the
equation  
\begin{equation}
 \label{dimensionless_forcing_eq}
\partial_t \textbf u+2 \textbf e_z \times \textbf u+\nabla P 
= - \epsilon r  \cos(  t+\varphi) \textbf e_z.
\end{equation}  
We look for the solution as a sum of Kelvin modes of axial and radial
wave numbers  $n$ and $l$ forced at the frequency $\omega_{0l}=1$ and
azimuthal wave number $m=1$ \cite{manasseh1994distortions}. In that
case Eq. (\ref{perturb_form2}) becomes 
\begin{equation}
\textbf u= 
\left( \sum_{n,l} 
a_{1nl}  
\textbf u_{1nl} (r,z) 
e^{i t+i \varphi }  \right)+c.c.
\end{equation}
The gyroscopic force on the right hand side of
equation~(\ref{dimensionless_forcing_eq}) is anti-symmetric with
respect to the mid-plane of the cylinder and only forces modes with an
axial component with the same parity, i.e. with an odd axial wave
number $n$. By using equation~(\ref{eq_uniform}) and the properties of
orthogonality~(\ref{eq::sp}), the amplitude $a_{[1,n,l]} $ is given by
\begin{equation}
a_{1nl}  = -\frac{i}{1-\omega_{0l}}{ {\frac{\Gamma}{\pi}}} 
\frac{\iiint \left(  \textbf u_{1nl} e^{-(i t+i \varphi)}\right) 
\cdot \textbf F \hbox{dV} }{ \langle \tilde{ \textbf  u}_{mnl},\tilde{
\textbf u}_{mnl} \rangle }  
\label{eq::amp}
\end{equation}
with $\omega_{0l}$ the eigenfrequency of the mode with $(m=1,n,l)$.
The forcing term $\textbf{F}$ (given by
(\ref{eq_dimensionless_force})) that appears beyond the Integral on
the right side of (\ref{eq::amp}) implies that the amplitude $a$ is
directly proportional to the precession ratio $\epsilon$.  An explicit
calculation of the terms $a_{1nl}$ can be found in
\cite{meunier2008rotating,liao2012flow,manasseh1994distortions}. The
forced mode is retrograde and standing in the turntable reference
frame.

For $\Gamma=0.5$ the eigenfrequency of the mode $(m,n,l)=(1,1,1)$ is
$\omega= 0.996 \Omega_{\rm{c} }$ so that our experimental
configuration corresponds to a nearly-resonant case with a single
forced Kelvin mode. Hence, we consider that the flow is mostly
composed of one Kelvin mode plus the background flow before any
instability. Near the resonance $\omega_{0l} \simeq 1$, the amplitude
given by~(\ref{eq::amp}) diverges and viscous and non-linear effects
must be considered for the computation of the amplitude of the Kelvin
modes \cite{gans1970,meunier2008rotating,liao2012flow}.
 

\section{\label{sec:experiment} Experimental setup}

\begin{figure}[h!]
\subfloat[]{\includegraphics[width=0.45\textwidth]{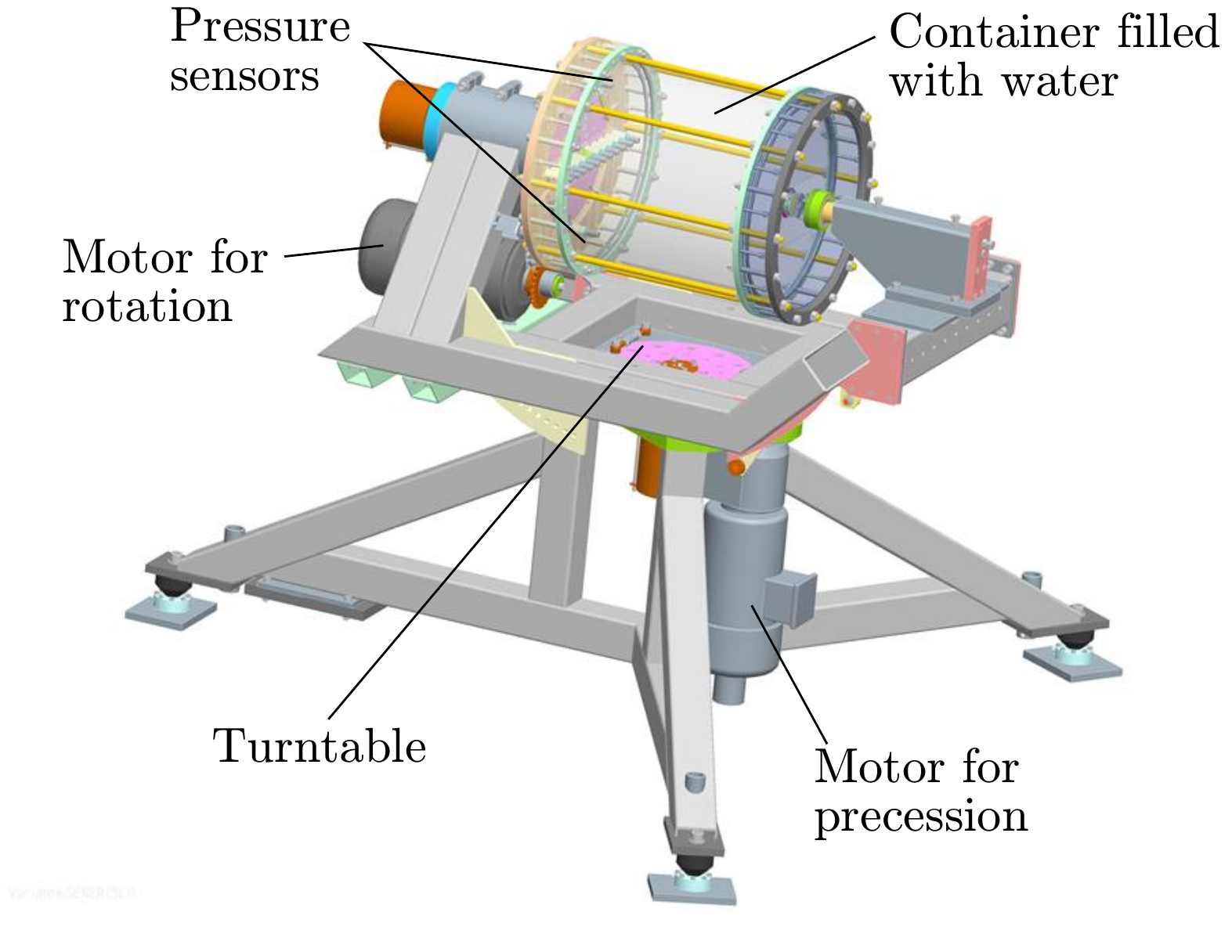}}
\hspace*{0.5cm}
\subfloat[]{\includegraphics[width=0.49\textwidth]{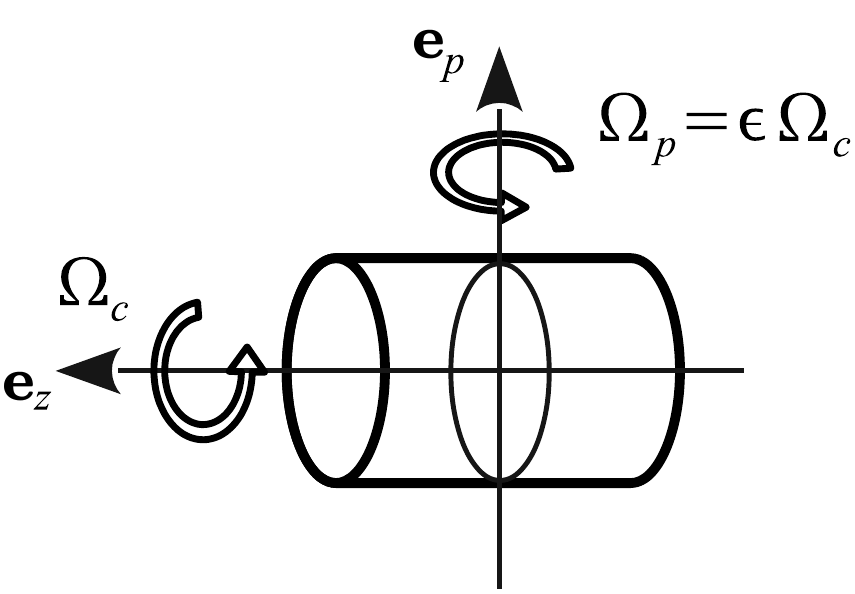}}
\caption{\label{fig_schema}
(a) Sketch of the experimental setup.
(b) Sketch of the reference frame.}
\end{figure}

The experimental setup is illustrated in Fig.~\ref{fig_schema}a. The
vessel is a cylinder of radius $R=163\,{\rm{mm}}$ and height
$H=326\,{\rm{mm}}$ filled with water. For a qualitative visualization
of the flow, a small amount of air is introduced so that the spatial
distribution of small gas bubbles within the rotating vessel enables a
first estimation of the basic structure of the flow.  The bubbles are
only used to visualize the flow pattern (reported in section
\ref{sec:flowpattern}) while all quantitative results presented in
sections \ref{sec:constEK} and \ref{sec:nonconstEK} have been obtained
with a significantly smaller gas fraction.  The container rotates
around its symmetry axis with a frequency $f_{\rm{c}}$ of up to
$10\,{\rm{Hz}}$ and is mounted on a turntable which can rotate with a
frequency $f_{\rm{p}}$ of up to $1\,{\rm{Hz}}$.  In all experiments
discussed here the rotation axis and the precession axis are
orthogonal ($\alpha=\pi/2$, Fig.~\ref{fig_schema}b).  A more detailed
description of the experiment can be found in
\cite{herault2015subcritical}.
 
\captionsetup[subfigure]{farskip=0.0cm,nearskip=0.0cm,margin=0.3cm,
                         singlelinecheck=false,format=plain,indention=0.45cm,
                         hangindent=0.0cm,justification=justified,
                         captionskip=0.1cm,position=top}    

In the following, we will refer to angular frequencies defined by
$\Omega_{\rm{c}}=2\pi f_{\rm{c}}$ and $\Omega_{\rm{p}}=2\pi
f_{\rm{p}}$ in order to match the notation in the theory section
below. The precessing system is now determined by three dimensionless
parameters: the aspect ratio $\Gamma=R/H$ which in this study is
always fixed at $\Gamma=0.5$, the Ekman number, ${\rm{Ek}}= \nu
({\Omega}_{\rm{c}} R^2)^{-1}$, and the precession ratio
$\epsilon=\Omega_{\rm{p}}/\Omega_{\rm{c}}$.  The mesured frequencies
$\Omega$ ($\Omega_{\rm{c}}$ included) will be rescaled by
$\Omega_{\rm{c}}$ so that $\Omega'=\Omega/\Omega_{\rm{c}}$. To lighten
the notations, we will omit the prime symbol in the next sections so
that the angular frequency $\Omega=1$ corresponds to the angular
frequency of the cylinder.  The essential physical parameters that
characterize the experimental set up are summarized in
Table~\ref{tab_param}.  Note that the reported Ekman numbers are one
order of magnitude smaller than those obtained in previous studies
\cite{lagrange2008instability,lin2014experimental}, which allows us to
investigate less viscous regimes. In addition to the direct
visualization of the flow, the pressure is measured at one of the
end-caps close to the motor (Fig.~\ref{fig_schema}a).  Two XPM5
miniature pressure sensors with a diameter of $3.6\,{\rm{mm}}$ are
mounted flush on one end-cap at a radius $r_{\rm{p}}=160\,{\rm{mm}}$
with an angle of $180^{\circ}$ between them. For these diametrically
opposed probes, we define the symmetrical component $P_{\rm{s}}$ and
the anti-symmetrical component $P_{\rm{a}}$ as the sum, respectively
the difference, of their measurement values:

\begin{equation}
\left\{
\begin{array}{ll}
P_{\rm{s}}=\left[P(r_{\rm{p}},\varphi,H/2)+P(r_{\rm{p}},\varphi+\pi,H/2) \right]/2  \\  \\
P_{\rm{a}}=\left[P(r_{\rm{p}},\varphi,H/2)-P(r_{\rm{p}},\varphi+\pi,H/2) \right]/2. 
\end{array}
\right.   
\label{eq_pressure}
\end{equation}
The  power-spectral density (PSD) of $P_{\rm{s}}$ is given by
\begin{equation}
E_{\rm{s}}(\Omega)= 
\frac{1}{T_{\rm{m}}} 
\overline{\tilde{P}_{\rm{s}}(\Omega)\tilde{P}_{\rm{s}}^{*}{{(\Omega)}}} 
\quad \hbox{with} \quad 
\tilde{P}_{\rm{s}}{ {(\Omega)}}= 
\int\limits_0 ^{T_{\rm{m}}} P_{\rm{s}}{{(t)}} e^{-i \Omega t} dt 
\end{equation}
with $T_{\rm{m}}$ the duration of the measurements and
$\overline{~\cdot~ }$ the average over the realizations. The same
procedure defines the power-spectral density $E_{\rm{a}}$ of
$P_{\rm{a}}$.

\begin{table}[t!]
\caption{\label{tab_param}Parameters and non-dimensional  numbers
  used in this study. The Ekman number is based on a viscosity
  of water at $15^{\circ}\mbox{ C}$  of
  $\nu=1.17 \times 10^{-6}\,{\rm{m}}^2/{\rm{s}}$.} 
\begin{ruledtabular}
\begin{tabular}{lccc}
Name & Notation & Definition & Experimental values \\[0.1cm]
\hline 
{\vphantom{$\hat{C}$}}Cylinder frequency   & $f_{\rm{c}}$ & $-$  &  $0-10\,{\rm{Hz}}$   \\
Precession frequency & $f_{\rm{p}}$ & $-$  &  $0-1\,{\rm{Hz}}$    \\
Radius		     & $R$	    & $-$  &  $163\,{\rm{mm}}$    \\
Height		     & $H$	    & $-$  &   $326\,{\rm{mm}}$   \\
Ekman number	     & ${\rm{Ek}}$  & $\nu (\Omega_{\rm{c}} R^2)^{-1}$ & $ 7\times 10^{-7}-7\times 10^{-6}$\\
Precession ratio     & $\epsilon$   & $f_{\rm{p}}/f_{\rm{c}}$          & $2.5\times 10^{-2}-6\times 10^{-2}$\\
Aspect ratio	     & $\Gamma$     & $R/H$	                       & $0.5$\\
\end{tabular}
\end{ruledtabular} 
\end{table}


\section{Experimental evidence of frequency detuning} 
\label{sec_application_experiment} 

\subsection{Flow pattern}
\label{sec:flowpattern}

\begin{figure}[b!]
(a) \hspace{4cm} (b) \hspace{4cm} (c) \hspace{4cm} (d) \hspace{4cm}
\includegraphics[height=4cm,width=18cm]{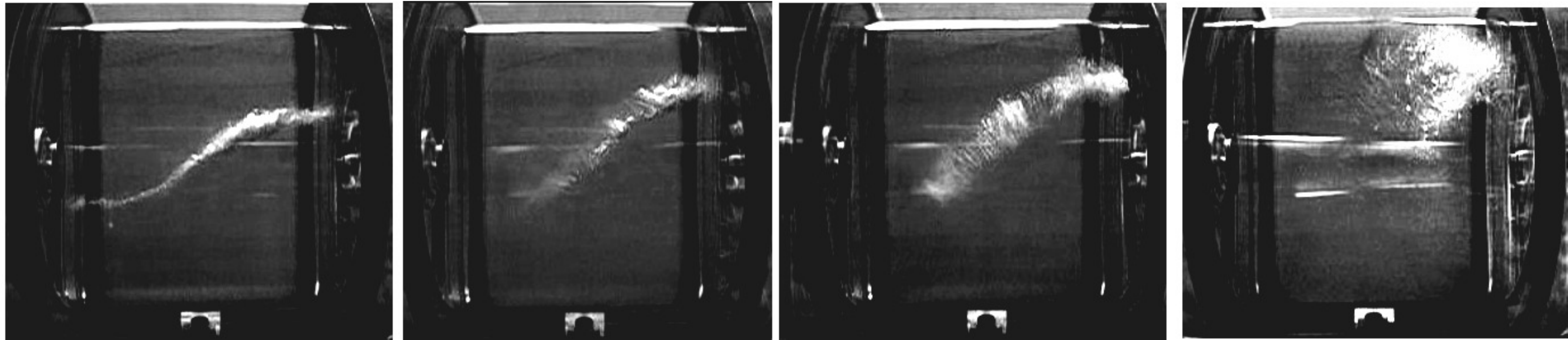}
\caption{\label{fig_pattern}Photographs of the precessing cylinder in 
  the turntable frame. The bright regions correspond to air  
  bubbles, where the pressure is  minimal. 
The Ekman number is ${\rm{Ek}}=7\times10^{-7}$
($f_{\rm{c}}=10\,{\rm{Hz}}$),
and the precession ratio $\epsilon=0.74\times 10^{-2}$ (a), 
  $\epsilon=3\times 10^{-2}$ (b), $\epsilon=3.4\times 10^{-2}$ (c) 
  and $4.8\times 10^{-2}$ (d).} 
\end{figure}

Before presenting quantitative results on the instability of the
precessing flow, we discuss the qualitative diversity of the flow
pattern for increasing precession ratio. We characterize the flow by
analyzing the region of minimal pressure which is visualized by means
of small air-bubbles that -- at least in the laminar regime -- align
along the minimum pressure line.  Although we do not observe any
measurable changes of the instability thresholds in the experiments
that include a small amount of gas, the air is removed to perform the
quantitative measurement of the wall-pressure presented in the
remainder of the paper.  Figure~\ref{fig_pattern} shows four snapshots
of characteristic flow patterns at ${\rm{Ek}} = 7 \times 10^{-7}$
corresponding to $f_{\rm{c}}=10\,{\rm{Hz}}$.  In the laminar regime
the superposition of the nearly-resonant Kelvin mode and the solid
body rotation yields an S-shaped tube that roughly corresponds to the
effective rotation axis of the fluid ($\epsilon=0.74\times 10^{-2}$,
Fig.~~\ref{fig_pattern}a). The pattern is standing in the turntable
frame, where the pictures are taken, and respects the centro-symmetry
of the forcing.  When the laminar flow becomes unstable around
$\epsilon=3 \times 10^{-2}$, we see erratic fluctuations resulting in
a blurring of the S-tube (Fig.~\ref{fig_pattern}b). Further increasing
$\epsilon$, the apparent width of the tube grows and we observe a
stronger mixing in the vicinity of the tube associated to the
spreading of bubbles (Fig.~\ref{fig_pattern}c).  Around
$\epsilon=4.8\times 10^{-2}$, the S-tube has dissolved and the bubbles
are trapped in the upper-right region close to the end-cap
(Fig.~\ref{fig_pattern}d). This localization of the bubbles close to
one side can be explained by the appearance of one or few modes
breaking the centro-symmetry of the laminar flow and thus creating a
region of lower pressure close to one end-cap. At this state, the flow
is chaotic but not yet turbulent.  However, it is important to state
that we never observe a phenomenon comparable to the resonant collapse
\cite{manasseh1992breakdown} which would lead to an intermittent fine
scale turbulence in the entire vessel.  An abrupt transition to a
turbulent state characterized by a homogeneous spreading of bubbles in
the entire vessel only occurs above a critical precession ratio that
is approximately twice as large as that for the onset of the
instability. This phenomenon has been characterized in
\cite{herault2015subcritical} (see Fig 2 in
\cite{herault2015subcritical}) and will not be discussed in this
study.


\subsection{Pressure measurement at constant Ekman number}
\label{sec:constEK}

\subsubsection{Frequencies}

\begin{figure}[b!]
\includegraphics[width=0.49\textwidth]{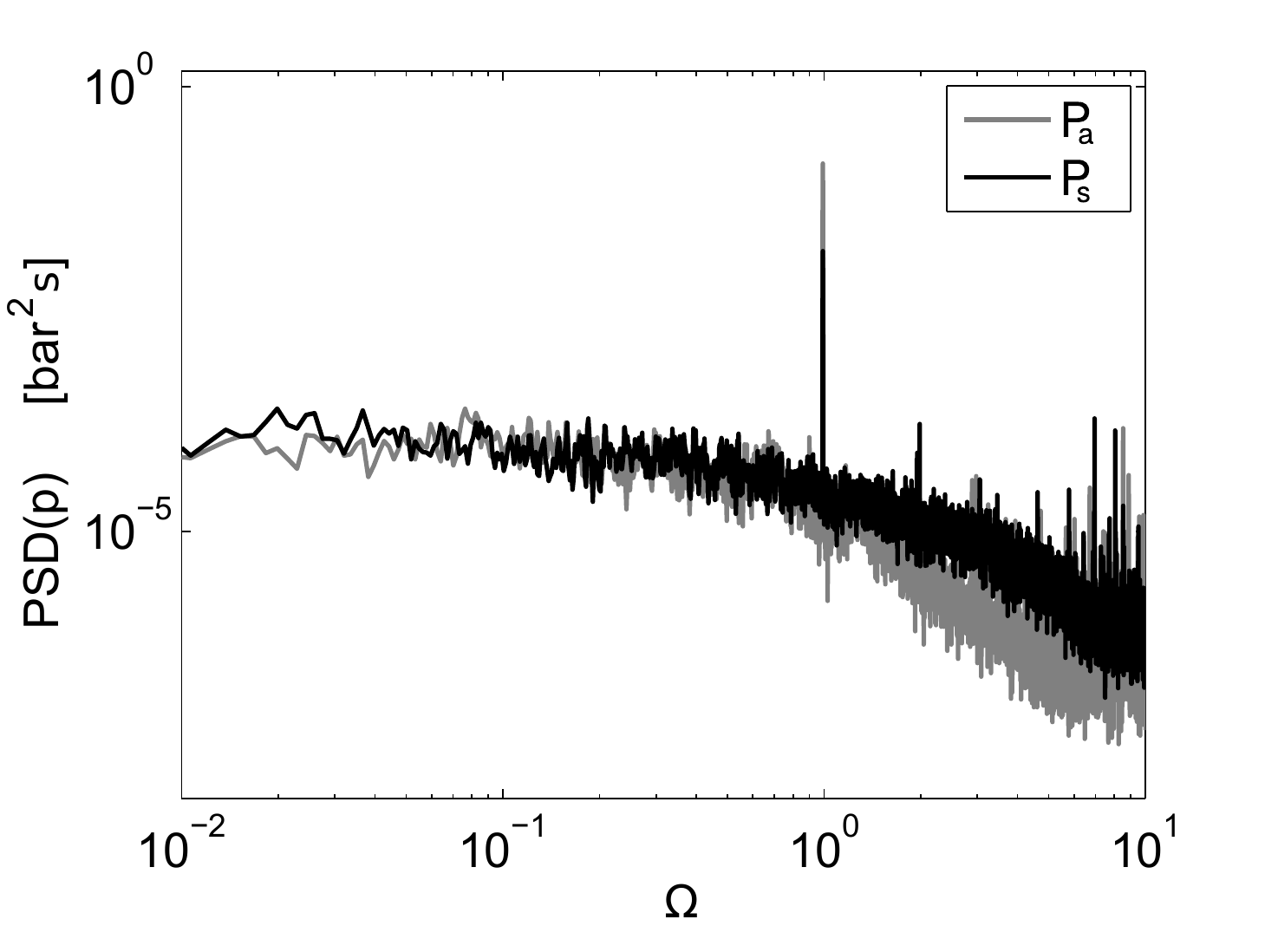}    
\caption{\label{fig_spectre_lam}Spectra of the
  pressure components $P_{\rm{a}}$ (anti-symmetric, grey) and $P_{\rm{s}}$
  (symmetric, black)  for $\epsilon=0.68 \times 10^{-2}$ and ${\rm{Ek}}=1.76
  \times 10^{-6}$ ($f_{\rm{c}}=4\mbox{ Hz}$, laminar regime).} 
\end{figure}

\begin{figure}[t!]
\subfloat[]{\includegraphics[width=0.49\textwidth]{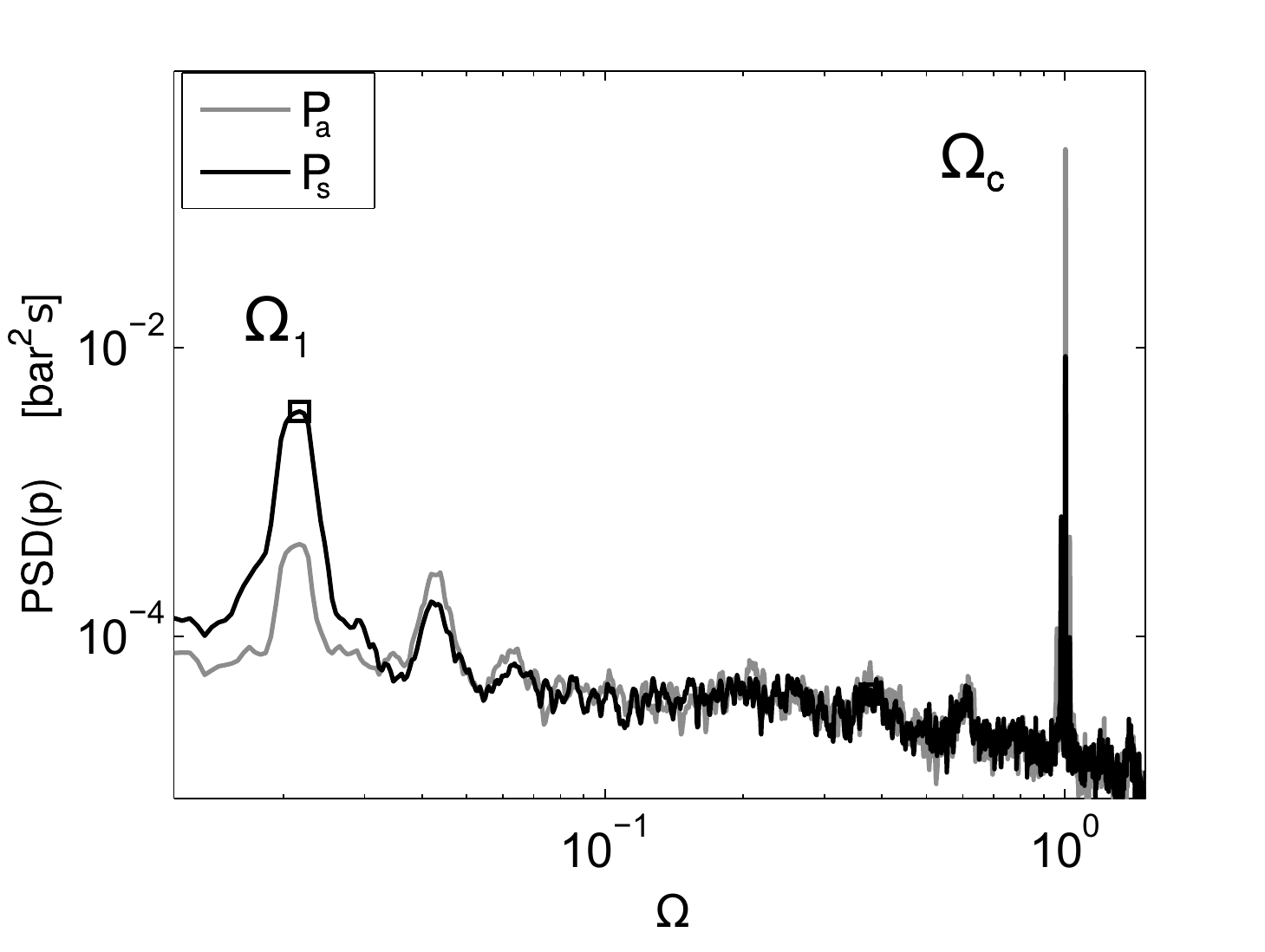}}
\subfloat[]{\includegraphics[width=0.49\textwidth]{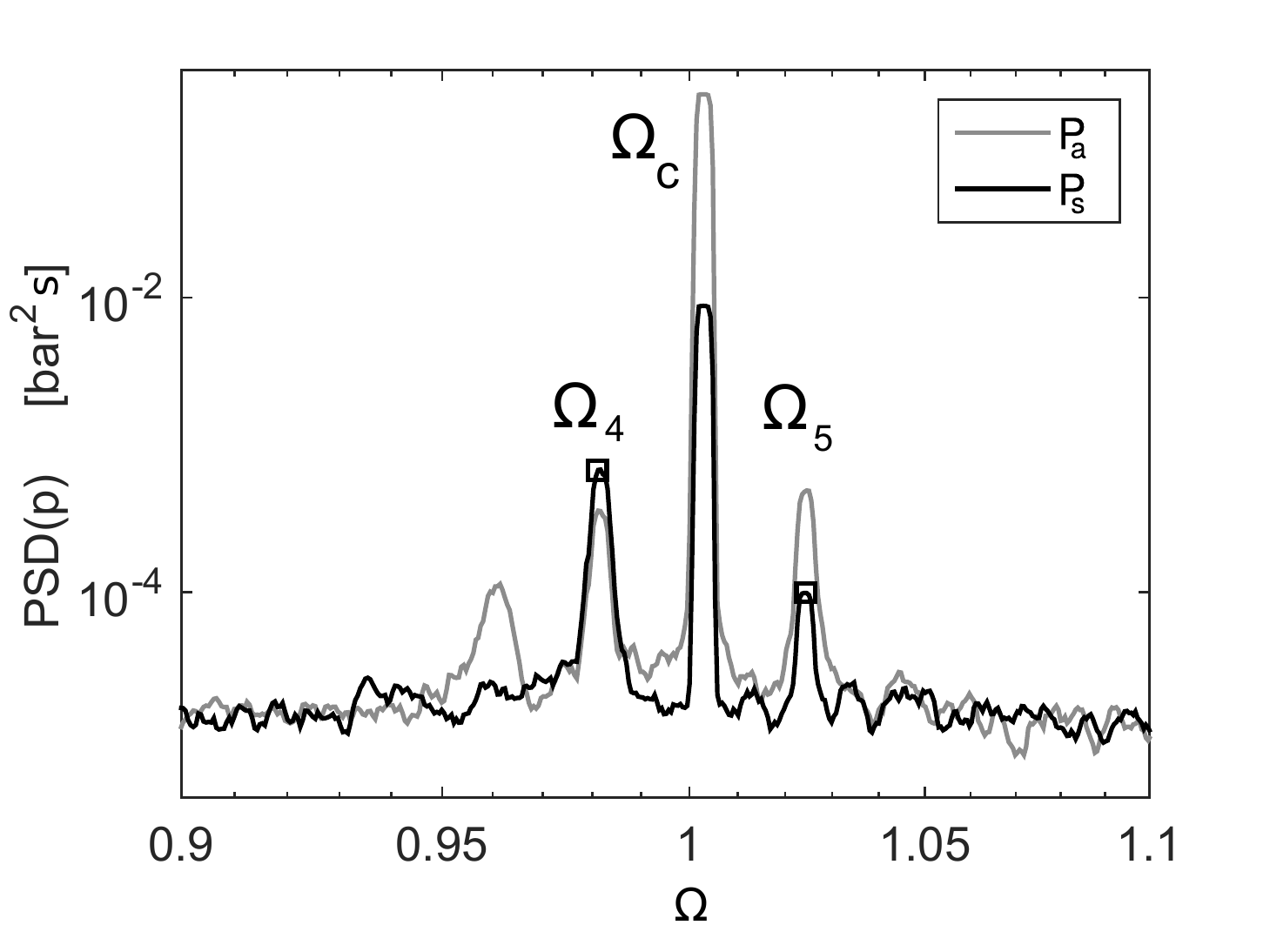}}
\caption{\label{fig_spectre_eps3}
(a) Spectra of the pressure components $P_{\rm{a}}$
(anti-symmetric,grey) and $P_{\rm{s}}$ (symmetric, black) for
$\epsilon=3 \times 10^{-2}$  above  the onset of  instability. (b)
Zoom in the vicinity of the forced  mode ($\Omega_{\rm{c}}$). Three peaks
in addition to the forced mode are well identified  and labelled by
$\Omega_1,\Omega_4$ and $\Omega_5$.  
}
\end{figure}

In the present section, we study the spectral information contained in
the pressure measurements.  The measurements are performed for
constant Ekman number ${\rm{Ek}}=1.76 \times 10^{-6}$
($f_{\rm{c}}=4\,{\rm{Hz}}$) and the precession ratio $\epsilon$ is
varied in the range $[0,6] \times 10^{-2}$.  Each run lasts at least
for $40$ minutes, or $9.6 \times 10^4 $ periods.  In the following,
time is measured in units of the rotation time of the cylinder so that
all (angular) frequencies are denoted in units of the angular
frequency of the cylinder, which, consistently, is set to unity,
$\Omega_{\rm{c}}=1$.  The power-spectra of the pressure components
$P_{\rm{a}}$ (anti-symmetric) and $P_{\rm{s}}$ (symmetric), defined by
equation~(\ref{eq_pressure}), are shown in Fig.~\ref{fig_spectre_lam}
for $\epsilon=0.68 \times 10^{-2}$.
In this case the flow is laminar and the spectrum is only
composed of peaks located at $\Omega_{\rm{c}}$ and its multiples. The most
intense peak belongs to the anti-symmetric component $P_{\rm{a}}$ at
$\Omega_{\rm{c}}$ and originates from the forced Kelvin mode with $m=1$.
Increasing the precession ratio to $\epsilon=3 \times 10^{-2}$ the
power spectra exhibit new features (Fig.~\ref{fig_spectre_eps3}a and
b). An intense peak appears at low frequency at $\Omega_1=0.021$
(Fig.~\ref{fig_spectre_eps3}a). The smaller peak on the right side of
$\Omega_1$ corresponds to the frequency $2\Omega_1$ and disappears for larger
$\epsilon$. A close inspection of the power spectrum in the vicinity
of $\Omega_{\rm{c}}=1$ shows the presence of two further peaks located on
both sides of the forced mode $\Omega_{\rm{c}}$
(Fig.~\ref{fig_spectre_eps3}b). The maxima of these peaks correspond
to the frequencies $\Omega_4 = 0.981$ and 
$\Omega_5 = 1.025$ (the choice for the index $4$ and $5$ is explained in next
paragraph). The four frequencies $\Omega_{\rm{c}}, \Omega_1, \Omega_4$ and $\Omega_5$
fulfil the relations 
\begin{equation}
\label{relation_res}
\left\{
\begin{array}{l}
{{\Omega}}_4+{{\Omega}}_1 \simeq {{\Omega}}_{\rm{c}} =1 \\ \\
{{\Omega}}_5-{{\Omega}}_1 \simeq {{\Omega}}_{\rm{c}}=1.
\end{array} \right. 
\end{equation}
The location of the frequencies $\Omega_1, \Omega_4,$ and $\Omega_5$ depends
monotonously on the precession ratio such that they always fulfill the
relations~(\ref{relation_res}). 

\begin{figure}[b!]
\subfloat[]{\includegraphics[width=0.49\textwidth]{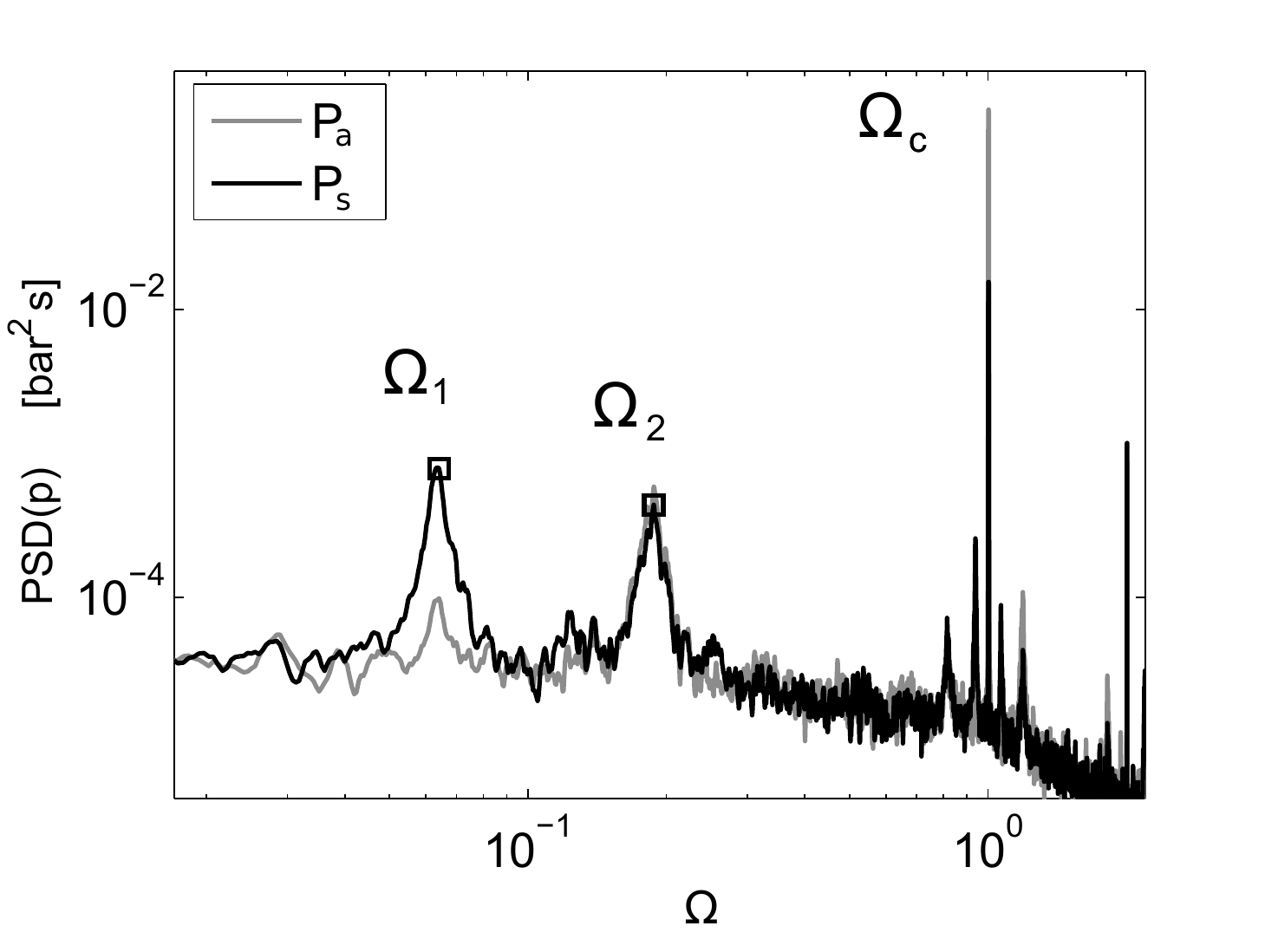}} 
\subfloat[]{\includegraphics[width=0.49\textwidth]{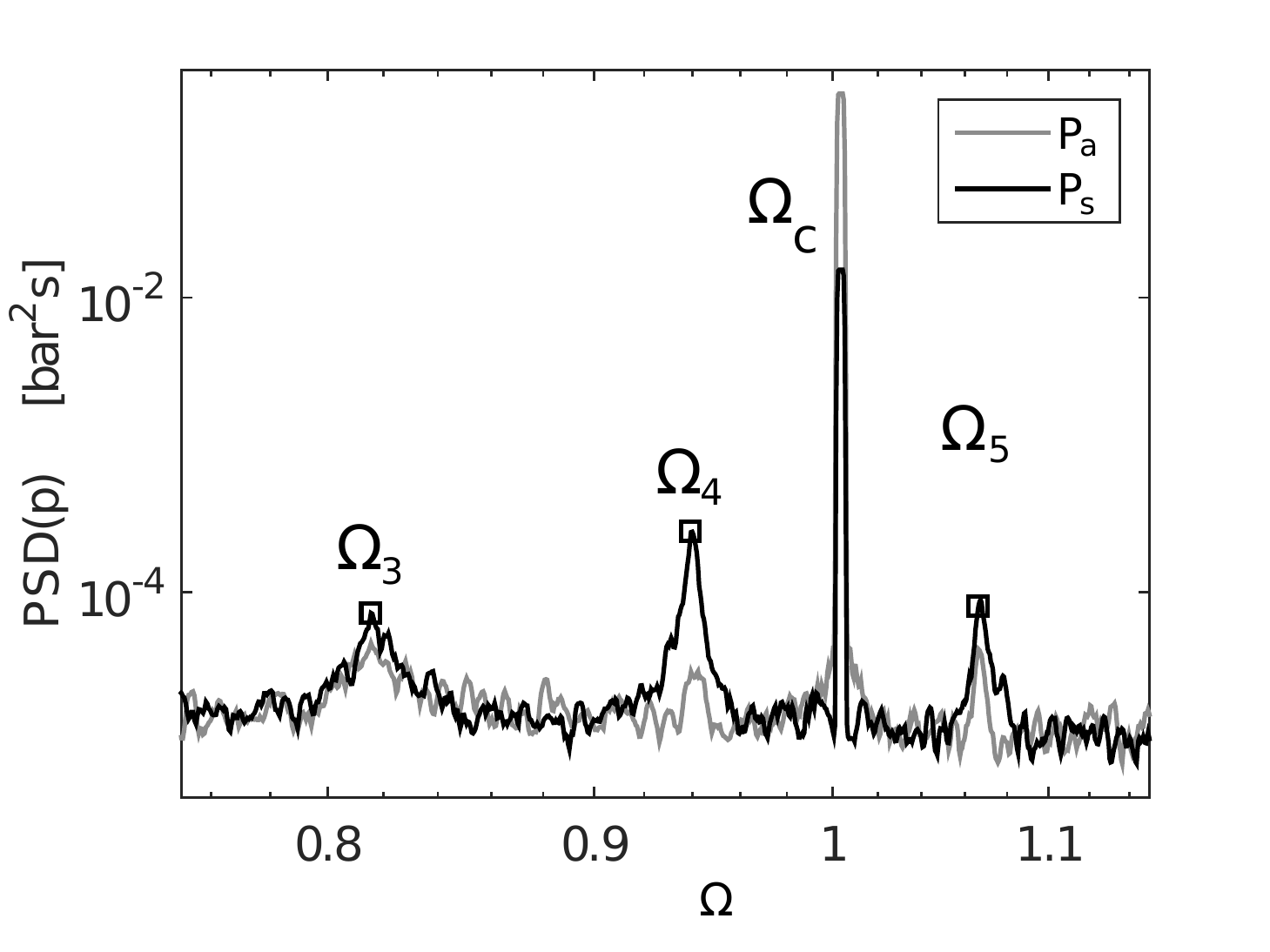}} 
\caption{\label{fig_spectre_eps375} (a) Spectra  of the
pressure components $P_{\rm{a}}$ (anti-symmetric, grey) and
$P_{\rm{s}}$ (symmetric, black) for $\epsilon=3.75 \times 10^{-2}$ and 
${\rm{Ek}}=1.77\times 10^{-6}$ past the onset of the
instability. (b)  Zoom in the vicinity of ${{\Omega}}={{\Omega}}_{\rm{c}}$. Beside the
forced mode ${{\Omega}}_{\rm{c}}$ five peaks are well identified labeled by  ${{\Omega}}_1,{{\Omega}}_2,{{\Omega}}_3,{{\Omega}}_4$ and
  ${{\Omega}}_5$.}  
\end{figure}

Further increasing the precession ratio we find another pair of
signals emerging at $\epsilon=3.75 \times 10^{-2}$
(Fig.~\ref{fig_spectre_eps375}). At the lower end of the spectrum we
now see two peaks at $\Omega_1 \approx 0.064$ and $\Omega_2 \approx
0.187$ (Fig.~\ref{fig_spectre_eps375}a) and three peaks in the
vicinity of ${{\Omega}}_{\rm{c}}$ with ${{\Omega}}_3\approx 0.81$,
$\Omega_4\approx 0.94$ and $\Omega_5\approx 1.06$. Comparable to the
already known frequencies ${{\Omega}}_1, {{\Omega}}_4, {{\Omega}}_5$
the frequencies $\Omega_2$ and $\Omega_3$ fulfil the relation
\begin{eqnarray}
\label{relation_res2}
\Omega_2+\Omega_3 & \simeq & \Omega_{\rm{c}}=1.
\end{eqnarray}

\captionsetup[subfigure]{margin=0.0cm,singlelinecheck=false,
                         format=plain,indention=0.0cm,
                         justification=justified,
                         captionskip=0cm,position=top}  

\captionsetup[subfigure]{margin=0.0cm,singlelinecheck=false,
                         format=plain,indention=0.0cm,
                         justification=justified,captionskip=0cm,
                         position=top} 

\begin{figure}[t!] 
\subfloat[]{\includegraphics[height=5.5cm,width=9.5cm]{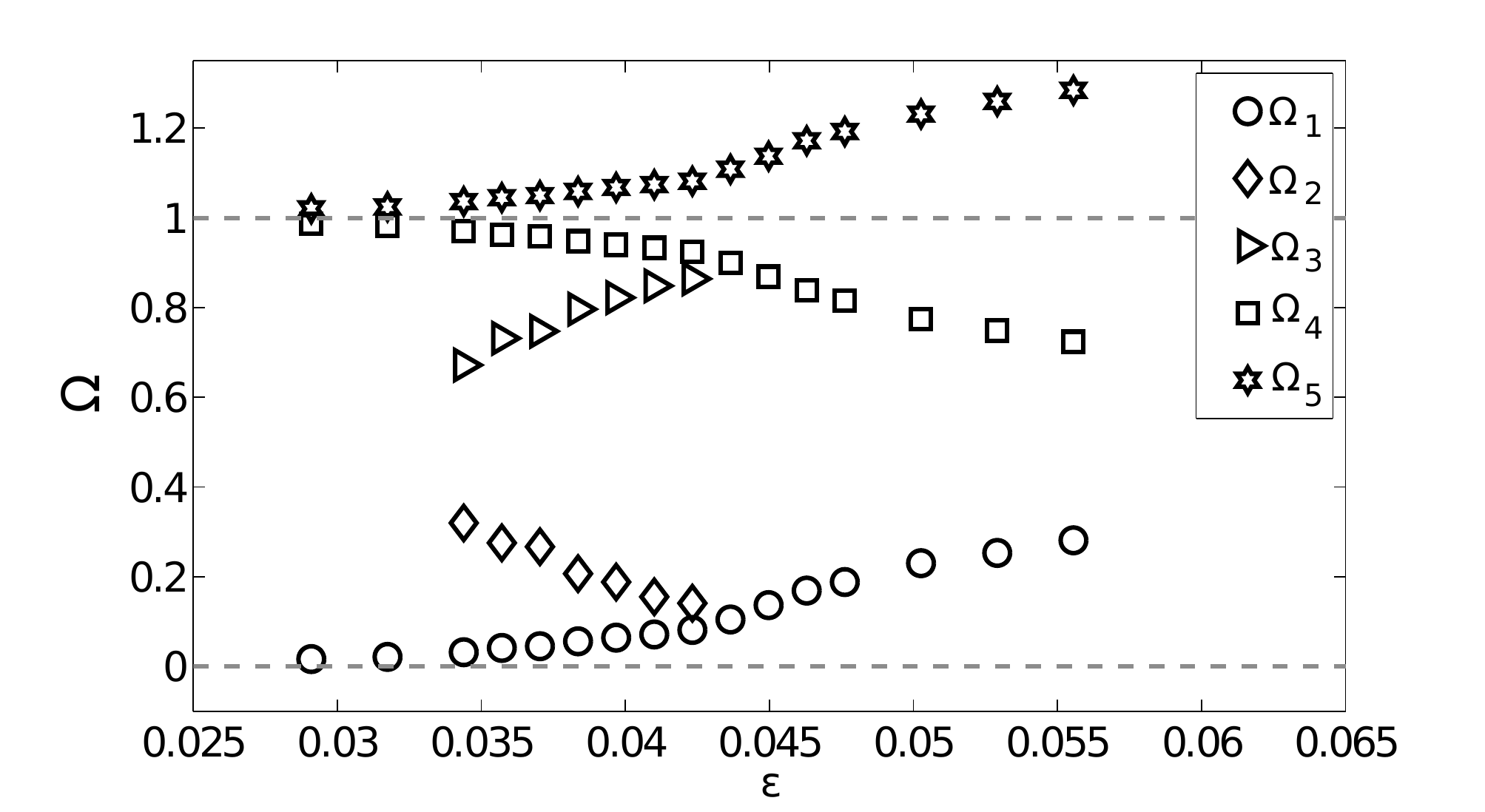}} 
\subfloat[]{\includegraphics[width=0.415\textwidth]{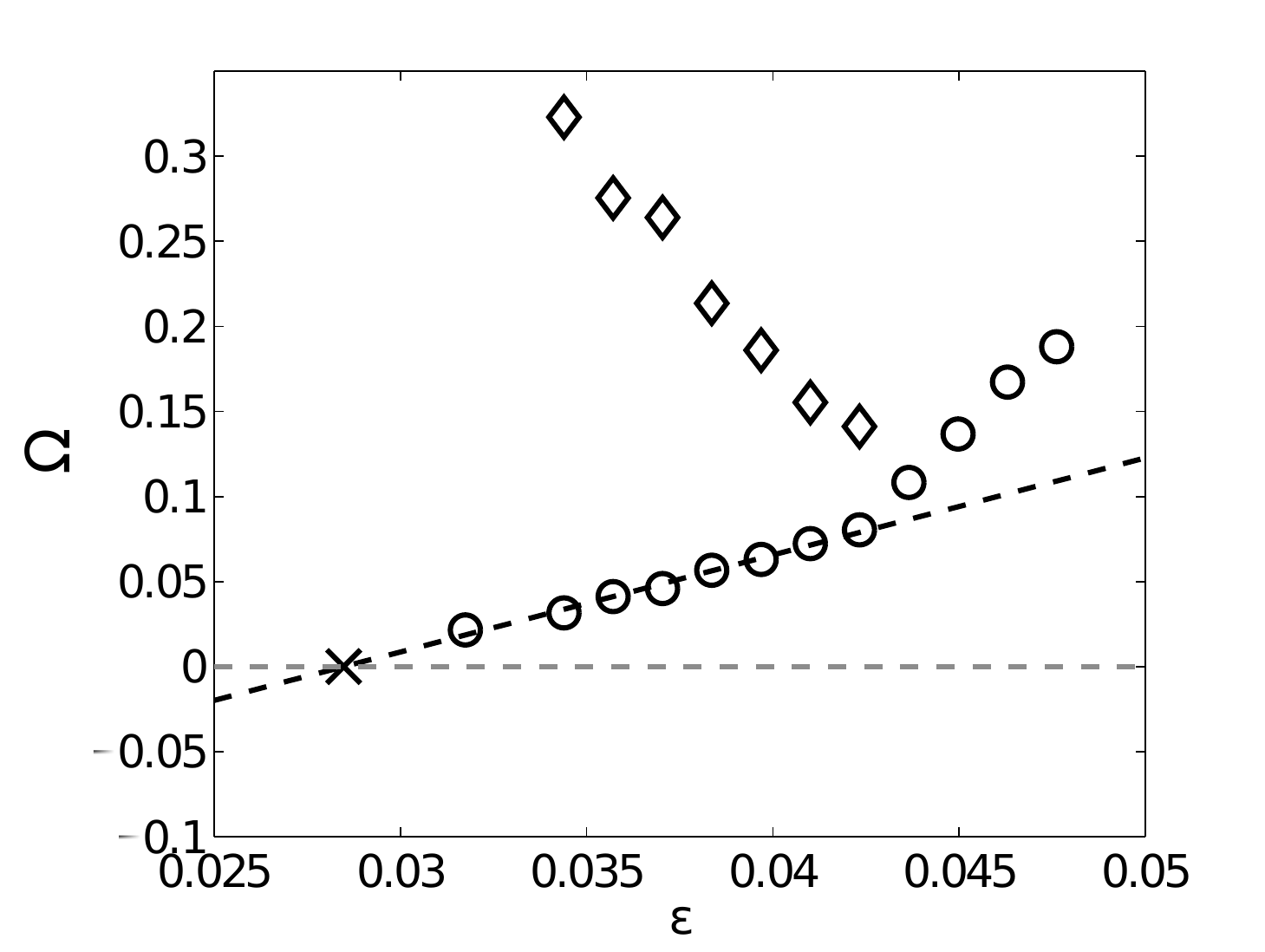}} 
\caption{\label{fig_peaks_freq4} 
    (a) Frequencies as a function of $\epsilon$ for
    ${\rm{Ek}}=1.76\times 10^{-6}$ ($f_{\rm{c}}=4\,{\rm{Hz}}$). The frequencies are sorted in an ascending
    order from the smallest ($\Omega_1$) to the largest ($\Omega_5$).
    (b) Zoom on the two smallest frequencies $\Omega_1$ and $\Omega_2$. Below
    $\epsilon^*=4.3\times 10^{-2}$ the frequency $\Omega_1$ follows a
    linear curve with
    $\Omega_1=C_{\rm{f}}(\epsilon-\epsilon_{\rm{c}})$.  
    The black cross corresponds to the estimated threshold
    $\epsilon_{\rm{c}}=2.85\times 10^{-2}$.}  
\end{figure}

\begin{figure}[b!]
\includegraphics[width=0.49\textwidth]{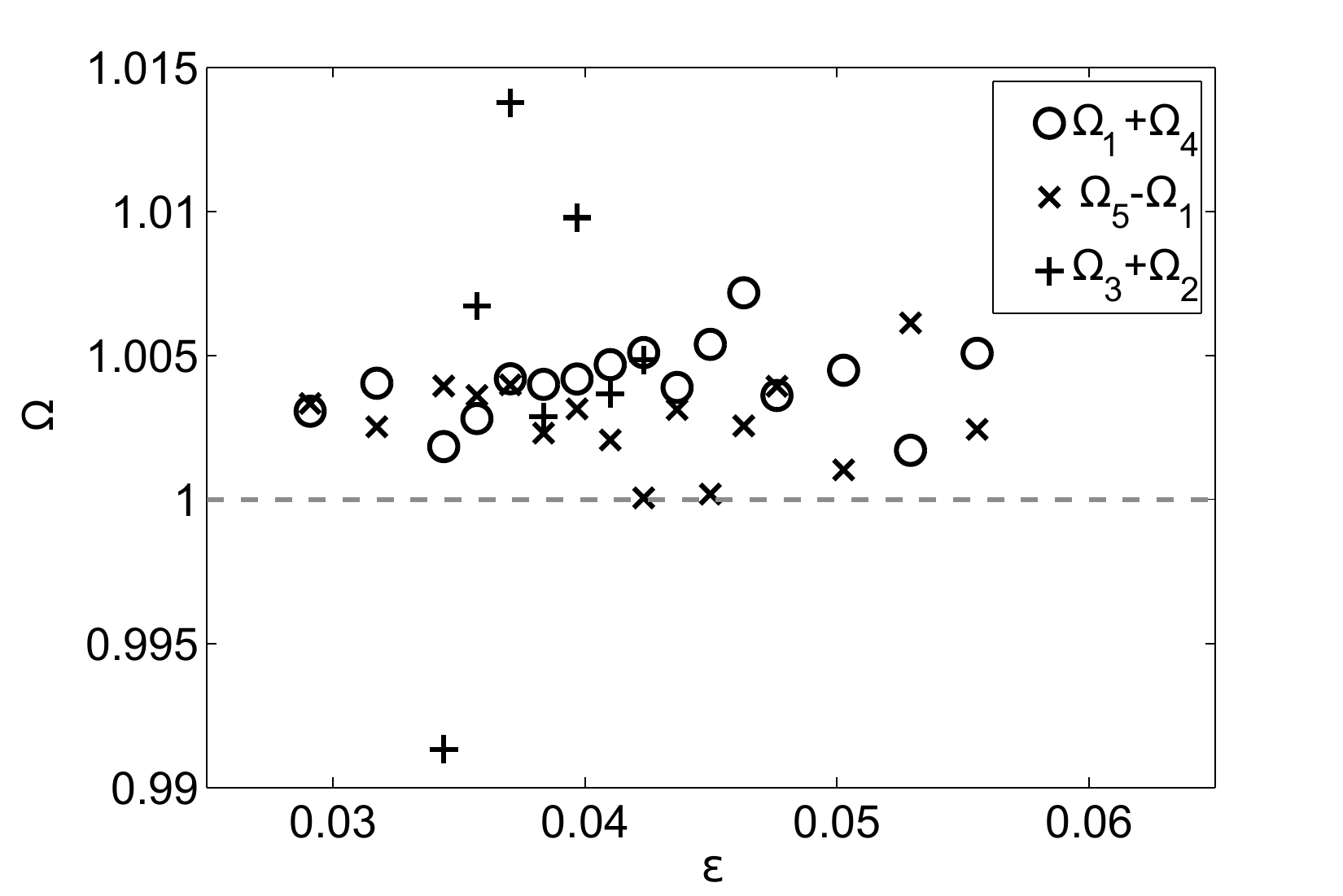}  
\caption{\label{fig_peaks_freq4b}   
The resonance conditions with $\Omega_1+\Omega_4$ (circles), $\Omega_5-\Omega_1$ (crosses)
and $\Omega_2+\Omega_3$ (plus) rescaled by $\Omega_c$.} 
\end{figure}

In total, there are now five frequencies indexed in an ascending order
from $1$ to $5$, that can be grouped into two frequency sets
$({{\Omega}}_1,\Omega_4,\Omega_5 )$ and $(\Omega_2,\Omega_3)$
satisfying (\ref{relation_res}) and (\ref{relation_res2}),
respectively. All frequencies show a systematic dependence on the
precession ratio $\epsilon$ shown in Fig.~\ref{fig_peaks_freq4} which
presents the frequencies, $\Omega_1$ (circles), $\Omega_2$ (diamonds),
$\Omega_3$ (triangles), $\Omega_4$ (squares) and $\Omega_5$ (stars) as
a function of $\epsilon$.  The relations~(\ref{relation_res}) and
~(\ref{relation_res2}) between different pairs of the observed
frequencies and the forcing frequency $\Omega_{\rm{c}}$ persist for
all precession ratios (Fig.~\ref{fig_peaks_freq4b}).

The remaining small differences, mostly smaller than $1\%$, are of the
order of the control accuracy of the rotation rate of the motors. The
individual signals occur only in a limited range of precession ratios
with the frequencies $\Omega_1,\Omega_4$ and $\Omega_5$ (respectively
$\Omega_2$ and $\Omega_3$) appearing at $\epsilon=2.85 \times 10^{-2}$
(resp. $3.34 \times 10^{-2}$) and disappearing at $5.5 \times 10^{-2}$
(resp.  $4.2 \times 10^{-2}$). The systematic study of the
power-spectra shows that the frequencies vary monotonically with
$\epsilon$, i.e. the frequencies $\Omega_1,\Omega_3$ and $\Omega_5$
increase with the precession ratio, whereas $\Omega_2$ and $\Omega_4$
decrease.  We further observe a qualitative change in the behavior of
the frequency variations. Initially, for $\epsilon < \epsilon^{*}
\approx 4.3\times 10^{-2}$, the frequencies $\Omega_1,\Omega_4$ and
$\Omega_5$ exhibit a basically linear variation with $\epsilon$ which,
e.g. for $\Omega_1$ can be described by

\captionsetup[subfigure]{margin=0.0cm,singlelinecheck=false,
                         format=plain,indention=0.0cm,
                         justification=justified,
                         captionskip=0cm,position=top}  

\begin{equation}
{{\Omega}}_1= C_f \left( \epsilon-\epsilon_c \right) \label{fit_f1}
\end{equation}
with $\epsilon_c=2.85 \times 10^{-2}$ and $C_f=5.7$ (see dashed line
in Fig.~\ref{fig_peaks_freq4}b). Similar relations can be given for
${{\Omega}}_4$ and ${{\Omega}}_5$, since ${{\Omega}}_{4,5} \pm
{{\Omega}}_1\simeq 1$.  We interpretate the coefficient $\epsilon_c$
as the threshold of the instability, by assuming that $\Omega_1$
departs from zero at the onset of the instability.  For $\epsilon>
\epsilon ^{*} \simeq 4.3\times 10^{-2} $, $\Omega_1,$ $\Omega_4$ and
$\Omega_5$ abruptly change their slope which may be associated to a
new instability.  Note that the signals corresponding to $\Omega_2$
and $\Omega_3$ approximately vanish close to $\epsilon^*$.  The
evolution of the frequencies suggests that the disappearance of
${{\Omega}}_2$ and $\Omega_3$ occurs when the frequencies $\Omega_1$
and $\Omega_2$ are almost equal at $\epsilon^{*}$. However,
measurements at $f_{\rm{c}}=1\,{\rm{Hz}}$ and
$f_{\rm{c}}=8\,{\rm{Hz}}$ show that the frequency $\Omega_2$ only
disappears slightly after the intersection with $\Omega_1$ (see
e.g. red and green symbols in Fig.~\ref{fig_peaks_freqR} below).


\subsection{Effect of the Ekman number}
\label{sec:nonconstEK}

\captionsetup[subfigure]{margin=0.0cm,singlelinecheck=false,
                         format=plain,indention=0.0cm,
                         justification=justified,captionskip=0cm,
                         position=top} 
\begin{figure}[b!]
\subfloat[]{\includegraphics[width=0.49\textwidth]{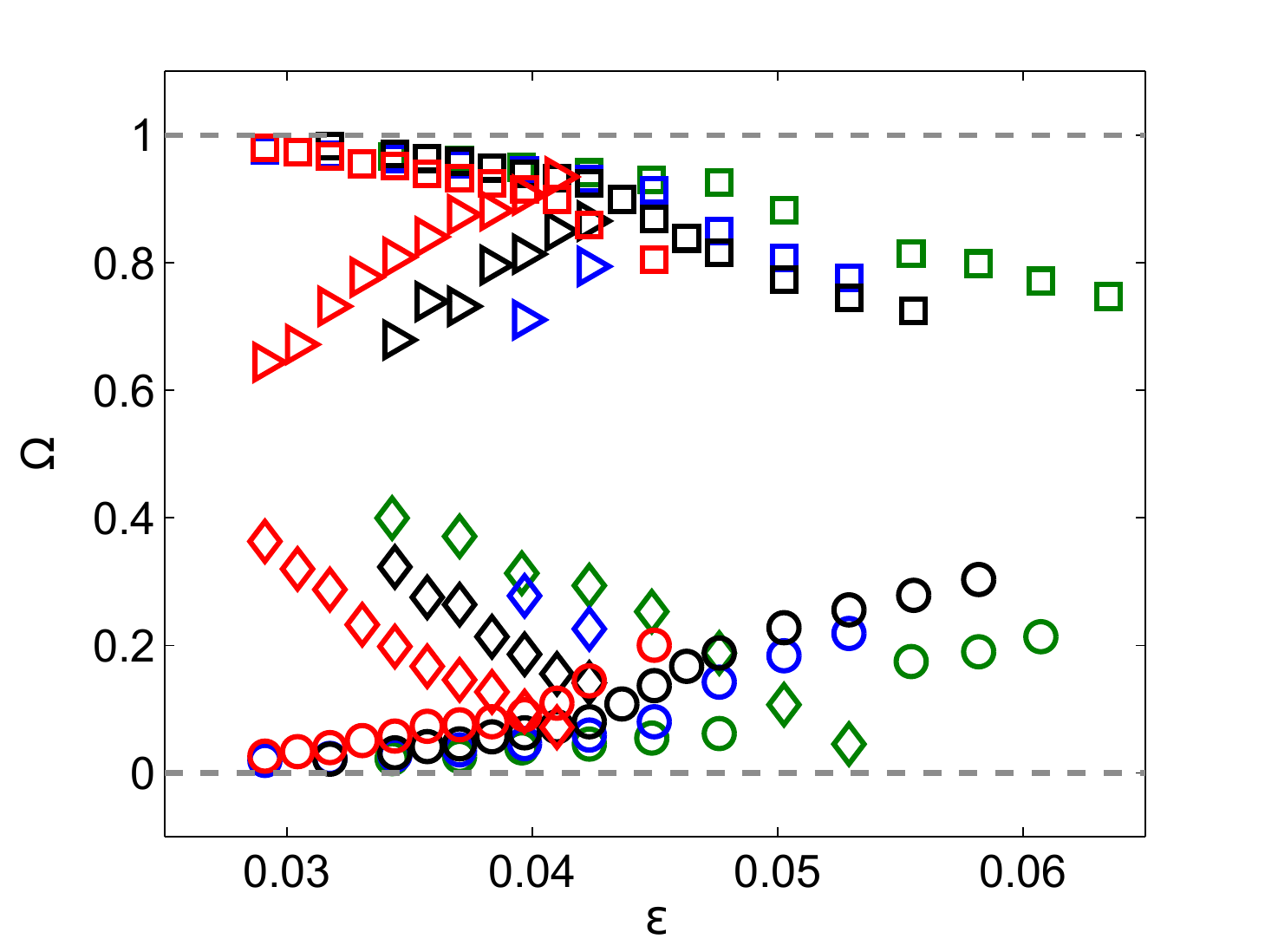}}  
\subfloat[]{\includegraphics[width=0.49\textwidth]{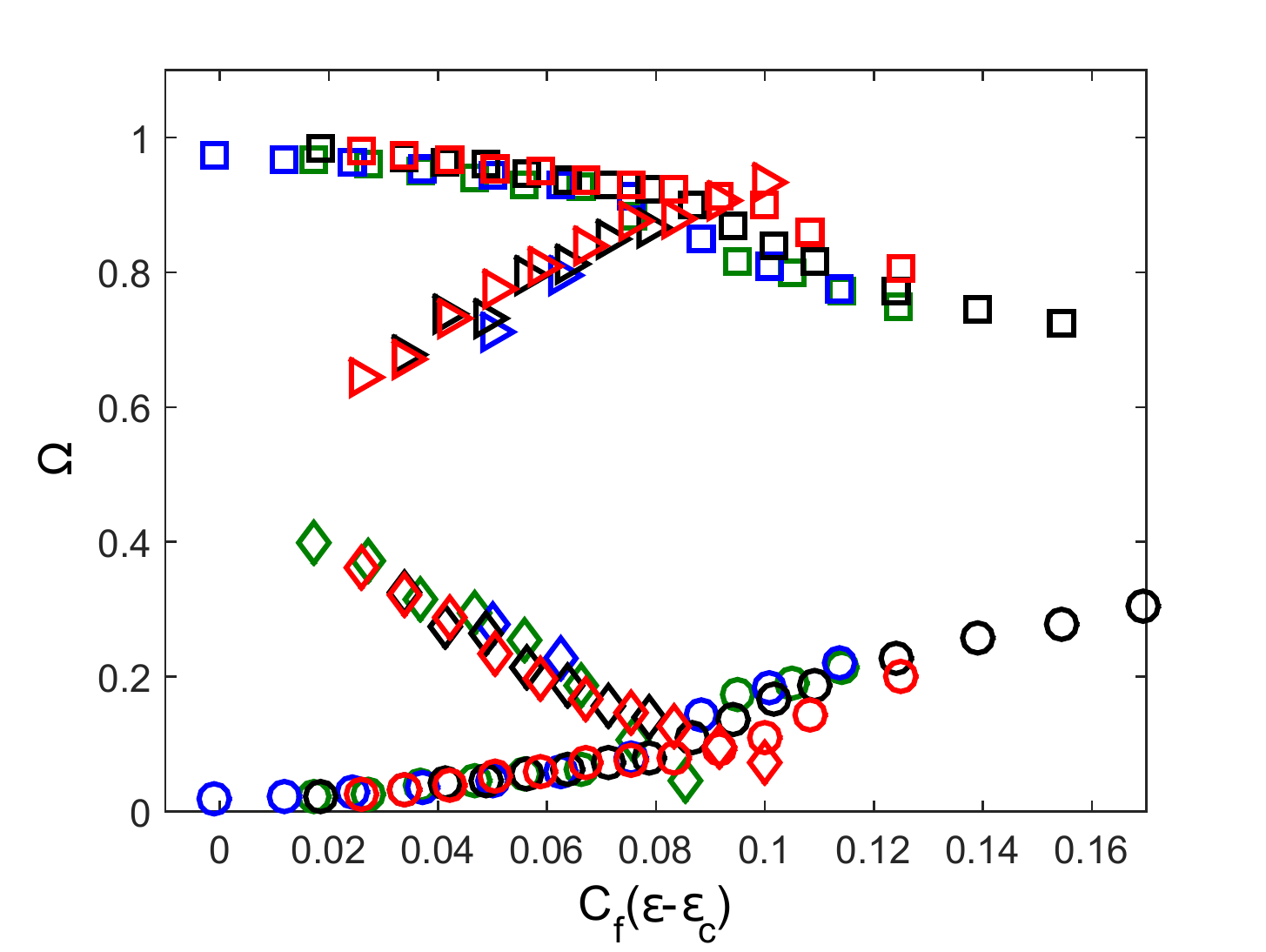}}  
\caption{(a) Evolution of the rescaled frequencies $\Omega_i$ as a
  function of $\epsilon$ for $\Omega_{\rm{c}}=1\,{\rm{Hz}}$ (green),
  $2\,{\rm{Hz}}$ (blue), $4\,{\rm{Hz}}$ (black) and
  $8\,{\rm{Hz}}$ (red). (b) The same frequencies with a rescaled
  abscissa $ C_f(\epsilon-\epsilon_{\rm{c}})$. In both plots $\Omega_1$
  is denoted by circles, $\Omega_2$ by diamonds, $\Omega_3$ by
  triangles, and $\Omega_4$ by squares.  
} 
\label{fig_peaks_freqR}
\end{figure} 

We have performed four series of measurements for different
${\rm{Ek}}=7 \times 10^{-6}$, $3.5 \times 10^{-6}$,
$1.76\times10^{-6}$ and $8.8 \times 10^{-7}$ that correspond to
rotation frequencies $f_{\rm{c}}=1$, $2$ , $4$ and $8\, {\rm{Hz}}$.
The results are shown in Fig.~\ref{fig_peaks_freqR}a with
$f_{\rm{c}}=1\,{\rm{Hz}}$ (green)\footnote{For the largest Ekman
number corresponding to $f_c=1 \mbox{ Hz}$ (green), the
signal-to-noise ratio is too small to systematically detect the peak
corresponding to the frequency $\Omega_3$, which is generally the peak
with the lowest amplitude. Hence, we did not report the frequency
$\Omega_3$ for ${\rm{Ek}}=7 \times 10^{-6}$ (no green triangles in
Fig.~\ref{fig_peaks_freqR}a).}, $2\,{\rm{Hz}}$ (blue), $4\,{\rm{Hz}}$
(black) and $8\,{\rm{Hz}}$ (red). For the sake of clarity, we omit the
frequencies $\Omega_5$.  In all cases the power spectra show a
behavior similar to the results presented in the previous
section. Basically, we find up to five different peaks that globally
exhibit the same tendency, but their thresholds of appearance and
disappearance and their slope vary with the Ekman number.  Generally,
the frequencies appear for lower $\epsilon$, when ${\rm{Ek}}$ is
decreased.  Moreover, the frequencies $\Omega_1$, $\Omega_4$ and
$\Omega_5$ always appear before $\Omega_2$ and $\Omega_3$.

\captionsetup[subfigure]{margin=0.0cm,singlelinecheck=false,
                         format=plain,indention=0.0cm,
                         justification=justified,
                         captionskip=0cm,position=top} 
\begin{figure}[t!]
\subfloat[]{\includegraphics[width=0.49\textwidth]{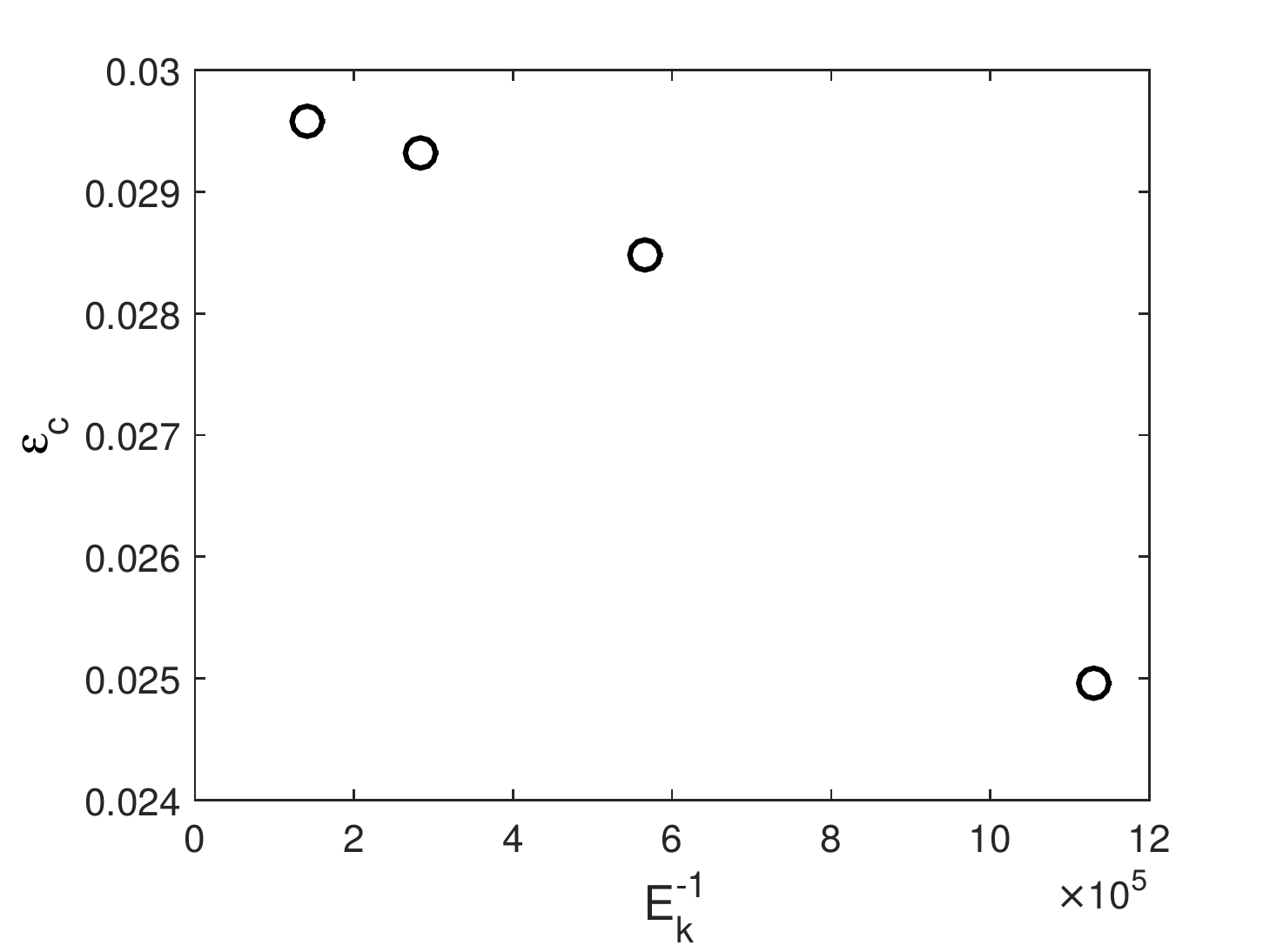}}  
\subfloat[]{\includegraphics[width=0.49\textwidth]{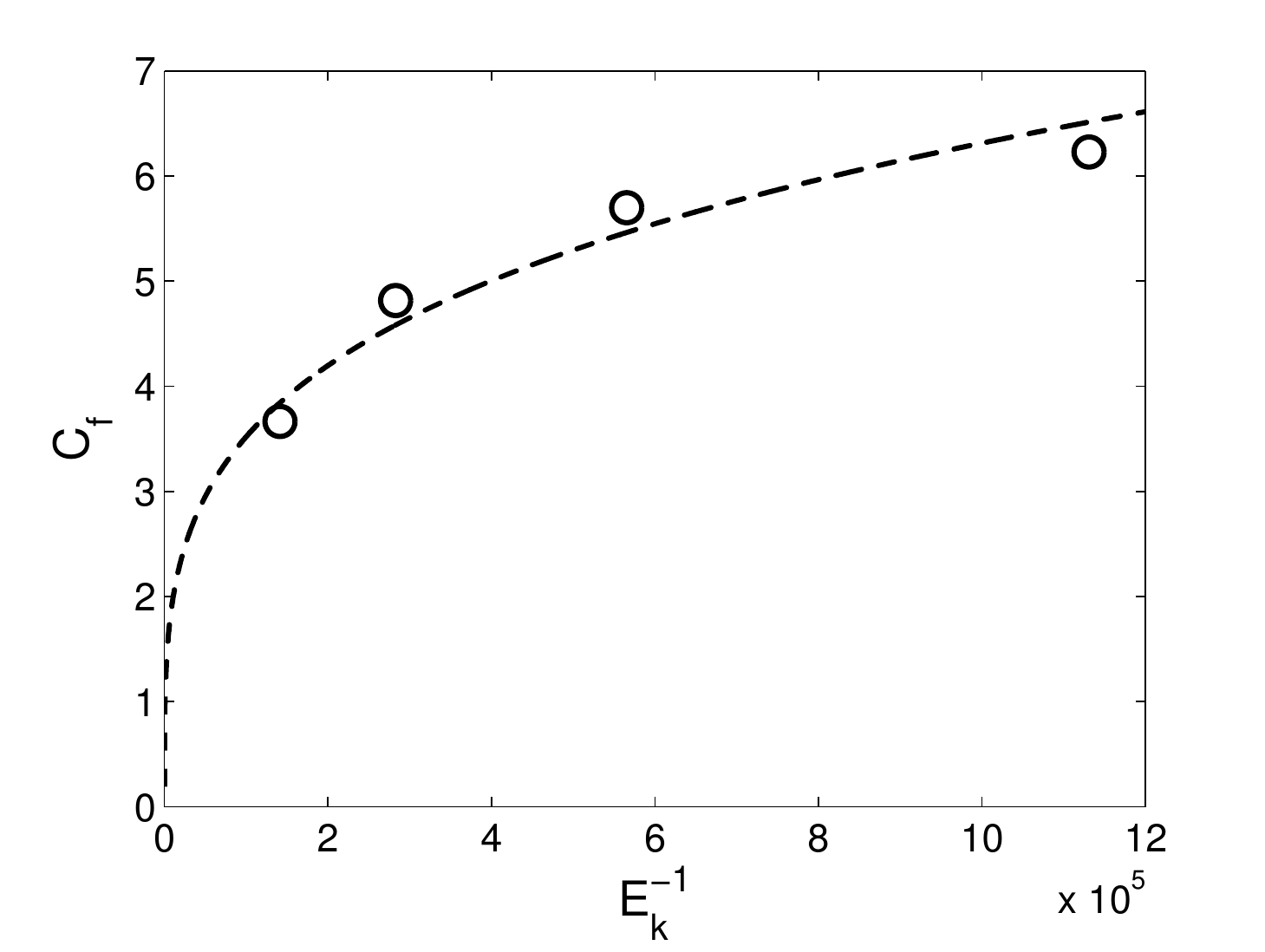}}  
\caption{Evolution of the threshold $\epsilon_{\rm{c}}$  for the
  appearance of the mode $\Omega_1$ (a) and  the 
  coefficient $C_{{f}}$ (b) as a function of the inverse of the 
  Ekman   number ${\rm{Ek}}^{-1}$. The coefficient $C_{{f}}$
  follows a power-law ${\rm{Ek}}^{-1/4}$.} 
\label{fig_coef_thre_exp}
\end{figure} 

In search for a common explanation of the underlying physical
behavior, we change the variables: $({\rm{Ek}},\epsilon) \rightarrow
C_f(\epsilon-\epsilon_c)$ with $C_f({\rm{Ek}})$ and the critical value
$\epsilon_{\rm{c}}({\rm{Ek}})$ for the onset of the instability.  The
frequencies are shown on Fig.~\ref{fig_peaks_freqR}b as a function of
$C_f(\epsilon-\epsilon_{\rm{c}})$ with $C_f$ and $\epsilon_{\rm{c}}$
calculated separately for each Ekman number. In that case, all curves
belonging to the frequencies $\Omega_1, \Omega_2, \Omega_3,$ and
$\Omega_4$, respectively, collapse, which was expected for $\Omega_1$
and $\Omega_4$ because $\Omega_1$ is used for the calculation of
$\epsilon_{\rm{c}}$.  However, it is surprising that the frequencies
$\Omega_2$ and $\Omega_3$ scale in the same way.  This strongly
suggests that the variation of the frequencies is controlled by the
same process, even if the nature of the instability may differ between
both sets (see section~\ref{sec_frist_analyse}
and~\ref{sec_parametric}).  The threshold $\epsilon_{\rm{c}}$
decreases when ${\rm{Ek}}$ decreases, i.e. when $\Omega_{\rm{c}}$
increases, which goes along with the reduction of dissipative effects
(Fig.~\ref{fig_coef_thre_exp}a).  The coefficient $C_f$ increases with
the inverse of ${\rm{Ek}}$ and an empirical fitting indicates a
scaling proportional to ${\rm{Ek}}^{-1/4}$
(Fig.~\ref{fig_coef_thre_exp}b) so that for a constant relative
precession ratio $\epsilon-\epsilon_c$ the detuning becomes stronger
when the dissipation is reduced.  Hence, we rule out the possibility
of viscous detuning, which should increase with ${\rm{Ek}}$, and imply
that the observed modification of all frequencies is provoked by a
change of the base flow.  For a constant relative precession ratio
$\epsilon-\epsilon_c$ we further conclude that the detuning becomes
stronger when the dissipation is reduced, which again rules out
viscous detuning, which is expected to be constant for a given Ekman
number \cite{kerswell1995viscous}.


\subsection{A preliminary analysis of the experimental results}
\label{sec_frist_analyse}

The first step is to identify the modes from the dispersion relation
\cite{lagrange2011precessional,lin2014experimental}.  One cannot
expect a perfect correspondence between the frequencies given by the
dispersion relation (\ref{eq_disp1}) and (\ref{eq_disp11}) and the
experimental frequencies because the experimental ones vary with the
precession ratio $\epsilon$.  Here, we assume that at the threshold of
appearance the frequencies $\Omega_2$ and $\Omega_3$ are close to
exact solutions of the dispersion relation (\ref{eq_disp1}) and
(\ref{eq_disp11}), i.e., the frequencies $\omega_2$ and $\omega_3$ of
Kelvin modes with solid body rotation.  This relies on the hypothesis
that the variation of the frequencies is a consequence of the
non-linear saturation of the instability
\cite{lagrange2011precessional}. Hence, we expect that the smaller the
distance from the threshold, the weaker is the detuning.

Despite the fact that pressure measurements do not give information
about the spatial structure of the modes it is suggestive to associate
the measured frequencies $\Omega_2$ and $\Omega_3$, which at the
threshold have the values $\Omega_2=0.37$ and $\Omega_3=0.64$, to
Kelvin modes with the wave numbers given by $(m_2,n_2,l_2)=(5,1,-1)$
and $(m_3,n_3,l_3)=(6,2,1)$.  According to the dispersion
relation~(\ref{eq_disp1}) and (\ref{eq_disp11}) the corresponding
eigenfrequencies have the values $\omega_{2}=-0.34$ and $\omega_{3}
=0.63$ so that the mode $\omega_2$ is prograde and the mode $\omega_3$
is retrograde.  These solutions have the simplest possible geometric
structure, i.e., the smallest azimuthal and axial wave numbers that
satisfy the spatial resonance conditions (see
equation~(\ref{Econd_resonance}) and~(\ref{Econd_resonance2}) below
and \cite{meunier2008rotating}).  Similar Kelvin modes have also been
identified in experiments conducted by \citet{lagrange2008instability}
even for aspect ratios that do not fullfil the resonance condition of
the forced Kelvin mode $(m,n,l)=(1,1,1)$ and the corresponding triadic
resonance turns out to be the most unstable combination
\cite{lagrange2011precessional}.

In contrast, a parametric resonance of free Kelvin modes with the
frequencies $\Omega_1$ and $\Omega_4$ or $\Omega_1$ and $\Omega_5$ is
not very likely.  This can be seen, if we tentatively assume that the
frequencies $\Omega_1,\Omega_4$ and $\Omega_5$ are associated with
Kelvin modes such that their frequencies at their threshold of
appearance is given by the dispersion relation (\ref{eq_disp1}) and
(\ref{eq_disp11}). The frequency $\Omega_1$ is almost equal to zero
with $\Omega_1 \simeq 0.025$ as the smallest value. Conversely, the
frequencies $\Omega_4$ and $\Omega_5$ are almost equal to $1$, with
$\Omega_4 < 1$ and $\Omega_5 > 1$.  Certainly such solutions are
allowed from the dispersion relation, however, these ``extreme'' cases
go along with specific requirements for the radial wave number
$\delta$ and the axial wave number $k$. In the limit of a vanishing
frequency $\omega\rightarrow 0$ the dispersion relation
(\ref{eq_disp1}) and (\ref{eq_disp11}) immediately yields $\delta\gg
k$, whereas on the other side the limit $\omega\rightarrow 1$ would
come along with $\delta\rightarrow\sqrt{3}k$.  Additionally, the
occurrence of a triadic resonance of two free Kelvin modes with
$\omega\rightarrow 0$ and $\omega\rightarrow 1$ involving the directly
forced mode further constrains the axial wave number $k$ by demanding
$|\Delta k| = |k_1-k_4|=\pi\Gamma$ (which is the axial wave number of
the forced mode), which, if both modes were free Kelvin modes, would
end up with $\delta_1\gg \delta_4$ as a necessary condition for the
radial wave numbers\footnote{
For example, looking for solutions of~(\ref{eq_disp1}) with
$\omega_1=0.025$ results in modes with $\delta_1 \approx 80 k_1$. On
the other side, the limit $\omega_4 \rightarrow 1$ corresponds to
Kelvin modes with $\delta_4 \simeq \sqrt 3 k_4$.  In order to
constitute a triadic resonance, the axial wave numbers of two free
Kelvin modes that are supposed to be in resonance with the forced mode
with $k=\pi \Gamma$ must fulfill $k_4= k_1\pm \pi \Gamma$.  Hence, the
ratio $ \delta_1/\delta_4$ is given by
\begin{equation} \frac{\delta_1}{\delta_4} \simeq \frac{80}{\sqrt{3}}
\frac{n_1 }{(n_1 \pm 1)}
\end{equation} where we switched to an integer wave number
$n_i=k_i/\pi\Gamma$.  If the denominator is $n_1 - 1$ with $n_1>1$,
the ratio ${\delta_1}/{\delta_4}$ will decrease with $n_1$ and will
reach the asymptotic limit ${\delta_1}\simeq
{80}/{\sqrt{3}}{\delta_4}$ for $n_1 \rightarrow \infty$. If the
denominator is $n_1 + 1$ with $n_1 \geq 1$, the ratio will increase
and the minimum will be $ {\delta_1}\simeq {80}/(2\sqrt{3}) {\delta_4}
$, i.e.  ${\delta_1}\simeq 23 {\delta_4}$ so that in both cases
$\delta_1/\delta_4\gg 1$.
}.

Consequently, if the modes $\Omega_1$ and $\Omega_4$ were free Kelvin
modes, they would have significant different radial wave numbers with
$\delta_1 \gg \delta_4$ (The same argument works for the mode
$\Omega_5$). The coupling between Kelvin modes with a large difference
of radial wave number is in principle possible, but the efficiency of
the coupling is very weak, and usually only parametric resonances with
$\delta_1 \simeq \delta_4$ are observed
\cite{kerswell1999secondary,lagrange2008instability}.  Moreover, a
mode with a large radial number is expected to be strongly damped by
viscous effects, inhibiting the instability. Thus, we conclude that
the frequencies $\Omega_1,\Omega_4,$ and $\Omega_5$ are very unlikely
to be the spectral signature of a triadic resonance between a forced
Kelvin mode and two free Kelvin modes.  Instead we propose a new kind
of instability
\cite{lagrange2008instability,lin2014experimental,albrecht2015triadic,herault2015subcritical,giesecke2015triadic}.
where $\Omega_1$ is associated with a geostrophic mode ($k=0$) that
results from the destabilisation of the (azimuthal) background flow.
There are a couple of (experimentally justified) reasons for our
interpretation.  In case of a solid body rotation, a geostrophic mode
must be steady in the cylinder reference frame, i.e. $\omega=0$.
However, when the background flow departs from the solid body
rotation, so that the background vorticity depends on the radius $r$,
geostrophic modes are expected to be unsteady.  The canonic example is
a Rossby wave which has an angular frequency given by $\omega- k U
\propto \beta$ where $\beta$ is the shear amplitude. In the same
manner, the frequency of a non-axisymmetric geostrophic mode in the
cylinder reference frame should increase with the shear amplitude
$\beta$, as shown in Eq.~(35) below.

\citet{kerswell1999secondary} has reported the presence of an
instability with a dominant geostrophic component, which displays some
similarities with the features of the frequency set
$(\Omega_1,\Omega_4,\Omega_5)$. In his model, the modes do not emerge
from a triad-type instability and the instability of the forced mode
involves three further participants: a dominant geostrophic mode and
two sub-dominant Kelvin modes with frequencies that are not solutions
of the dispersion relation.  These modes emerge from non-linear
interactions between the geostrophic mode and the forced mode and
their frequencies result from the linear combination of those two
modes. Hence, the frequencies $\Omega_4$ and $\Omega_5$ belong to
modes that are forced out of resonance, explaining the apparent
mismatch between the observed frequency and their genuine
eigenfrequency.  Furthermore, Kerswell \cite{kerswell1999secondary}
found that this instability emerges before the parametric resonance,
if the resonant triad is imperfectly tuned and the amplitude of the
forced mode is relatively large. Since both conditions are satisfied
in our experiment, this scenario may explain the presence of the
frequency group $(\Omega_1,\Omega_4,\Omega_5)$ before the frequency
set associated with the parametric resonance $(\Omega_2, \Omega_3)$.

An alternative explanation for the low frequency mode $\Omega_1$ could
be a secondary or tertiary instability of the triadic parametric
resonance as they have been found experimentally by
\cite{lagrange2008instability,lagrange2011precessional} and
numerically by \cite{marques2015}.  In these cases the low frequency
mode results from a slow drift in phase space between the forced mode
and the two free Kelvin modes with $m=5$ and $m=6$, thus reflecting a
slow near-heteroclinic cycle.  However, these low frequency modes have
only been found in a narrow region of parameter space so that most
probably we can rule out this interpretation for our measurements,
because we always observe $\Omega_1$ in a wide region of the parameter
space $({\rm{Ek}},\epsilon)$ {\it{before}} (and thus it rules out a
secondary instability) and {\it{after}} the
occurrence of $\Omega_2$ and $\Omega_3$ due to the parametric
resonances. Recently, Lopez and Marques \cite{lopez2018} found a low frequency state for a
wider range of rotation frequencies (keeping ${\rm{Po}}$ fixed).
However, due to their small nutation angle $\alpha=3^{\rm{\circ}}$,
precession and rotation frequency cannot be independently controlled
and the aspect ratrio $\Gamma=1.333$ used in their study entails  
quite different resonance frequencies, making a direct comparison
rather difficult.

Nevertheless, the presence of a geostrophic mode and
the associated instability remains speculative, and further
measurements would be needed to confirm our hypothesis.


\section{Theoretical Model of parametric resonances in rotating flows with shear}\label{sub_sec_motivation_background}

\subsection{Motivation and background flow}
\label{sec_dispersion_relation_shear}

In the following, we will discuss the effect of the frequency detuning
on the parametric resonance between Kelvin modes.  We focus on the
modes with the frequencies $\Omega_2$ and $\Omega_3$ which we
previously associated with Kelvin modes with $(m_2,n_2,l_2)=(5,1,-1)$
and $(m_3,n_3,l_3) = (6,2,1)$, and we aim at explaining the detuning
of $\Omega_2$ and $\Omega_3$ and the disappearance of the resonant
triad in the same framework.  We derive a model that gives the
variation of the frequencies (i.e. the solutions of
Eqs.~(\ref{eq_disp1}) and (\ref{eq_disp11})) with the amplitude of the
modification of the background shear flow.

First, we address the form of the modification of the background flow,
which we fix a priori since our measurements give no direct
information about this secondary flow. We consider an axisymmetric
stationary azimuthal perturbation $\beta U_\beta(r)=\beta r
\Omega_\beta(r)$ in the cylinder frame with $0<\beta<1$, the amplitude
of the secondary flow.  Evidently, this flow cannot satisfy no-slip
boundary conditions at the end-caps. We consider $V_\beta=0$ at the
lateral wall at $r=1$ and we assume that a boundary layer exists
between the end-caps and this flow. We choose a quadratic form for
$\Omega_{\beta}$ given by
\begin{equation}
\Omega_\beta(r)= r^2-1 
\label{perturbation_form}
\end{equation} 
which corresponds to a retrograde flow for $\beta>0$.  This functional
form is the first possible polynomial correction to the rotation rate
because any monomials $r^n$ with $n$ odd must vanish in order to
satisfy the axi-symmetry. It also corresponds to the Chebyshev
polynomial of the first kind $T_2(r)=r^2-1$ for $r\in[0, 1]$.
Interestingly, Meunier \textit{et al.} \cite{meunier2008rotating} have
observed a geostrophic mode (see their Fig. 17a) with a similar
structure for the laminar regime with a forced Kelvin mode with $m=1$
and $l=1$ close its resonance.

\begin{figure}[h!]
\includegraphics[width=0.49\textwidth]{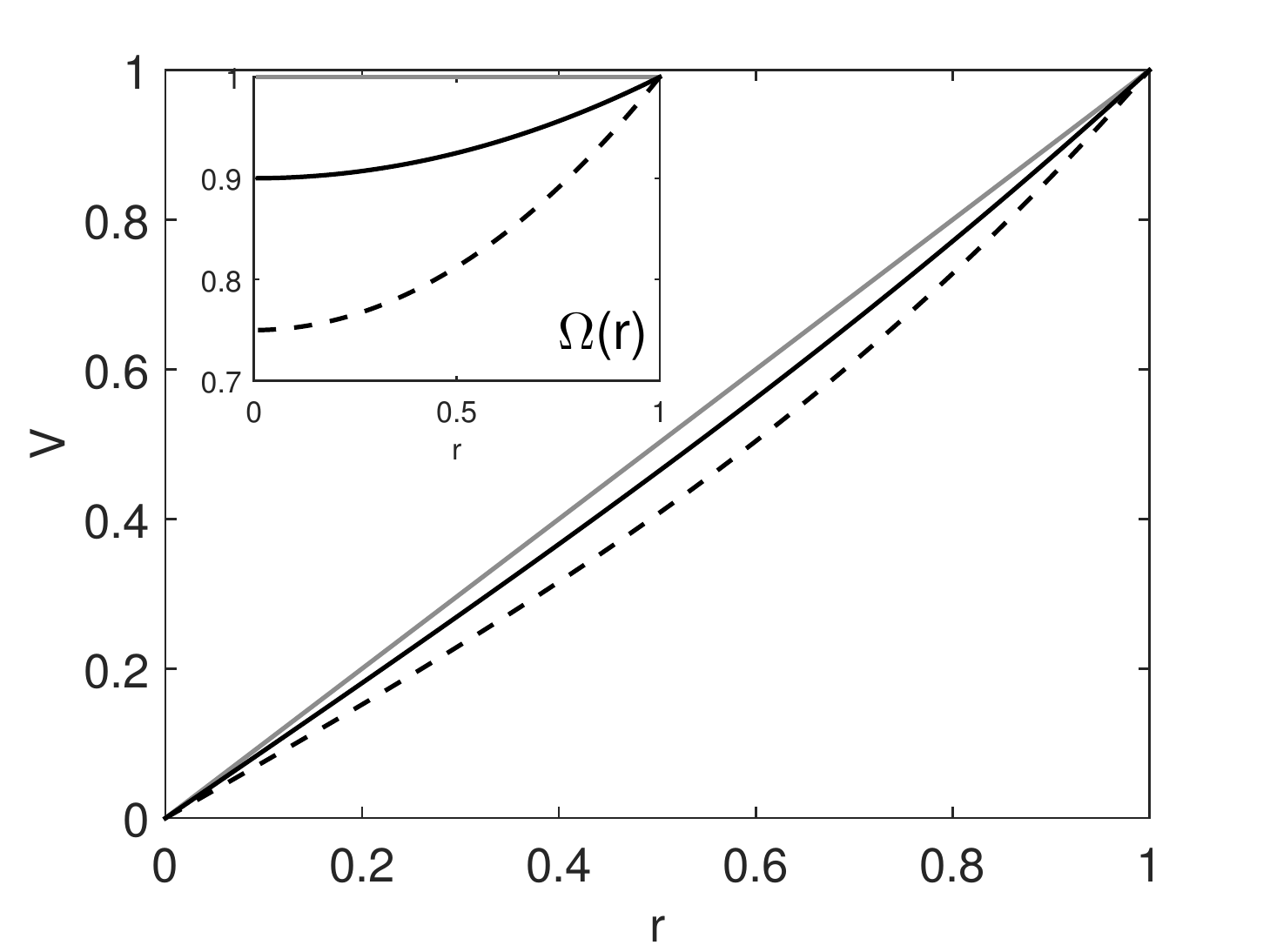}  
\caption{\label{fig_error} Radial dependence of the nondimensional
total azimuthal velocity $V_\varphi=r+\beta
U_{\beta}=(1+\beta\Omega_\beta)r$ with $\beta=0$ (grey line),
$\beta=0.1$ (black curve), $\beta= 0.25$ (dashed curve). The inserted
figure shows the associated angular frequencies $\Omega(r)=1+\beta
\Omega_\beta$.
}     
\label{fig_prof}
\end{figure}

The associated $z$-component of the vorticity is given by
$\zeta_\beta={ {2}} \Omega_\beta +r \partial_r \Omega_\beta = 4r^2-2$.
This flow is stable against centrifugal and shear-induced
instabilities \cite{chandrasekhar1970hydrodynamic}.  Figure
\ref{fig_prof} shows three paradigmatic azimuthal velocity profiles
for $\beta=0$, $0.1$ and $0.25$.
 
\subsection{Dispersion relation of the Kelvin modes}
\label{sub_sec_Dispersion}

Gunn and Aldridge \cite{Gunn1990} have calculated the dispersion
relation of Kelvin modes with a non-uniformly rotating fluid applying
a perturbation method.  In this section, we address this issue with a
spectral decomposition of the solutions of equation~(\ref{lin_eq}) for
small values of $\beta$.  By using the orthogonal base formed by the
solution obtained with $\beta=0$ (section \ref{subsec_kelvin_betaz})
for $m$ and $n$ given, the solutions of equation~(\ref{lin_eq}) are
searched with the ansatz
\begin{equation}
(\tilde{ \textbf  u}, \tilde p)=   \sum_{l=-N/2}^{N/2}  a_l (\tilde{ \textbf  u}, \tilde p)_{0l}, 
\label{kelvin_mode_dec}
\end{equation}
with $a_l=a_{[m,n,l]}$ the amplitude of the Kelvin mode for $\beta=0$
and $(\tilde{ \textbf u}, \tilde p)_{0l}$ taken as solutions of
Eq.~(\ref{eq_uniform}) for a given azimuthal and axial wave number $m$
and $n$. The sum is over the $N$ modes ($N$ being even) composed of
the $N/2$ first retrograde modes and $N/2$ first prograde modes. They
satisfy automatically the mass conservation and the boundary condition
at $r=1$ and $z= \pm \Gamma^{-1}/2$. Equation (\ref{lin_eq}) becomes
\begin{equation}
\sum_l a_l \left[ i \left( \omega-\omega_{0l} \right)
  { {\mathcal{I}}} +
  \beta  {\mathcal{L}}_\beta   \right] (\tilde{ \textbf  u}, \tilde
p)_{0l} =0 \label{eq::32}
\end{equation}
$ \mathcal{L}_0 (\tilde{ \textbf u}, \tilde p)_{0l}= -\omega_{0l}
(\tilde{ \textbf u}, \tilde p)_{0l}$ and ${\mathcal{I}}$ the matrix
defined by equation~(\ref{eq::matI}) in appendix~\ref{app::a1}. In
order to obtain a linear system coupling the coefficients, the
equation is projected on the Kelvin modes $(\tilde{ \textbf u}, \tilde
p)_{0j}$ via the scalar product $\langle (\tilde{ \textbf u},\tilde
p)_{0j}, \cdot \rangle$.  
The resulting decomposition consists of $N$ linear equations given by
\begin{equation}
i \left( \omega-\omega_{0j} \right) e_{jj}  a_j 
- i \beta \sum_{l}  a_{l}
q_{jl} =0 
\end{equation}
with  $q_{jl}= 
i \langle {{(\tilde{ \textbf  u},\tilde p)}}_{0j}, 
{\mathcal{L}}_\beta {{(\tilde{\textbf u}, 
\tilde p)}}_{0{ {l}}} \rangle$ 
and $e_{jj}=\langle(\tilde{ \textbf  u}, 
\tilde p)_{0j},(\tilde{\textbf u}, 
\tilde p)_{0j}\rangle$.  
This system can be written as a general eigenvalue problem    
\begin{equation}
\omega \mathcal{E} \textbf a_l= 
\left( \mathcal{E}\mathcal{D}_0  
+\beta \mathcal{Q} \right)\textbf a_l
\label{eig_pb}
\end{equation}
with $\mathcal{D}_0$ (resp. $ \mathcal{E}$) the diagonal matrix of
elements $\omega_{0j}$ (resp. $e_{jj}$). The elements of the matrices
$ \mathcal{Q}$ are given by $q_{ {jl}}$. The solutions of the equation
(\ref{eig_pb}) are the generalized eigenmodes described by the set of
eigenvectors $\textbf a_l$ and the corresponding eigenfrequencies
$\omega_l$ indexed by $l$.

The modified background flow $\Omega_\beta$ given by
(\ref{perturbation_form}) can be decomposed into a shear component
(the quadratic term) and a solid body rotation (the constant term). If
we only consider the constant term, the frequency detuning would be
$\omega=\omega_0+\beta(m-\omega_0)$ because for this simple case the
matrix $\mathcal{Q}$ is diagonal and equal to $\mathcal{E}
\left(-\mathcal{D}_0+m \mathcal{I}\right)$. Hence, the off-diagonal
terms in $\mathcal{Q}$ are only due to the shear component. It is
worth noting that the dispersion relation for the geostrophic mode
($k=0$) becomes
\begin{equation}
\omega_l \mathcal{E} \textbf a_l=  \beta \mathcal{Q}  \textbf a_l
\label{eig_pbGeo}
\end{equation}
because in that case $\mathcal{D}_0$ is a null matrix. The frequencies
$\omega_l $ are the eigenfrequencies of $\mathcal{E} ^{-1
}\mathcal{Q}$ multiplied by $\beta$. Hence, the shear removes the
frequency degeneracy, i.e. all the frequencies depart from
$\omega_0=0$ and increase linearly with $\beta$.

 \subsection{Effect of shear on forced Kelvin modes}

We now solve Eq.~(\ref{eig_pb}) numerically using the method described in
the appendix \ref{sub_sec_num} including the benchmarking of our
results against the ones obtained by a code based on the Chebyshev
pseudo-spectral method \cite{antkowiak2007vortex}.

\begin{figure}[b!]
\subfloat[]{\includegraphics[width=0.49\textwidth]{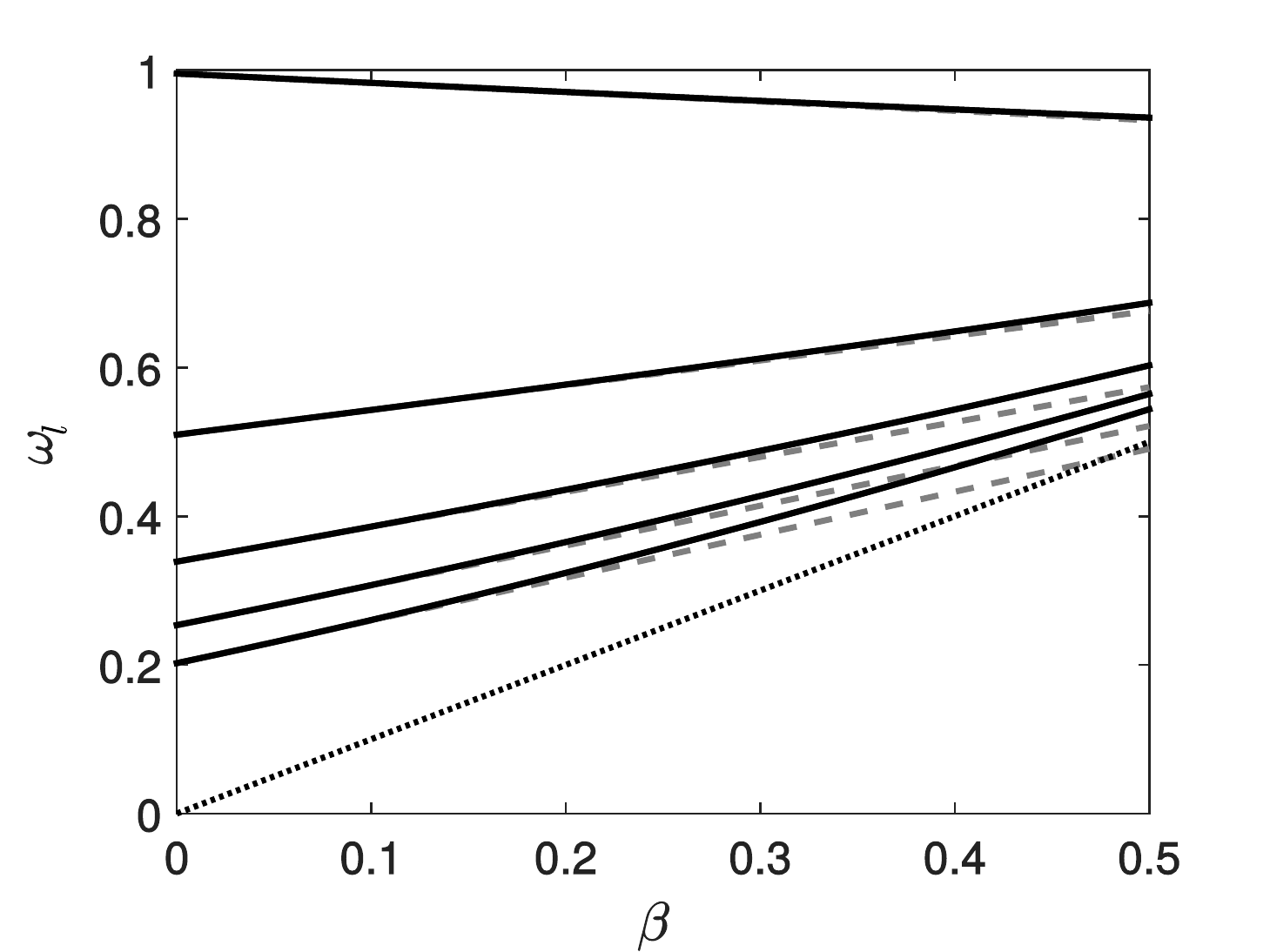}} 
\subfloat[]{\includegraphics[width=0.49\textwidth]{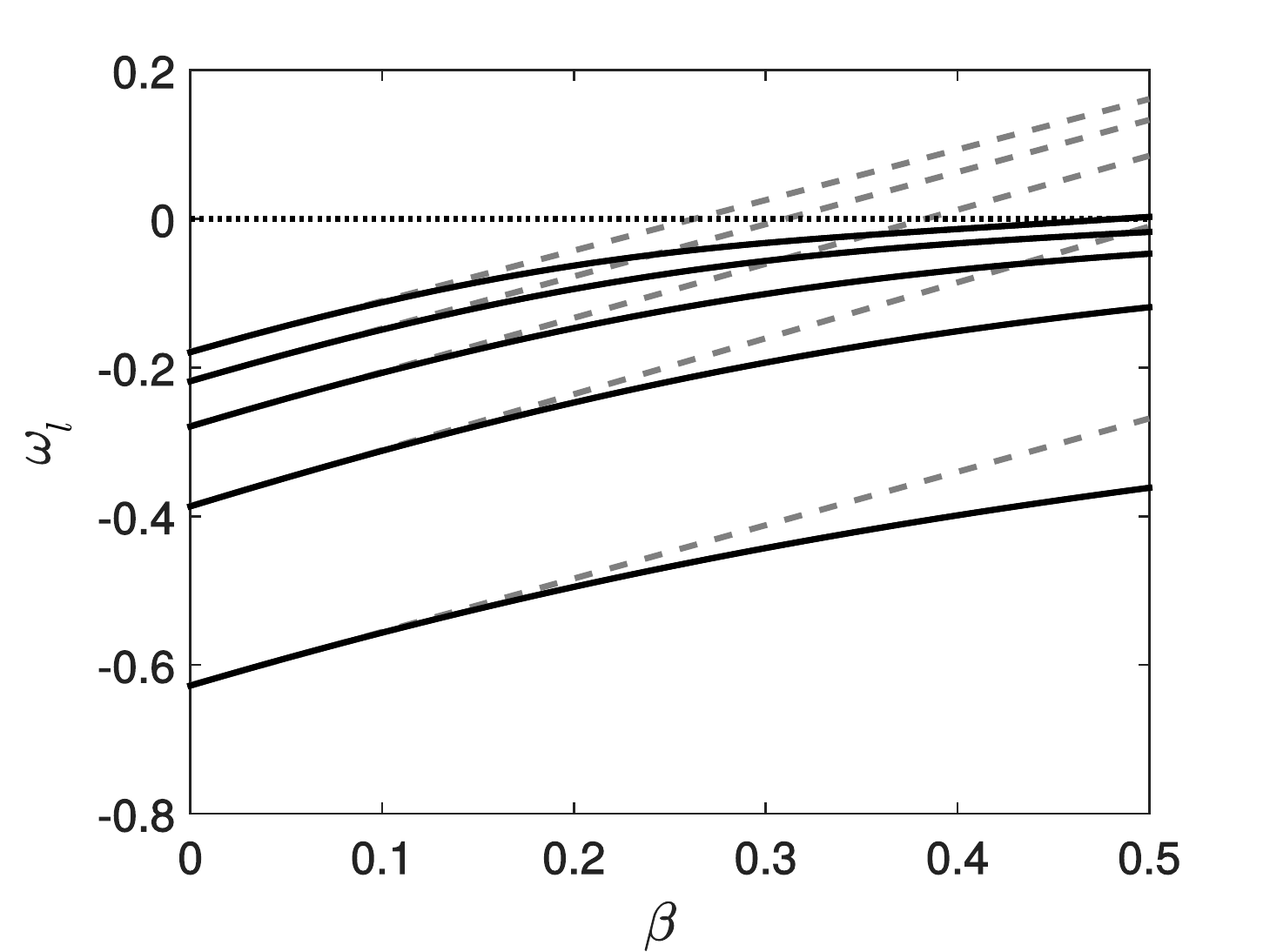}}  
\caption{Evolution of the first five eigenfrequencies $\omega_l$
  (black) of the retrograde modes with $l \in[1;5]$ (a) and prograde modes with
  $l \in[-5;-1]$ (b) with 
  $m=1$ and $n=1$ as a function of the shear amplitude
  $\beta$. Dashed lines correspond to the tangent at the
  origin. The dotted line corresponds to the line, where
  $\omega{ {_l}}= \beta$ for (a)  and $\omega_{ {l}}= 0$ for (b). 
} 
\label{fig_m1_inviscid}
\end{figure}

We start with the frequencies of the Kelvin mode with $(m,n)=(1,1)$
and vary $\beta$ in the range $[0,0.5]$.  We focus on the first five
retrograde and prograde modes ($\vert l \vert<5$).  The evolution of
the frequencies is represented in Fig.~\ref{fig_m1_inviscid} for the
retrograde (a) and prograde (b) modes. For the retrograde
(respectively the prograde) modes, the largest (resp. smallest)
frequencies correspond to the mode with the smallest radial wave
number for a given $\beta$.  Initially, the frequencies increase
linearly as a function of $\beta$ implying that the retrograde modes
accelerate and the prograde modes decelerate.  This effect can be
partially explained by the Doppler effect
$\omega=\omega_0+\beta(m-\omega_0)$ caused by the slowdown of the
solid body rotation.  The factor $(m-\omega_0)$ being positive for the
considered Kelvin modes, yields the increase of the frequency (Doppler
shift).  An exception is the first retrograde mode with $l=1$, which
is only weakly impacted by the modification of the solid body rotation
because $\omega_0\simeq m=1$ (it is almost standing in the turntable
reference frame).  For increasing $\beta$ we observe a deviation of
the linear detuning, which would correspond to the tangents at the
origin represented by the dashed lines in Fig.~\ref{fig_m1_inviscid}.
The departure of the retrograde and prograde modes from the linear
behavior occurs approximately for $\beta > 0.2$.  The initial prograde
mode may even become retrograde as, e.g., the prograde mode with
$l=-5$ for $\beta>0.4$.  For large $\beta$, the frequencies $\omega{
{_l}}$ of the retrograde modes become tangent to $\beta$ (dotted
line).
 
\captionsetup[subfigure]{margin=0.0cm,singlelinecheck=false,
                         format=plain,indention=0.0cm,
                         justification=justified,
                         captionskip=0cm,position=top} 
%

 
\subsection{Effect of shear on  free Kelvin modes}
\label{sec_effect_shear_unst}

\captionsetup[subfigure]{margin=0.0cm,singlelinecheck=false,
                         format=plain,indention=0.0cm,
                         justification=justified,captionskip=0cm,
                         position=top} 
\begin{figure}[t!] 
\centerline{
\subfloat[]{\includegraphics[width=0.49\textwidth]{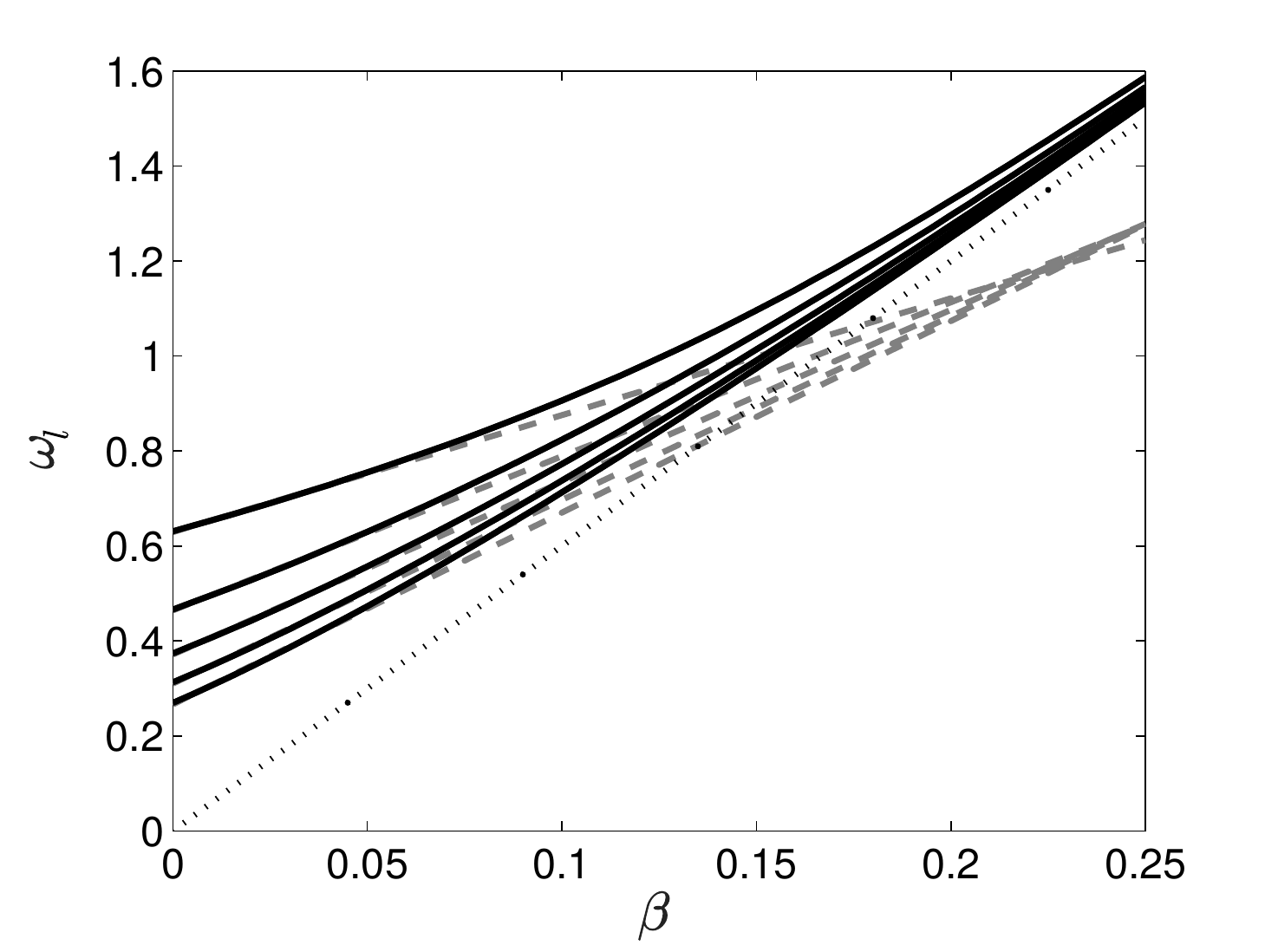}}  
\subfloat[]{\includegraphics[width=0.49\textwidth]{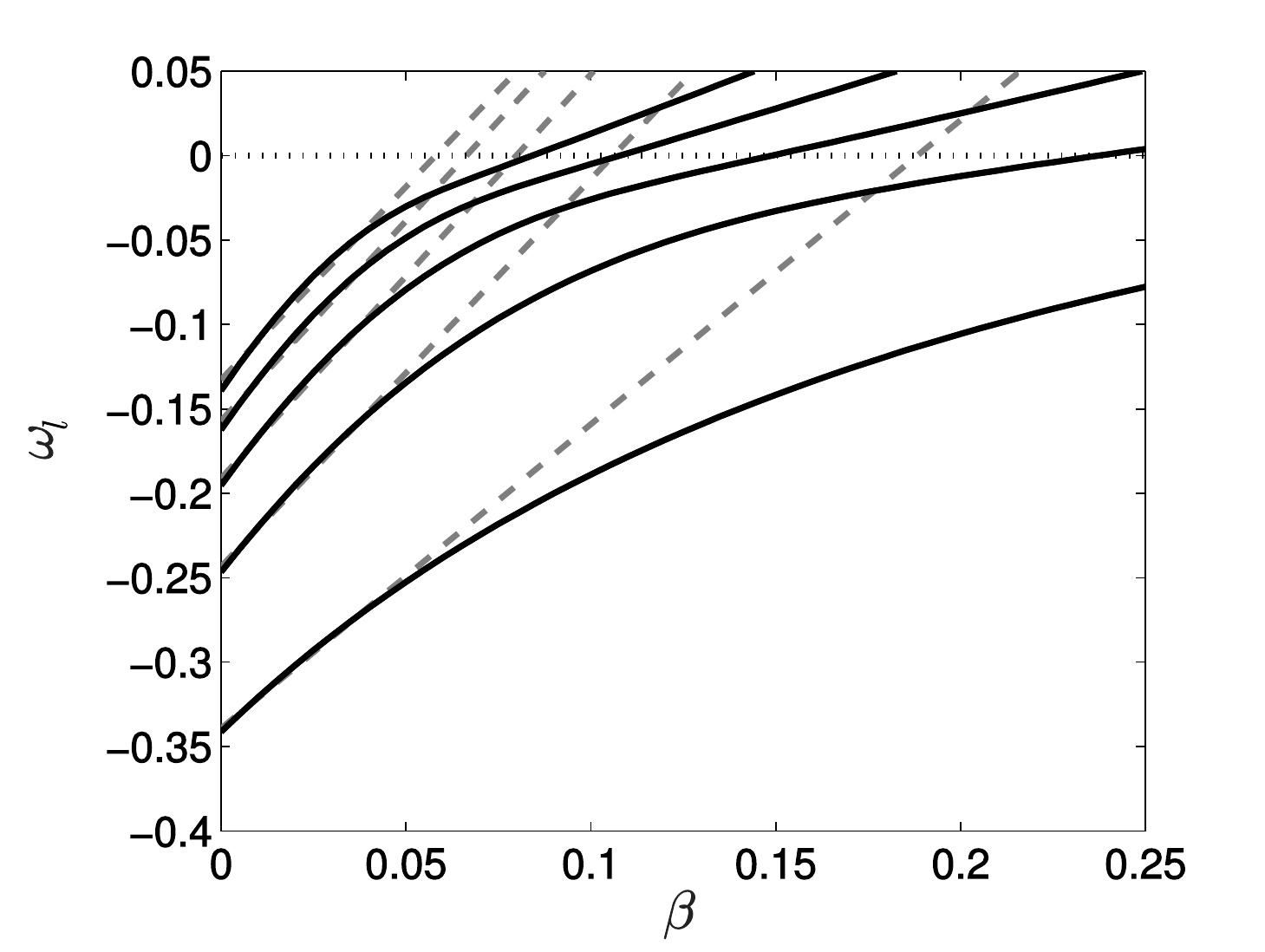}}  
}
\caption{Dependence of the first five eigenfrequencies $\omega_{ {l}}$ 
  (solid black line) of the retrograde modes with $(m,n)=(6,2)$  (a)
  and the prograde  modes with $(m,n)=(5,1)$ on the shear
  amplitude $\beta$. Dashed lines correspond to the tangent at the 
  origin. The dotted line in (a) corresponds to the critical line, 
  where $\omega_{ {l}}=6 \beta$.}  
\label{fig_m5_m6}
\end{figure}

\begin{figure}[t!]  
\centerline{ 
\subfloat[]{\includegraphics[width=0.535\textwidth]{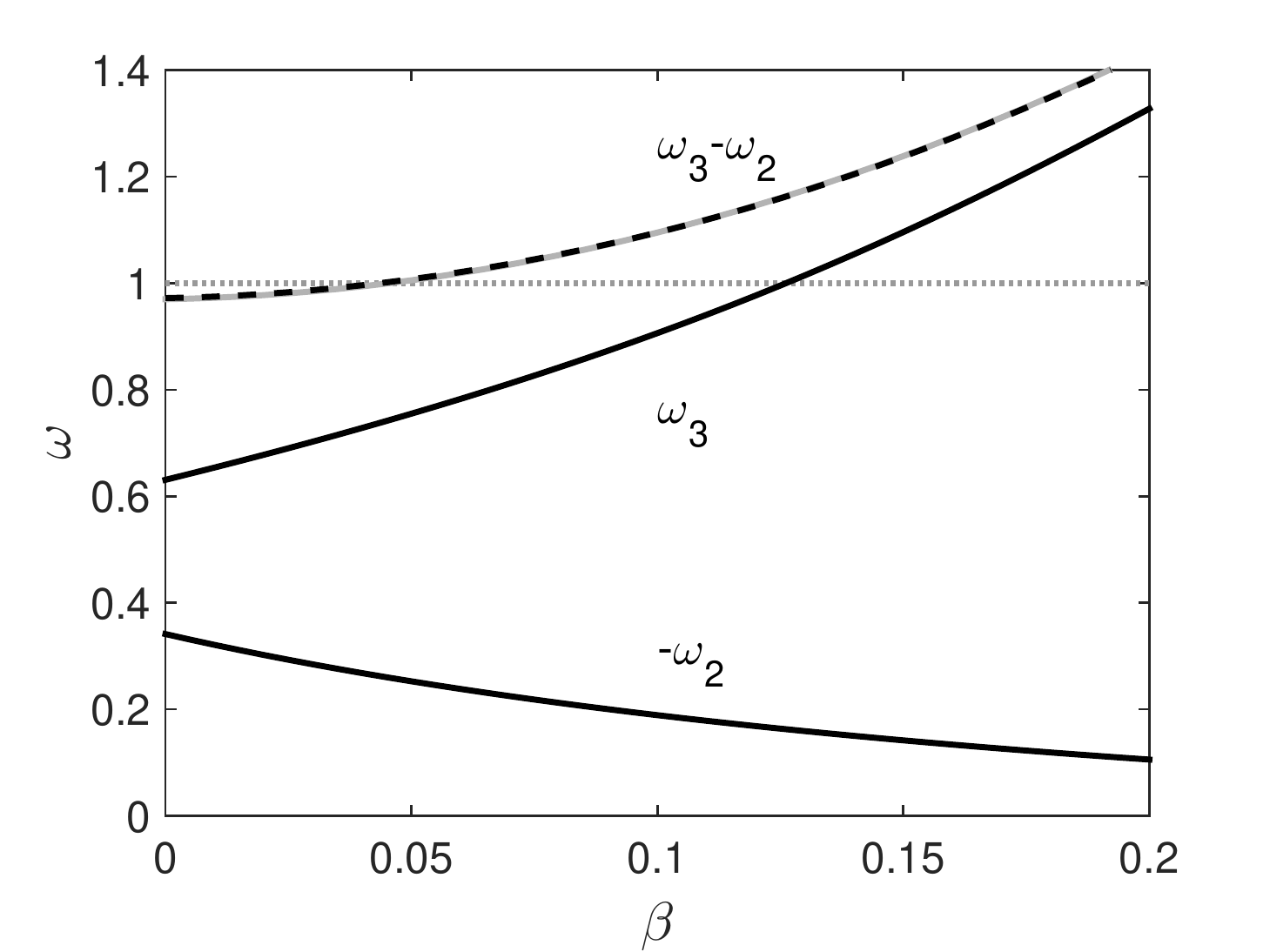}} 
\subfloat[]{\includegraphics[width=0.535\textwidth]{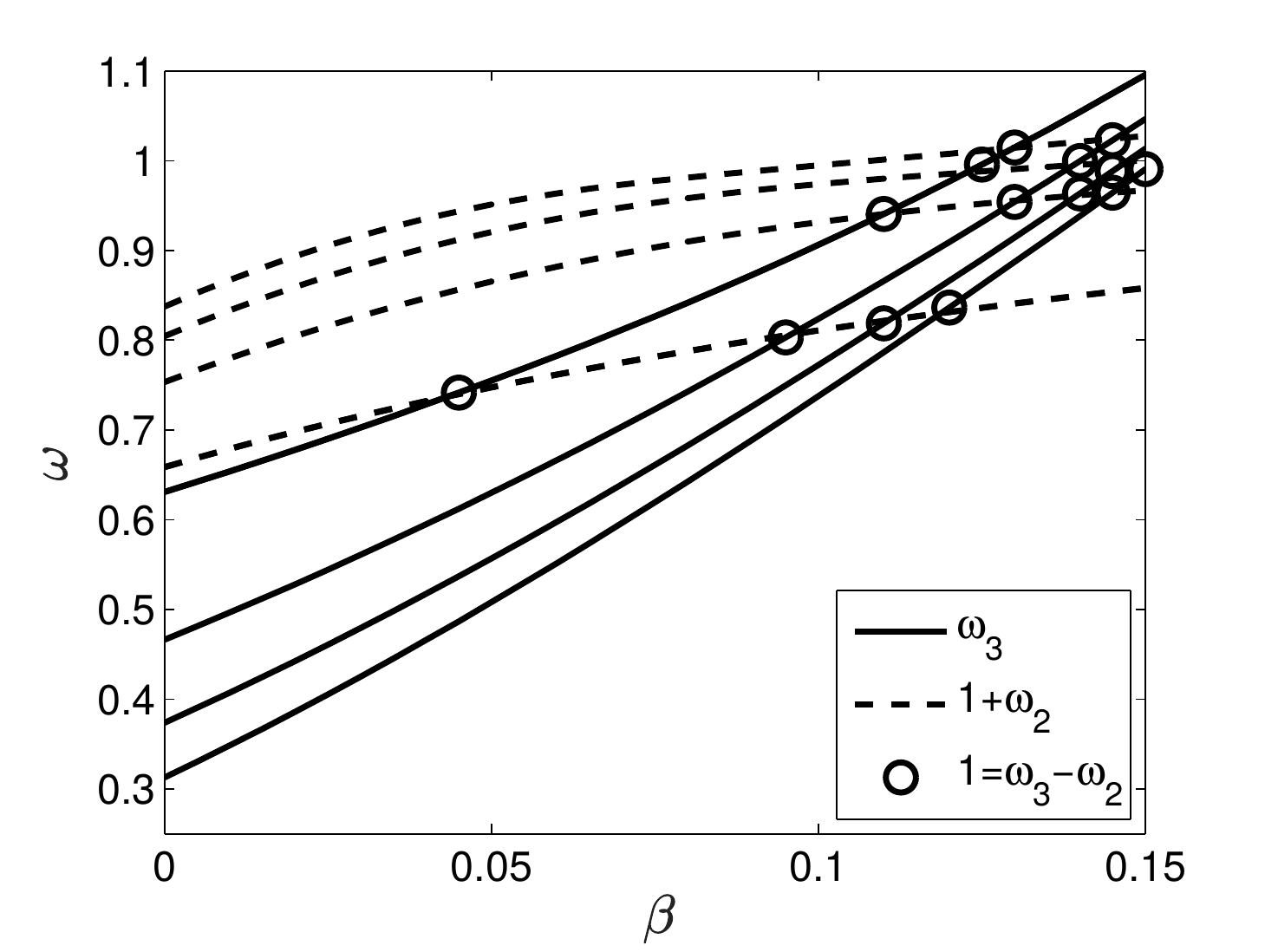}} 
}
\caption{(a) Frequencies $\omega_{3}$ and  $\omega_{2}$ of free 
  Kelvin modes, corresponding to  $(m,n ,l)=(6,2,1)$
  and $(5,1,-1)$,  as a function of $\beta$. The dashed line
  represents the difference $\omega_{3}-\omega_{2}$, such that
  its intersection with the dotted grey line corresponds to an
  exact resonance. The thick grey curve (below the dashed line)
  corresponds to the polynomial expansion of $\omega_{3}-\omega_{2}$
  given by (\ref{eq::polyapprox}). (b) The black curves show to
  the frequencies $\omega_{ {3}}$ of the first four retrograde
  modes with $(m,n)=(6,2)$. The dashed lines correspond to the
  shifted frequencies $\omega_{{2}}+1$ of the first four prograde
  modes with $(m,n)=(5,1)$. The circles indicate the exact resonance
  between the modes, where $\omega_{3}-\omega_{2}=1$.}  
\label{fig_m5_m6_res}
\end{figure}

We focus on the modes $(m,n)=(6,2)$ and $(5,1)$ which we assigned to
the unstable Kelvin modes with the frequencies $\Omega_2$ and
$\Omega_3$ observed in the experiment. We recall that the mode $(6,2)$
is retrograde whereas the mode $(5,1)$ is prograde. The dependence of
the first five retrograde modes with $m=6{ {,n=2}}$ and prograde modes
with $m=5{ {,n=1}}$ on $\beta$ is shown in Fig.~\ref{fig_m5_m6}a and
b.  A visible departure from the tangent at the origin occurs for
$\beta {\ {\gtrsim}} 0.07$ for the retrograde mode~\ref{fig_m5_m6}a,
and for $\beta { {\gtrsim}} 0.05$ for the prograde
mode~\ref{fig_m5_m6}b. Hence, the linear detuning is only valid for
small values of $\beta$. This property is important for the conditions
of the parametric resonance, as we will see below. The frequencies of
the retrograde modes approach $\omega_l=6 \beta$ for large $\beta$.
In the following we concentrate on the behavior of the first prograde
and the first retrograde mode.  Figure~\ref{fig_m5_m6_res}a shows the
absolute value of the frequencies of the first prograde mode
$(5,1,-1)$ (${{\omega_2}}$), the first retrograde mode $(6, 2,1 )$
(${{\omega_3}}$) and their difference $\omega_3-\omega_2$. The
frequencies follow a similar behavior as the observed frequencies
$\Omega2$ and $\Omega3$ in Fig.~\ref{fig_peaks_freq4}.  At $\beta=0$,
the difference is equal to $\omega_{3}-\omega_{2}=0.97$ and the exact
resonance, i.e.  $\omega_{3}-\omega_{2}=1$, occurs at
$\beta=0.044$. In the range $\beta \in [0,0.2]$, the difference $
\omega_{3}-\omega_{2} $ is well approximated by a polynomial expansion
(grey curve in Fig. \ref{fig_m5_m6_res}a)
\begin{equation}
\omega_{3}- \omega_{2} 
\simeq  \delta \omega_{(0)} +\beta \delta \omega_{(1)} 
+ \beta^2\delta \omega_{(2)}+\Delta
\label{eq::polyapprox}
\end{equation}
with $\delta\omega_{(0)}=0.97$ , $\delta \omega_{(1)} =0.172$, $\delta
\omega_{(2)} =10.8$, and a residual $\Delta$ of order $|\Delta| \simeq
3 \times 10^{-3}$.  The quadratic term becomes significant for $\beta
= \delta \omega_{(1)}/\delta \omega_{(2)} = 0.017$, i.e. already
before the exact resonance at $\beta=0.044$.  This expansion shows
that the quadratic detuning is required in the vicinity of the exact
resonance, since the linear detuning underestimates the departure from
the resonance even for small amplitudes of $\beta$.  We point out that
the exact resonant interaction at $\beta=0.044$ of the first prograde
$[(m,n,l)=(5,1,-1)]$ and the first retrograde $[(m,n,l)=(6,2,1)]$
modes with the forced mode occurs well before a sequence of possible
exact resonances between higher radial wave numbers. Figure
\ref{fig_m5_m6_res}b shows the frequencies $\omega_{3}$ (black curves)
and $1+\omega_{2}$ (dashed curves) for the first four retrograde and
prograde modes such that their intersection corresponds to a resonance
(black circles). We see that for $\beta>0.1$, a large number of
resonances occurs between modes with larger radial wave number. These
resonances could trigger instabilities, which would be an efficient
mechanism to transfer energy from large ($m=1$) to small scales like
observed in the resonance collapse. This phenomenon suggests that the
shear of the background flow could play an important role during the
resonant collapse. A similar idea was proposed in \citet{Gunn1990}.

\captionsetup[subfigure]{margin=0.0cm,singlelinecheck=false,format=plain,
                         indention=0.0cm,justification=justified,
                         captionskip=0cm,position=top} 

\section{Effect of the shear on the parametric resonance of Kelvin modes}
\label{sec_parametric}
\subsection{Formulation of the parametric resonance of Kelvin modes  }

In this section, we study the effect of the detuning on the condition
of parametric resonance. We consider a parametric instability of a
single forced Kelvin mode $\gamma \textbf u_0$ of amplitude $\gamma$
and wave numbers $(m,n,l)=(1,1,1)$, via wave-interactions with two
infinitesimal perturbations $( \xi \textbf u^{(a)} ,\xi \textbf
u^{(b)})$ with $\xi \ll 1$ with the azimuthal and axial wave numbers
$(m_a,n_{a})$ and $(m_b,n_{b})$.  According to
\cite{kerswell1999secondary}, we assume that the total velocity field
$\textbf u$ can be decomposed into
\begin{equation}
\textbf u=      \gamma  \textbf u_0(\textbf r,t)+     \xi
\textbf u^{(a)}(\textbf r,t)+\xi \textbf u^{(b)} (\textbf r,t)  
\label{perturbation_velo}
\end{equation}
with 
\begin{equation}
\left\{
\begin{array}{ll}
\textbf u_0&=      \textbf u_{111} (r,z)
e^{i(\varphi+ t) }  +c.c.  \\ \\
\textbf u^{(a)}&=    \sum\limits_j a_{j}(t)  \textbf  u_{m_an_aj} (r,z)
e^{i m_a \varphi }+c.c. \\  \\
\textbf u^{(b)}&=     \sum\limits_l b_{l}(t)  \textbf  u_{m_bn_bl}  (r,z)
e^{i m_b \varphi} +c.c.
\end{array} \right. 
\label{eq::ansatz_triad}
\end{equation} 
The modes $\textbf u^{(a)}$ and $\textbf u^{(b)}$ are decomposed as
sums of Kelvin modes with different radial wave number indexed
respectively by $j$ and $l$, and associated with the complex
amplitudes $a_j$ and $b_{{l}}$. The velocity components
$u_{mnl}(r, z)$ are given by Eq.~(\ref{perturb_form3}), and
satisfy the boundary conditions (\ref{BC2}).  For $\beta=0$, the
amplitude of the modes $a_{j}$ and $b_{l}$ are associated to their
eigenfrequencies $\omega_{0j}^{(a)}$ and $\omega_{0l}^{(b)}$ (see
section \ref{sub_sec_Dispersion}).  Following the approach of Kerswell
\cite{kerswell1999secondary}, we now consider two Kelvin modes
satisfying the spatial resonance conditions so that their axial and
azimuthal wave numbers fulfil the relations
\begin{equation}
\left\{
\begin{array}{ll}
m_a-m_b = & 1  \\ \\
\vert n_{a}- n_{b}\vert =  &  1 \end{array}  \right.
\label{Econd_resonance}
\end{equation}
There is no restriction for the resonance in terms of radial wave
numbers. An exact resonance between two modes indexed by $j$ and $l$
occurs if their frequencies satisfy the temporal resonance condition
\begin{equation}
{{\omega_{0j}^{(a)}-\omega_{0l}^{(b)}=1}}.
\label{Econd_resonance2}
\end{equation}
However, the relations (\ref{Econd_resonance}) and
(\ref{Econd_resonance2}) require particular conditions in terms of a
suitable aspect ratio in order to be fulfilled simultaneously for all
contributing modes. In the following, we assume that the frequency
condition~(\ref{Econd_resonance2}) can be relaxed in order to achieve
near-resonance. The azimuthal wave number and the index of the axial
wave number must be integers (if $n$ is not an integer, it violates
the boundary condition at the end-caps). Thus, only the condition on
the frequency can be changed into a near-resonance condition by
introducing the parameter
\begin{equation}
\Delta \omega_0= \frac{1}{2} 
\left( \omega_{0j}^{(a)}-\omega_{0l}^{(b)}-1 \right).
\label{eq::detuning}
\end{equation}
We now consider the dimensionless inviscid Navier-Stokes equation
expressed in the cylinder reference frame including the non-linear
terms but with $\beta=0$
\begin{equation}
\partial_t \textbf u=-2 \textbf e_z \times \textbf u-\nabla p
+ \textbf u \times \left( \nabla \times \textbf u \right).
\label{eq:euler}
\end{equation}
Inserting ansatz~(\ref{eq::ansatz_triad}) in (\ref{eq:euler}) and
projecting respectively onto the modes $\textbf u^{(a)}$ and $\textbf
u^{(b)}$ we obtain two equations for the amplitudes $a_j$ and $b_l$
given by
\begin{equation} \left\{
\begin{array}{ll}
\dot a_j =i \omega^{(a)}_{0j} a_j+ c_{jl}^{(a)} \gamma b_l e^{it}, 
\\ \\
\dot b_l =i \omega^{(b)}_{0l} b_l+ c_{lj}^{(b)} \gamma^{*} a_j e^{-it}  
\end{array} \right.
\label{eq_parametric1}
\end{equation}
with $c_{jl}^{(a)}$ and $c_{lj}^{(b)}$ denoting coupling terms given
by \cite{kerswell1999secondary}   
\begin{equation}
\left\{
\begin{array}{ll}
c_{jl}^{(a)}= \displaystyle{\frac{1}{e_{jj}^{(a)}} {2
    \pi}{\Gamma} \left(\frac{\omega_{0l}^{(b)}
      -n_{ {b}}}{ \omega_{0l}^{(b)}} \right) \left
    \langle \textbf u_{0j}^{(a)} \cdot \left(   \textbf
      u_{0l}^{(b)}\times \textbf u_0 \right) \right \rangle }, 
\\ \\ 
c_{lj}^{(b)}=  \displaystyle{\frac{1}{e_{ll}^{(b)}}
{2\pi}{\Gamma} \left(\frac{\omega_{0j}^{(a)}
-n_{ {a}}}{\omega_{0j}^{(a)}} \right) 
\left\langle \textbf u_{0l}^{(b)}{} \cdot 
\left(   \textbf u_{0j}^{(a)}\times  \textbf u_0 {}^{*}\right) \right \rangle. }   
\end{array} \right. 
\label{eq_coupling_term1}
\end{equation}
where $\textbf{u}^{(a)}_{0j}$ (respectively $\textbf{u}^{(b)}_{0l}$)
is the unperturbed eigenfunction given by (\ref{perturb_form3}) and
(\ref{kelvin_mode0}) with the indices $j$ (respectively l)
representing the $j$th (respectively $l$th) solution of the dispersion
relation and the indizes $m$ and $n$ being supressed for the sake of
clarity (they have to be fixed a priori).  Furthermore, we applied the
relation between velocity and vorticity for inertial waves, $\nabla
\times \textbf u=\left( 2 \pi n \Gamma/\omega \right) \textbf u$
\cite{kerswell1999secondary}.  The system (\ref{eq_parametric1})
reduces to a linear system with constant coefficients via the change
of variables $(a_j,b_l)=(\tilde a_j e^{it/2}, \tilde b_l e^{-it/2})e^{
i \sigma t}$ with $(\tilde a_j , \tilde b_l )$ two constants and
$\sigma$ the eigenvalue of system~(\ref{eq_parametric1}), with
$\sigma_r$ its real component and $\sigma_i$ the imaginary component.
After some algebra, we obtain the solution
\begin{equation}
\sigma= \frac{\omega_{0j}^{(a)}+\omega_{0l}^{(b)}}{2} \pm 
\sqrt{(\Delta \omega_0)^2-C \vert \gamma \vert^2}
\label{eq_sig_para}
\end{equation}
with the product $C=c_{lj}^{(a)} c_{jl}^{(b)}$   given by  
\begin{equation}
C=  \displaystyle{\frac{4 \left(\pi \Gamma \right)^2}{e_{{ {jj}}}^{(a)} e_{ {ll}}^{(b)}}  
 \frac{ \left(\omega^{(b)}_{0l} -n_{ {b}}+2 \Delta \omega_0
   \right)\left(\omega^{(b)}_{0l}-n_{ {b}}  \right)}{ -\omega_{0l}^{(b)} 
   \omega_{0j}^{(a)}   }}  \times \left \vert \left \langle \textbf u^{(a)}_{0j}
   \cdot \left(   \textbf u^{(b)}_{0l} \times \textbf u_0 \right) \right
 \rangle \right \vert^2. 
 \label{eq_coupling_term}
\end{equation}

Note that for perfectly tuned resonance the relations
(\ref{eq_coupling_term1}) and (\ref{eq_coupling_term}) giving the
growthrate of the instability are equivalent to Eq.~(5.5) and
Eq.~(5.6) in \citet{kerswell1999secondary}.  If $C$ is negative,
the imaginary part of $\sigma$ always vanishes and the solution
remains stable. Hence, the minimum requirement for the instability is
$C > 0$. For an exact resonance, i.e. $\Delta \omega_0=0$, the
coefficient $C$ is positive if the product $\omega_{0l}^{(b)}
\omega_{0j}^{(a)}$ is negative, which is always fulfilled when one
mode is prograde ($\omega_{0l}^{(b)}<0$) and the other one is
retrograde ($\omega_{0j}^{(a)}>0$).  In that case, each mode has a
positive feedback on the other mode, thus enforcing the instability.
The growth rate is then given by $\sigma_i=\pm \sqrt{\vert C \vert}
\vert \gamma \vert$, i.e. the flow is unstable for all $\gamma$ in the
inviscid case. Out of resonance $\Delta \omega_0 \neq 0$, the growth
rate becomes imaginary if $\Delta \omega_0 ^2-C \vert \gamma \vert^2$
becomes negative. This requires that the coefficient $C$ must be
positive, so that
\begin{equation}
\left\{
\begin{array}{lll}
2\Delta \omega_0 >-(\omega_{0l}^{(b)}-n_b) & 
\hbox{if} & (\omega_{0l}^{(b)}-n_b) >0 
\\ \\
2\Delta \omega_0 <-(\omega_{0l}^{(b)}-n_b) & 
\hbox{if} & (\omega_{0l}^{(b)}-n_b) <0.
\end{array}\right.
\end{equation}  
    
The parametric resonance occurs if the eigenvalue $\sigma$ has a
non-vanishing imaginary contribution, i.e. when the amplitude of
$\vert\gamma\vert$ is larger than $\gamma_c=\vert
\Delta\omega_0\vert/\sqrt{\vert C\vert}$.

We now consider the mode $a$ given by $(m_a,k_{a},l)=(6,2,1)$ and the
mode $b$ given by $(m_b,k_{b},j)=(5,1,-1)$ where $\Delta \omega_0=-1.5
\times 10^{-2}$.  The frequencies of the modes are given by
$\omega^{(a)}=\sigma_r^{+}+0.5$ and $\omega^{(b)}= \sigma_r^{-}-0.5$,
while the growthrate is $- \sigma_i$: the mode is stable if $\sigma_i
\geq 0$ and it is unstable otherwise.

Initially, the frequencies are given by
$\omega^{(a)}=\omega_{0j}^{(a)} $ and $\omega^{(b)}=\omega_{0l}^{(b)}$
for $\gamma=0$.

\begin{figure}[t!]
\centerline{
\subfloat[]{\includegraphics[width=0.49\textwidth]{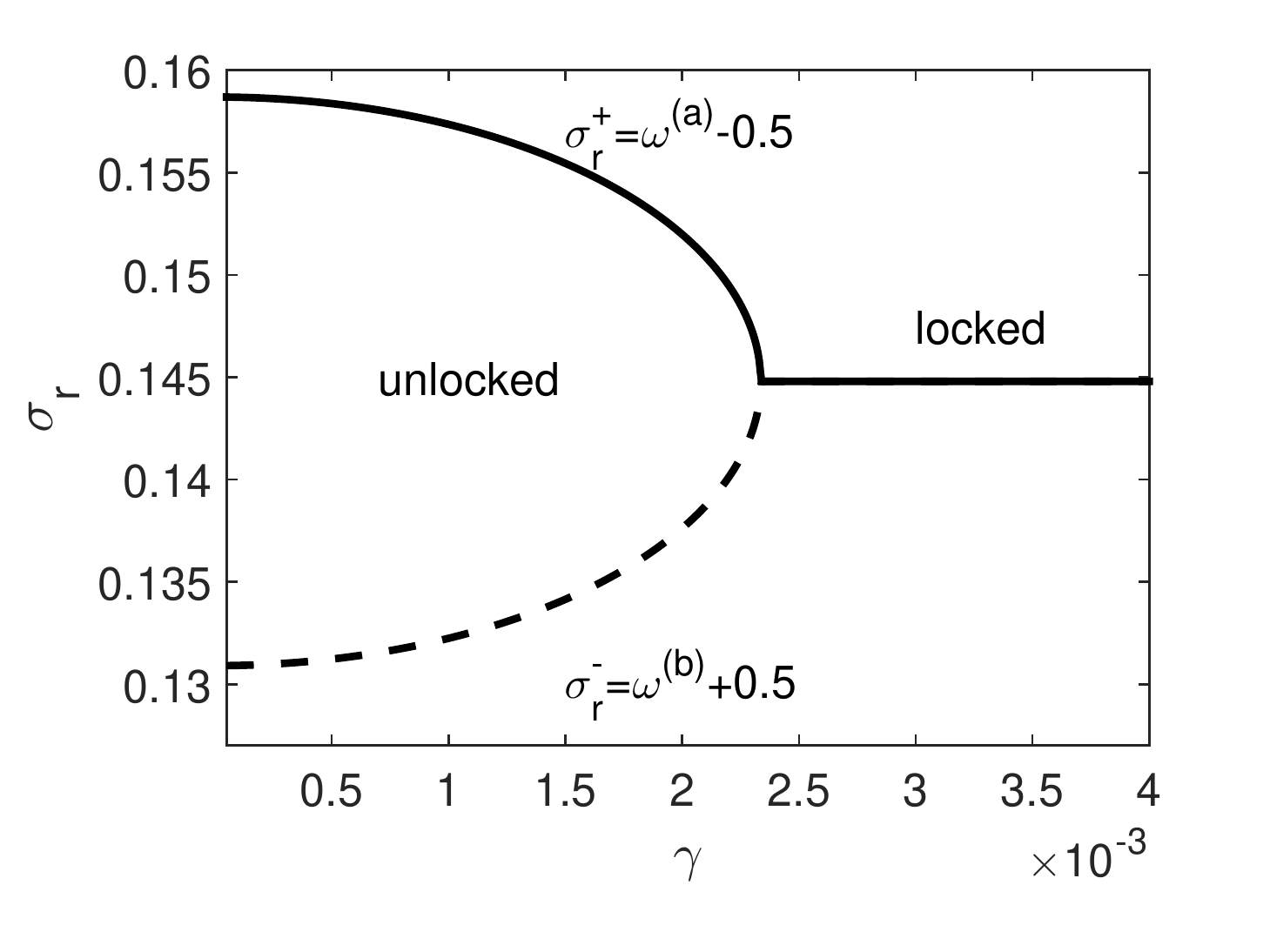}} 
\subfloat[]{\includegraphics[width=0.49\textwidth]{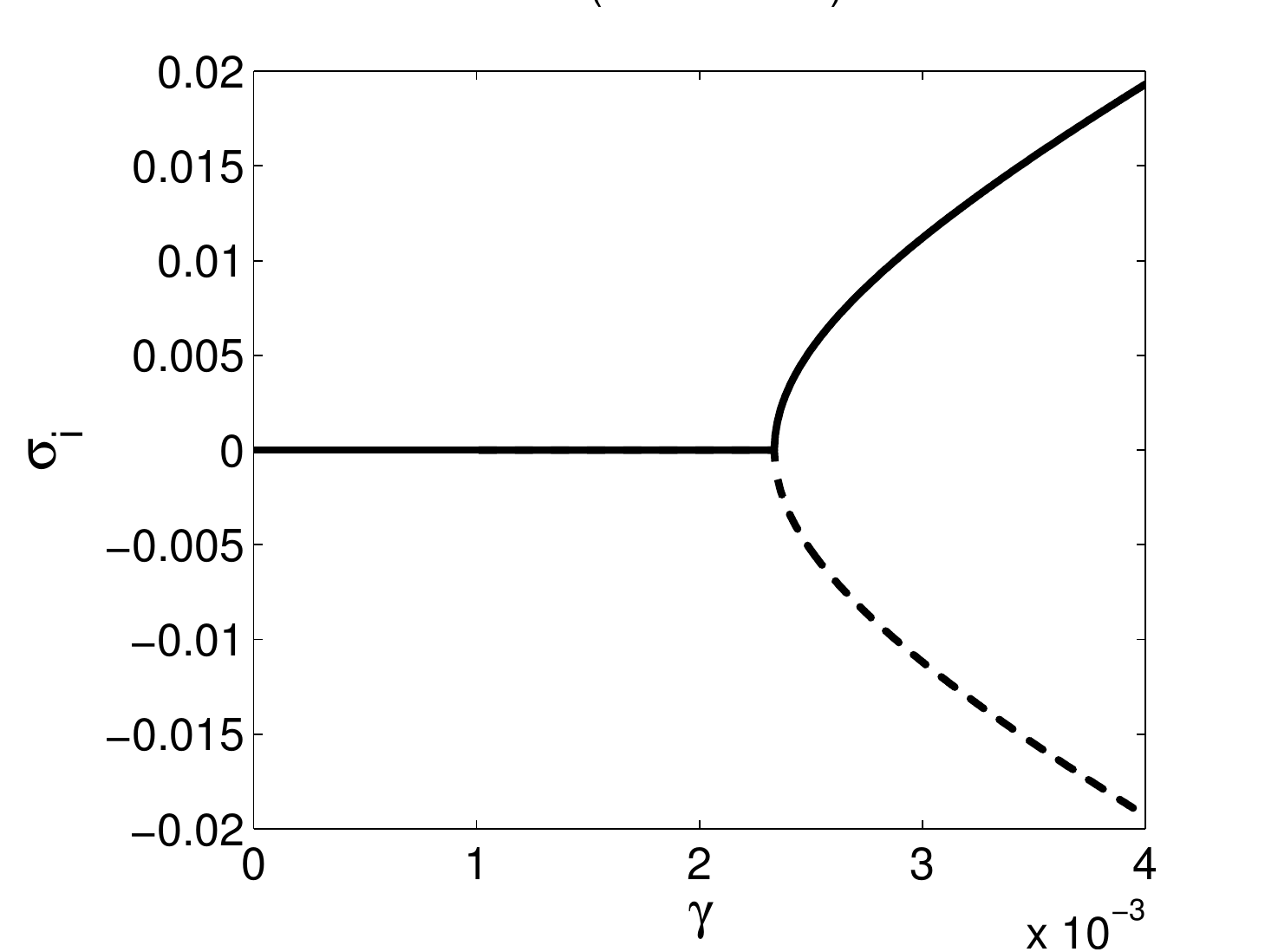}}  
}
\caption{Frequencies $\sigma_r$ (a) and growth-rates $\sigma_i$ (b) of
  the eigenmodes given by the equation~(\ref{eq_sig_para}).   
\label{fig_parametric_sans_shear}
}
\end{figure}

The behavior of the eigenvalues $\sigma$ in dependence of the
amplitude $\gamma$ of the unstable mode is shown in
figure~\ref{fig_parametric_sans_shear}a and b.  Below the critical
amplitude $\gamma_c=2.3 \times 10^{-3}$, the growth rate $\sigma_i$
vanishes and we have two solutions with frequencies such that the
resonance condition $\sigma_r^+-\sigma_r^- =
\omega^{(a)}-\omega^{(b)}-1$ is not equal to zero. Both frequencies
converge, eventually merging at the onset of the instability at
$\gamma_c$. Exactly at the critical amplitude, $\sigma_r^+$ and
$\sigma_r^-$ collapse. Past this bifurcation we have two solutions for
$\sigma_i$ and the solution becomes unstable with the two frequencies
of the modes being locked with a difference equal to $1$, so that
$\omega^{(a)}=0.645$ and $\omega^{(b)}=-0.355$.  This simple model
explains the basic properties of the parametric instability but does
not reproduce the frequency drifting and the disappearance of the
instability for large $\epsilon$, which requires the formalism
introduced in section \ref{sec_dispersion_relation_shear}.

\captionsetup[subfigure]{margin=0.0cm,singlelinecheck=false,
                         format=plain,indention=0.0cm,
                         justification=justified,
                         captionskip=0cm,position=top} 
 

\subsection{Effect of the shear on the parametric instability}

Now we consider the effect of shear with $\beta \neq 0$. Unlike in the
previous section, the eigenmodes are no longer the original Kelvin
modes because the shear couples different Kelvin modes (section
\ref{sec_dispersion_relation_shear}).  The velocity decomposition
(\ref{perturbation_velo}) is now extended to
\begin{equation}
\textbf u=      \gamma  \textbf u_0(\textbf r,t)+     \xi
\textbf u^{(a)}(\textbf r,t)+\xi \textbf u^{(b)} 
(\textbf r,t) +\beta r \Omega_\beta(r) {\textbf e}_\varphi
\label{perturbation_velo_ext}
\end{equation}
with $\textbf u_0, \textbf u^{(a)}$ and $\textbf u^{(b)}$ again given
by the decomposition (\ref{eq::ansatz_triad}) and the last term on the
right hand side corresponding to the modification of the background flow. 
After some algebra, the linear system coupling the $N$  amplitudes of
$a_j$ and $b_l$ becomes

\begin{equation}
\left( \begin{array}{ccc}
\mathcal{E}^{(a)} &&0 \\ 0 && \mathcal{E}^{(b)}
\end{array} \right)
\frac{d \textbf x}{dt}=
\left( \begin{array}{ccc}
i \mathcal{N}^{(a)} &&\gamma e^{it} \mathcal{C}^{(a)} 
\\ 
\gamma^{*} e^{-it} \mathcal{C}^{(b)} && i \mathcal{N}^{(b)} 
\end{array} \right)\!\textbf x 
\qquad\mbox{ with }\qquad
\left\{
\begin{array}{ll}
\textbf x ^T=\left(  a_1 ,\dots ,a_N , b_1 , \dots , b_N  \right) 
\\ \\  
\mathcal{N}^{(i)} =\left( \mathcal{E}\mathcal{D}_0  
+\beta \mathcal{Q} \right) ^{(i)}  
\end{array} \right.
\label{Eq_parametric20}
\end{equation} 
where $\mathcal{C}^{(a)}$ (respectively $\mathcal{C}^{(b)}$) is the
matrix with the elements $c_{jl}^{(a)} e_{jj}^{(a)}$
(resp. $c_{lj}^{(b)}e_{{{ll}}}^{(b)}$) given by (\ref{def:ell}) and
(\ref{eq_coupling_term1}). When the matrices $\mathcal{C}^{(a)}$ and
$\mathcal{C}^{(b)}$ are set to zeros, we recover the linear dispersion
of Kelvin modes given in section \ref{sec_dispersion_relation_shear},
and when the elements of $\mathcal{Q}$ are set to zeros,
i.e. $\mathcal{N}=\mathcal{E}\mathcal{D}_0$, we obtain $2N$ equations
describing the parametric resonance for Kelvin modes with $\beta=0$.
The system (\ref{Eq_parametric20}) reduces to a linear system with
constant coefficients with the change of variables
\begin{equation}  
\textbf x  = \left( \begin{array}{ccc}
\mathcal{I}  e^{it/2} &&0 \\ 0 && \mathcal{I}  e^{-it/2} 
\end{array} \right)  
\tilde{\mbox{$\textbf x$}}e^{i \sigma t}
\label{Eq_parametric22}
\end{equation} 
and the eigenvalue $\sigma$ is the solution of the general eigenvalue
problem 
\begin{equation}
\sigma \left( \begin{array}{ccc}
\mathcal{E}^{(a)} &&0 \\ \\ 0 && \mathcal{E}^{(b)}
\end{array} \right)
\tilde{\mbox{$\textbf x$}}  =
\left( \begin{array}{ccc}
- \frac{1}{2}\mathcal{E}^{(a)} +  \mathcal{N}^{(a)}   &&-i \gamma
\mathcal{C}^{(a)} \\ \\ 
-i \gamma^{*}  \mathcal{C}^{(b)} &&    
\frac{1}{2}\mathcal{E}^{(b)}+\mathcal{N} ^{(b)}  
\end{array} \right) \tilde{\mbox{$\textbf x$}} .
\label{Eq_parametric23}
\end{equation}
This system is solved with the same procedure as described in section
\ref{sec_dispersion_relation_shear}.  


\subsection{Application to the modes $(m,n)=(5,1)$ and $(6,2)$}

Finally, we give a brief outline of an application of the theoretical
framework developped in the previous section and a possibility to
confirm the theory from our experimental measurements. We calculate
the growthrates of free Kelvin modes in dependence of the amplitude of
the modification of the background flow. Assuming that $\Omega_1$ is a
geostrophic mode, we can linearly relate the shear amplitude $\beta$
to the precession ratio $\epsilon$, which is at least qualitatively
confirmed in our measurements.

The matrices $\mathcal{E}$, $\mathcal{N}$ and $\mathcal{C}$ are
calculated for prograde modes with $(m_{2},n_{2})=(5,1)$ and
retrograde modes with $(m_{3},n_{3})=(6,2)$.  Since we do not know the
amplitude of the forced mode, we consider $\gamma$ as a free control
parameter in addition to the parameter $\beta$, the amplitude of the
perturbation of the solid body rotation.

\captionsetup[subfigure]{margin=0.0cm,singlelinecheck=false,
                         format=plain,indention=0.0cm,
                         justification=justified,
                         captionskip=0cm,position=top} 
\begin{figure}[b!] 
\centerline{
\subfloat[]{\includegraphics[width=0.49\textwidth]{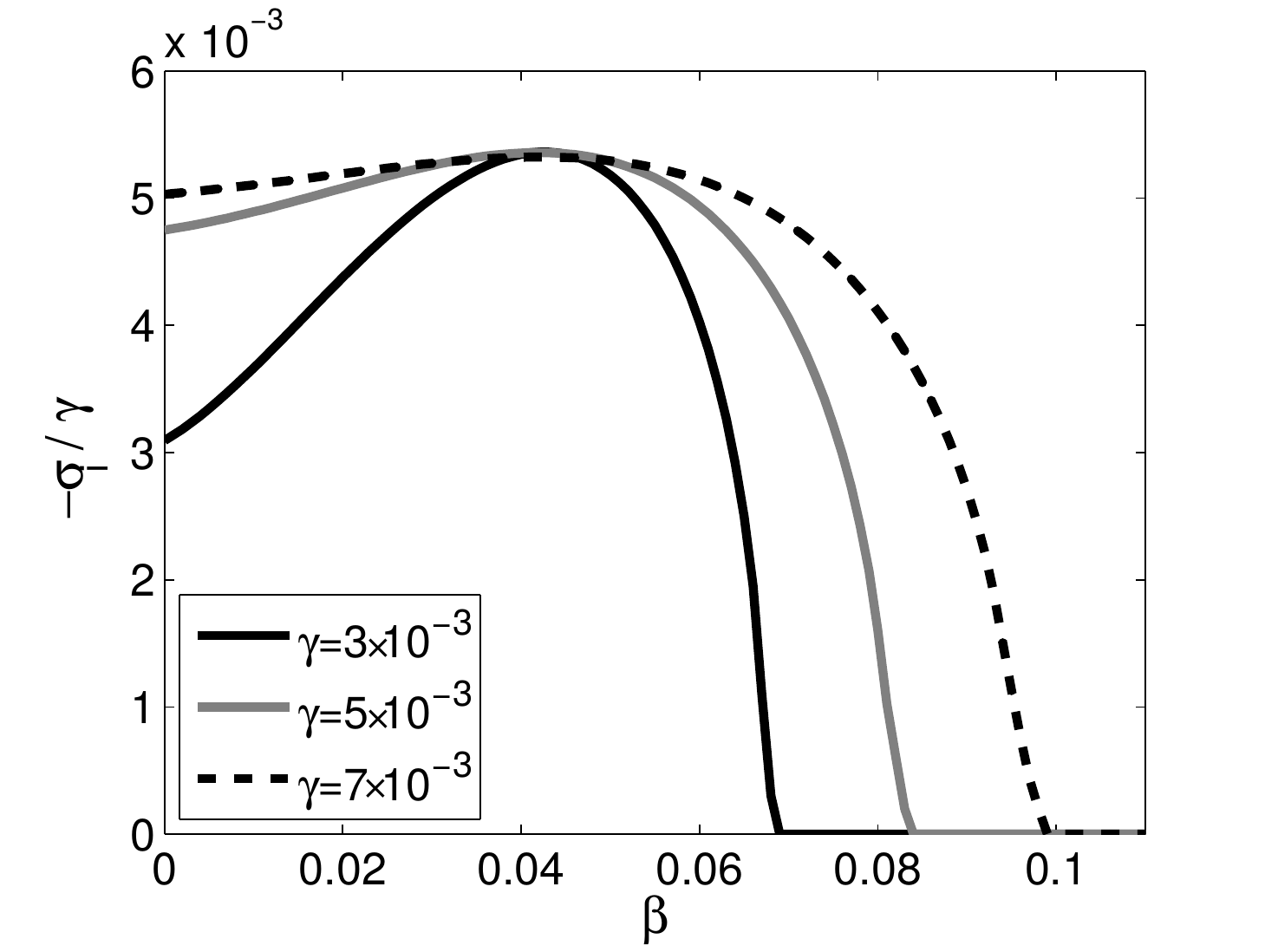}}  
\subfloat[]{\includegraphics[width=0.49\textwidth]{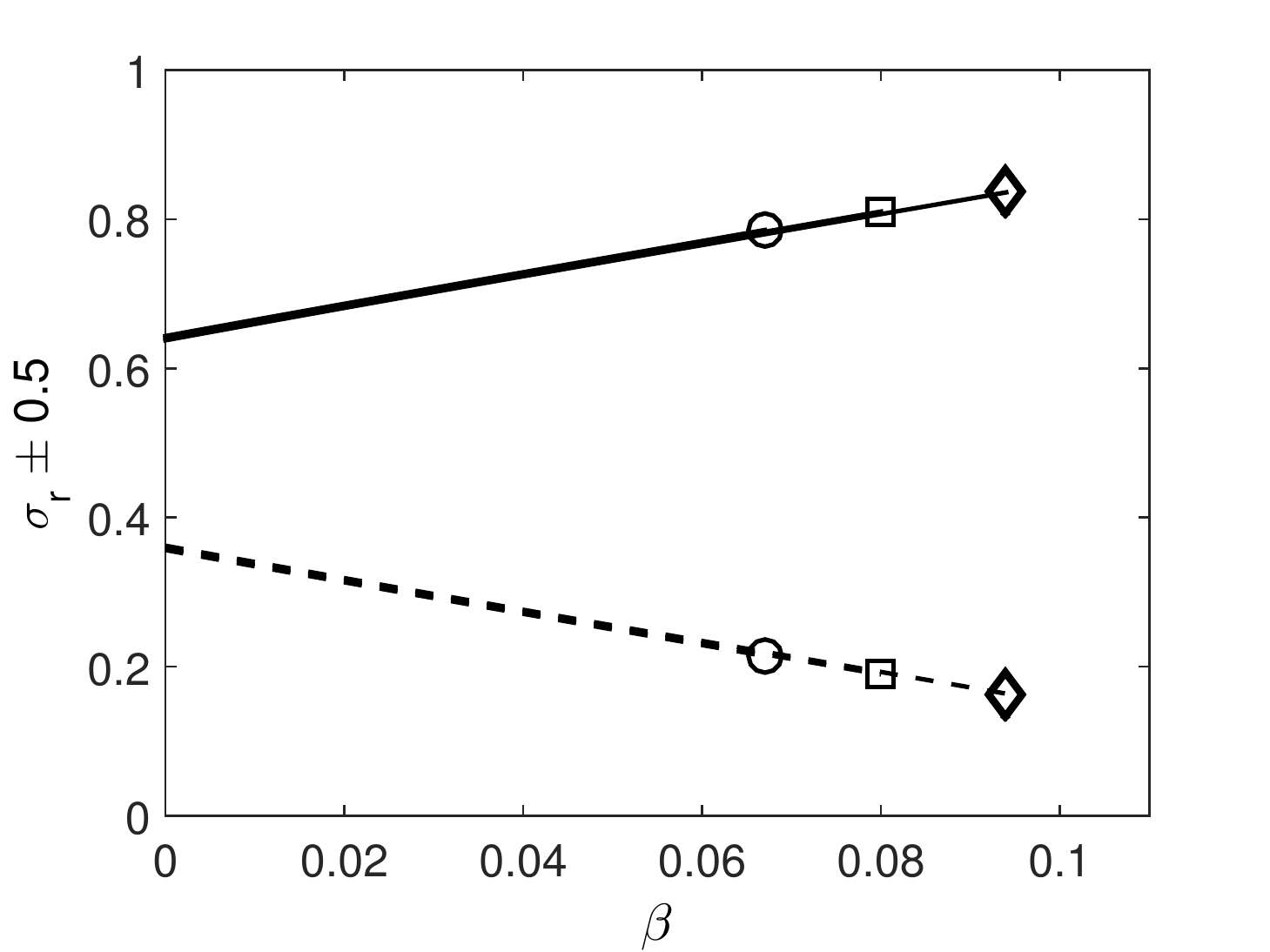}}   
}
\caption{(a) Rescaled growth-rates $-\sigma_i/ \gamma$   of the
  eigenmodes given by the system (\ref{Eq_parametric20}) as a
  function of $\beta$ for an amplitude of the forced mode $\gamma$
  equal to $3\times 10^{-3}$ (black line), $5\times 10^{-3}$ (grey
  line) and    $7\times 10^{-3}$ (black dashed line). (b)
  Corresponding    frequencies $\sigma_r \pm \frac{1}{2}$   in the
  cylinder reference 
  frame. The symbols denote the
  frequencies of marginal stability where $\sigma_i=0$, i.e. the
  parametric resonance switches off (circles, squares and diamonds
  correspond to $\gamma=3\times 10^{-3},5\times 10^{-3},7\times
  10^{-3}$). 
} 
\label{fig_parametric_shear}
\end{figure}

We calculate the eigenvalues $\sigma= \sigma_r+i \sigma_i$ of system
(\ref{Eq_parametric23}) for three values $\gamma=[3,5,7]\times
10^{-3}$ which are above the critical value $\gamma_{\rm{c}}$ required
for the onset of the instability. The parameter $\beta$ is varied in
the interval $\beta\in[0,0.15]$.  The results yield one unstable mode
in dependence of $\beta$ and $\gamma$, which for $\beta=0$ is formed
by the Kelvin modes $(6,2,1)$ and $(5,1,-1)$.  The corresponding
growthrates rescaled by $\gamma$ are shown in
Fig.\ref{fig_parametric_shear}a.
 
\begin{figure}[t!] 
\subfloat[]{\includegraphics[width=0.49\textwidth]{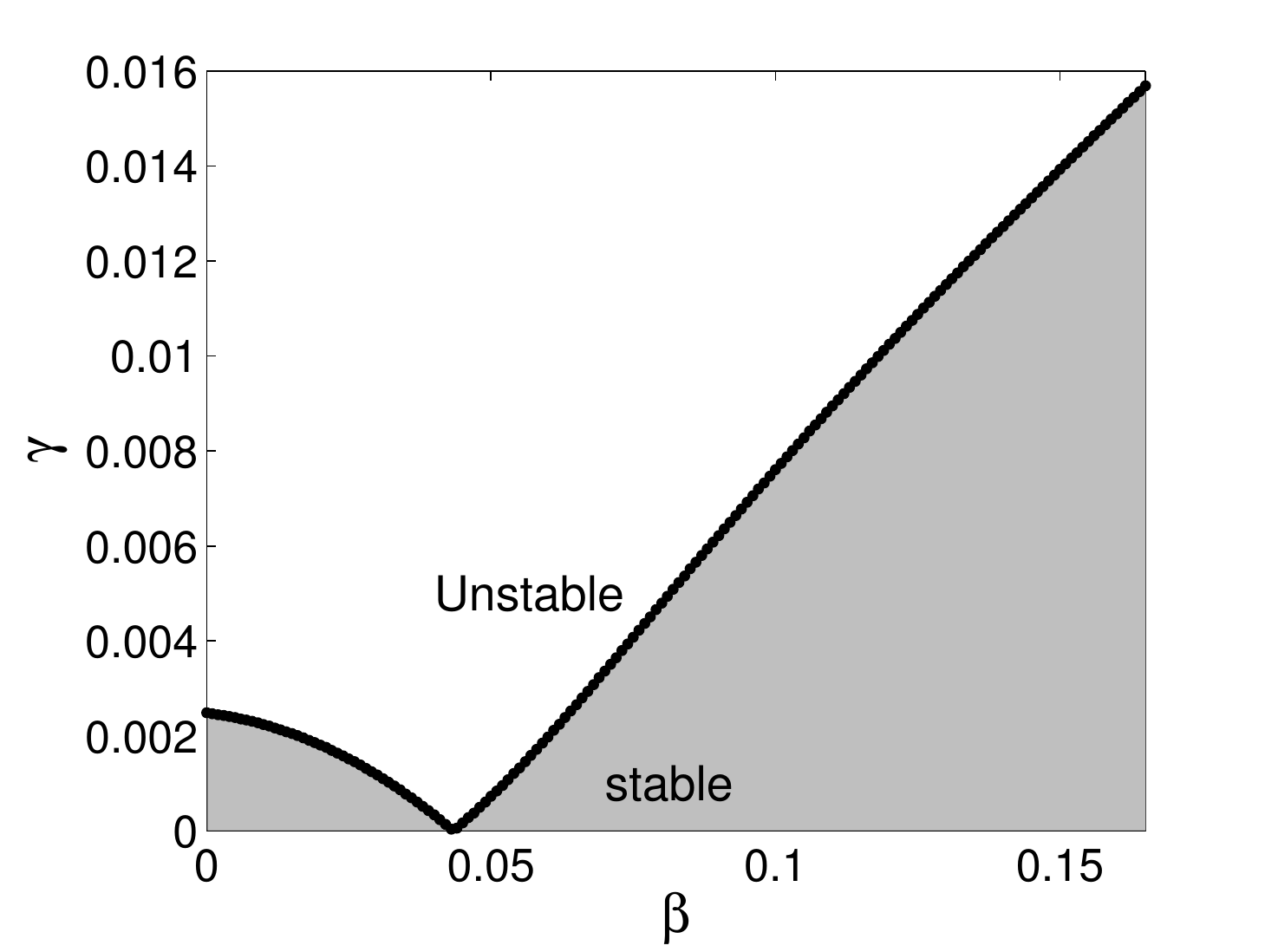}}  
\subfloat[]{\includegraphics[width=0.49\textwidth]{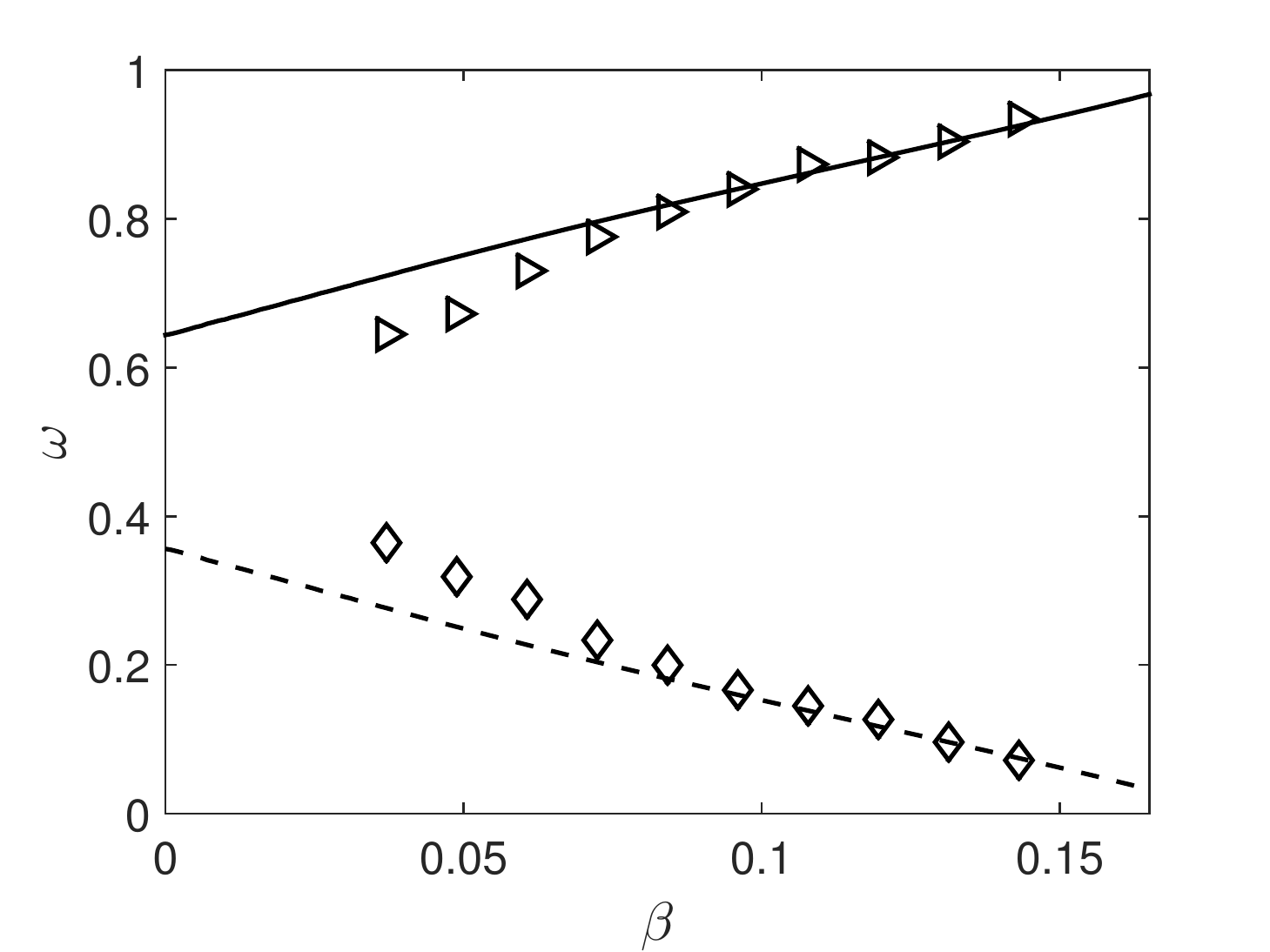}}  
  \caption{(a) Parameter space $(\beta,\gamma)$ with the domain of
existence of the parametric instability (white region) separated by
the marginal stability curve with $\sigma_i = 0$ (black curves).  (b)
Frequencies corresponding to the marginal stability curve. The
diamonds correspond to the experimental frequencies $\Omega_2$ and the
triangles correspond to $\Omega_3$ for
${\rm{Ek}}=8.8\times 10^{-7}$ as a
function of ${ {\beta=}}\alpha C_f(\epsilon-\epsilon_{\rm{c}} )$ with
$C_f=6.23$, $\epsilon_c=2.5 \times 10^{-2}$ and $\alpha=1.43$. }
\label{fig_parametric_shear2}
\end{figure}

The positive growth-rate, $-\sigma_i$, has a maximum at the exact
resonance $\beta=0.044$, (see section \ref{sec_effect_shear_unst}),
which increases linearly with $\gamma$ so that $\max(-\sigma_i)= {
{5.3 \times 10^{-3}}}\gamma$.  The growthrate becomes zero when the
background flow is strong enough to detune the parametric
resonance. Qualitatively, the regime with resonance becomes broader
(i.e. allows larger $\beta$, i.e. stronger detuning) for increasing
$\gamma$.  The eigenfrequencies $\sigma_r$ in the reference frame of
the cylinder are shown in Fig.~\ref{fig_parametric_shear}b. The
frequencies vary almost linearly with $\beta$ and the amplitude of
$\gamma$ only modifies the maximum $\beta$ at which the parametric
resonance vanishes and the frequency locking ceases. The circles
($\gamma=3 \times 10^{-3}$), squares ($\gamma=5 \times 10^{-3}$) and
diamonds ($\gamma=7 \times 10^{-3}$) denote the exact locations at
which unlocking of the frequencies occurs, i.e.  when $\sigma_i$
vanishes. The marginal stability curves are shown in
Fig.~\ref{fig_parametric_shear2}a as a function of $\gamma$ and
$\beta$ (grey color for stable region and white for unstable). The
unstable region describes a resonance tongue \cite{arnold}, which for
large $\beta$ gives a threshold $\gamma_c \sim \beta$.  This can be
explained by a heuristic model when assuming that the coupling term
$C$, given by Eq.~(\ref{eq_coupling_term}), increases linearly with
$\Delta \omega$. Fixing $\sigma_i=0$, the threshold $\gamma_c$ varies
with $\Delta \omega/\sqrt{\vert C \vert }$ (Eq.~\ref{eq_sig_para}) and
we obtain $\gamma_c \sim \sqrt{\Delta \omega}$.  On the other hand, we
have shown in section \ref{sec_effect_shear_unst} that $\Delta \omega
\sim \beta^2$ for large $\beta$, thus we obtain $\gamma_c \sim \beta$.

To compare numerical and experimental data, the parameters $\gamma$
and $\beta$ of the dispersion relation (\ref{Eq_parametric23}),
quantifying respectively the amplitude of the forced mode and the
amplitude of the shear, have to be expressed as a function of the
precession ratio $\epsilon$ and the Ekman number $\rm{Ek}$. However,
both parameters cannot be determined directly by the local
wall-pressure measurements.  To overcome this limitation we assume
that the instability saturates via frequency detuning of the unstable
modes \citep{WaleffePhD,lagrange2011precessional}.  In our model, the
saturation occurs when the shear amplitude $\beta$ reaches the
marginal stability curves for a given amplitude
$\gamma$. Consequently, we reduce the number of parameters to one, by
only considering the parameter $\beta$ as a function of $\gamma$ all
along the marginal stability curve (thick line in
Fig. \ref{fig_parametric_shear2}a).  Figure
\ref{fig_parametric_shear2}b shows the curves of the theoretical
frequencies $\omega_2$ (dashed curve) and $ \omega_3$ (thick curve)
obtained from the dispersion relation (\ref{Eq_parametric23}) as a
function of $\beta$ along the marginal stability curve.  In section
\ref{sec:constEK}, we showed that the experimental frequencies
$\Omega_2$ and $\Omega_3$ vary with $C_f( \epsilon-\epsilon_c)$ for
all Ekman numbers and precession ratios $\epsilon$, with a relatively
small scatter.  Since the frequencies are functions of $\beta$ (from
theory) and $C_f( \epsilon-\epsilon_c)$ (from experiment), it is
suggestive to express the shear amplitude $\beta$ as a function of
$C_f( \epsilon-\epsilon_c)$.  The frequencies $- \Omega_2$
(respectively $\Omega_3$) are monotonically decreasing
(resp. increasing) functions of $\beta(\gamma_c)$ and/or $C_f(
\epsilon-\epsilon_c)$, so we conclude that the shear amplitude $\beta$
is an increasing function of $C_f( \epsilon-\epsilon_c)$. Hence, the
coefficient $C_f$ quantifies the amplitude of the shear for a given
distance from the onset of instability $\epsilon-\epsilon_c$. We
emphasize that the coefficient $C_f$ increases with $\rm{Ek}^{-1/4}$
(Fig. \ref{fig_coef_thre_exp}b), i.e. the shear increases when the
Ekmann number decreases, a feature in agreement with previous
observations \cite{meunier2008rotating,lagrange2008instability}.  The
functional form of $\beta$ finally may be deduced by assuming that the
mode associated with the frequency $\Omega_1$ is a geostrophic mode,
as suggested in section \ref{sec_frist_analyse}. From the dispersion
relation of the geostrophic modes given by equation (\ref{eig_pbGeo}),
the frequency $\Omega_1$ can be written as
\begin{equation}
\Omega_1= \beta/\alpha. 
\label{eq:assume}
\end{equation}

The coefficient $\alpha^{-1}$ corresponds to one of the
eigenfrequencies of the matrix $\mathcal{E}^{-1}\mathcal{Q}$ (see
section~\ref{sub_sec_Dispersion}) and is independent of $\rm{Ek}$ and
$\epsilon$ by construction. Using Eq.~\ref{fit_f1}, the shear
amplitude $\beta$ can be written in the form
\begin{equation}
\beta=\alpha C_f(\epsilon-\epsilon_c).
\label{eq:param}
\end{equation}
Once the parameter $\alpha$ is estimated, we are able to evaluate the
shear amplitude $\beta$ for any Ekman number and precession ratios so
that all sets are equivalent when expressed as a function of
$C_f(\epsilon-\epsilon_c)$ (Fig. \ref{fig_peaks_freqR}).  As an
example we choose the frequencies from the set of measurements at the
smallest Ekman number examined within the present study
(${\rm{Ek}}=8.8\times 10^{-7}$) for which we have $\epsilon_c=2.5
\times 10^{-2}$ and $C_f=6.23$.  The coefficient $\alpha$ is
determined empirically so that the distance between experimental and
numerical data is minimized.  The results are shown in
Fig.~\ref{fig_parametric_shear2}b, with $\Omega_2$ (diamonds) and
$\Omega_3$ (triangles) plotted as a function $\beta$ given by
(\ref{eq:param}) with $\alpha=1.43$. We observe a good agreement, even
if close to the threshold the experimental frequencies differ slightly
from the theoretical solutions (difference of $12\%$).

Of course, this application is based on a simple hypothesis and the
model essentially relies only on two parameters.  Whereas the
frequency of the (hypothetic) geostrophic mode (the inverse of the
eigenvalue $\alpha^{-1}$) can be computed from experimental data, it
remains impossible to determine the amplitude $\gamma$ of the forced
mode from our pressure measurements without elaborate measurements of
the flow (PIV, Doppler probes), which could confirm our scenario.


\section{Conclusion}
  
In the present study, we have investigated experimentally and
theoretically the stability of a Kelvin mode driven by precession.
The instability is detected experimentally above a critical precession
ratio via the appearance of peaks in the temporal power spectrum of
pressure fluctuations measured at the end-walls of a precessing
cylinder. All frequencies satisfy resonance conditions so that the sum
or the difference of the frequencies is equal to the frequency of the
forced Kelvin mode.  Two sets have been identified: a triad with the
frequencies $(\Omega_{\rm{c}}, \Omega_2, \Omega_3)$ and a second set
consisting of four modes $(\Omega_{\rm{c}}, \Omega_1, \Omega_4,
\Omega_5)$.  All frequencies vary with the precession ratio, while
they continue to fulfil the temporal resonance condition with the
forced mode. The frequency variations are significant, and cannot be
explained by a viscous detuning, since the observed detuning increases
when the Ekman number decreases. Similar to observations by
\citet{lin2014experimental}, each set of frequency disappears for
larger precession ratio.

The first set corresponds to two free Kelvin modes characterized by
$(\omega_2,m_2,n_{2},l_2)=(-0.34,5,1,-1)$ and
$(\omega_3,m_3,n_{3},l_3)=(0.64,6,2,1)$ which result from a parametric
instability of the forced mode.  Comparable free Kelvin modes were
found in different experiments and simulations. These modes constitute
the most unstable modes in the precessing system with aspect ratio
$\Gamma=R/H=0.5$, which is close to the exact $(m=1,n=1,l=1)$
resonance at $\Gamma=0.502559$ and there are nearly no doubts about
this interpretation. 
The behavior may be different for another aspect ratio, where other
free Kelvin modes emerge \cite{giesecke2015triadic}, and where the forced
mode is no longer resonant, which would result in a weaker amplitude and a
different location in the stability diagram shown in
Figure~\ref{fig_parametric_shear2}a.

We developed a consistent framework to explain the behavior of the
frequencies $\Omega_2$ and $\Omega_3$, and we showed that the
parametric resonance is able to lock the frequencies to maintain the
resonant condition, despite the detuning due to background shear flow.
However, the instability is suppressed when the amplitude of the
forced mode is not sufficiently large for allowing the frequency
locking.  We have investigated theoretically the modification of the
dispersion relation due to a non-uniformly rotating background flow.
Our numerical results are in good agreement with the experimental
observations: the frequency of a retrograde Kelvin mode
(resp. prograde Kelvin mode) increases (resp.  decreases) with the
coefficient $\beta$ that parameterizes the slowdown of the background
flow. We show that a slight modification of the background solid body
rotation ($\beta=0.044$) is sufficient to trigger an exact resonance
between the Kelvin modes.  Furthermore, we have studied the effect of
the detuning on the condition of parametric resonance.  The frequency
difference $\omega_3-\omega_2$ increases quadratically with $\beta$
and a small perturbation of the solid body rotation ($\beta \simeq
0.15$) is sufficient to unlock the frequencies so that the parametric
resonance is suppressed. This feature explains the disappearance of
the frequencies $\Omega_2$ and $\Omega_3$, above a critical precession
ratio. At the critical point, the amplitude of the forced mode is no
longer strong enough to lock the frequencies of the unstable Kelvin
modes.

\bigskip

The identification of the modes of the second set and the nature of
this instability are more speculative. At the threshold of appearance,
the frequency $\Omega_1$ is close to zero and the frequencies
$\Omega_4$ and $\Omega_5$ are close to $\Omega_{\rm{c}}=1$.  This
frequency set always appears at a lower critical precession ratio than
required for the triadic resonances discussed above.  With almost
certainty we can rule out the possibility that these modes form a
resonant triad and we suggest that $\Omega_4$ and $\Omega_5$
correspond to Kelvin modes resulting from the interaction of the
directly forced mode and a geostrophic mode $\Omega_1$.  In that case,
the non-vanishing frequency of the geostrophic mode results from a
background shear flow that removes the frequency degeneracy at
$\omega_0=0$ in case of the background flow being an unperturbed solid
body rotation.  Following this hypothesis, the second set could be an
experimental hint of the instability involving a geostrophic mode,
predicted numerically in \cite{kerswell1999secondary}.

\bigskip

The \textit{ad-hoc} parameterization of the shear amplitude
$\beta$ as a function of the precession ratio $\epsilon$ is based on
the hypothesis that the mode associated with the frequency $\Omega_1$ is a
geostrophic mode. From the dispersion relation, we know that the
frequency of the geostrophic mode increases linearly with $\beta$.
Assuming $\Omega_1\propto \beta$ , we observe a good agreement between
numerical prediction and the experimental results.  However, without
experimental confirmation of the spatial structure of the instability,
the presence of a geostrophic mode remains speculative, although we
find similarities between our observations and the frequency variation
of the instability of a geostrophic mode described by Kerswell
\cite{kerswell1999secondary}.

Our experimental and theoretical results point out that the saturation
mechanism of the instability of a Kelvin mode is related to a
modification of the background flow, which changes the dispersion
relation of the Kelvin modes. The conditions of resonance are
particularly sensitive to the amplitude of the background flow: a
slight modification can tune or detune a resonance. Moreover, our
calculations suggest that a sequence of small scale unstable modes
could emerge after the disappearance of the first resonant triad. We
did not detect these modes in the pressure measurements, since the
power spectrum becomes flat for moderate precession ratio (see also
\cite{lin2014experimental}). This regime is not yet turbulent
\cite{herault2015subcritical}, although the flow cannot be described
by a superimposition of free Kelvin modes. The nature of this regime
remains enigmatic, and further numerical and experimental studies are
required to characterize the features of this non-linear regime.  At
current stage, we presume that the instability of small scale modes
may be inhibited by the viscous dissipation.
 
The development of the azimuthal shear is of relevance for the large
precession dynamo experiment currently under construction at HZDR,
where a precession-driven flow of liquid sodium is expected to
generate a magnetic field \cite{Stefani2014}.  Simulations have shown
that the modification of the azimuthal shear profile is accompanied by
further instabilities, which are of great importance for the dynamo
process \cite{Giesecke2018}.  A deeper understanding of these
processes and especially a robust detection of the azimuthal flow
profile in an opaque medium like a liquid metal are therefore of great
importance for this experiment. The kind of
``spectroscopy'' of Kelvin modes as developed for this study, could be
applied using the measured detuning of the parametric
resonance as starting point to deduce a model of the
radial profile of the azimuthal shear, which, of course will require a
more sophisticated approach than the simple (quadratic) form for the
rotation law, as assumed in Eq.~(\ref{perturbation_form}).

So far the ultimate confirmation of the emergence of a geostrophic mode
by directly resolving the flow structure is lacking and other
possibilities for explaining our observations might be possible. For example, 
an altenative explanation may interpretate the frequency $\Omega_1$ as a `beat'
frequency indicative of a near-commensurate quasi-periodic response,
which, however, would require different assumptions, e.g., 
about the origin of this near-commensurate quasi-periodic frequency. 
At this stage, we are not able to give a final answer to this question
so that -- although the scenario connected with a geostrophic mode is
quite plausible -- the presence and the origin of the modes $\Omega_1,
\Omega_4,$ and $\Omega_5$ must remain an unsolved issue.

\begin{acknowledgments}This study has been conducted in the framework of the 
project DRESDYN (DREsden Sodium facility for
DYNamo and thermohydraulic studies), which provides
a platform for experiments dedicated to geo- and astrophysical
problems at Helmholtz-Zentrum Dresden-Rossendorf (HZDR).
The authors would like to thank Bernd Wustmann
for the mechanical design of the precession experiment.
Discussions with R. Kerswell are gratefully acknowledged. 
\end{acknowledgments}

\appendix

\section{Operators and solutions}

 
\subsection{Linear operators}\label{app::a1}
 
The operators used in equation~(\ref{lin_eq})   read as follows 

\begin{equation}\label{eq::matI}
{\mathcal{I}}= \left(
\begin{array}{ccccccc}
1 &&	0 &&	0 &&	  0 \\ \\
\vphantom{\displaystyle\frac{1}{r}}0 &&	1 &&	0 &&	 0 \\ \\
0 &&	0 &&	1 &&	 0 \\ \\
\vphantom{\displaystyle\frac{1}{r}}0 &&	0 &&	0 &&	 0 \\
\end{array} \right)
\end{equation}
 
\begin{equation}
{\mathcal{L}}_0= \left(
\begin{array}{ccccccc}
0 &&	-2 &&	0 &&	  \partial_r \\ \\
2 &&	0 &&	0 &&\displaystyle	 i \frac{m}{r} \\ \\
0 &&	0 &&	0 &&	 i k \\ \\
\displaystyle\frac{1}{r}+\partial_r &&\displaystyle	i \frac{m}{r} &&	i k &&	 0 \\
\end{array} \right)
\end{equation}

\begin{equation}
{\mathcal{L}}_\beta= \left(
\begin{array}{ccccccc}
i m \Omega &&	-2 \Omega  &&	0 &&	  0 \\ \\
\zeta  &&	i m \Omega &&	0 &&	 0 \\ \\
0 &&	0 &&	i m \Omega &&	0 \\ \\
0 &&	0 &&	0&&	 0 \\
\end{array} \right).
\end{equation}

\subsection{Numerical convergence}  
\label{sub_sec_num}

Equation (\ref{eig_pb}) is solved via a \textit{MATLAB} code. The
number of involved Kelvin modes is fixed to $N=100$, i.e the first
$50$ prograde and retrograde modes. We have checked the convergence by
also using $N=200$. The roots of the dispersion relations
(\ref{eq_disp1}) and (\ref{eq_disp11}) are solved with a residual
smaller than $10^{-10}$. The scalar products $m_{jl}$ and $e_{ll}$ are
calculated via a Gauss-Legendre quadrature rule, which is based on a
collocation method using the Legendre polynomials as weighting
functions. We used $10^3$ collocation points for the integration of
the Bessel functions, which corresponds to $20$ points per lobe for
the Kelvin modes with $l=\pm 50$. The generalized eigenvalue problem
(\ref{eig_pb}) is solved by using the routine \textit{EIG}, which
calculates the eigenvalues $\omega_l$ and eigenvectors $\textbf
a_l$. We have checked that the residual for the first Kelvin modes,
i.e $\vert{l}\vert<5$, is smaller than $10^{-10}$. Another way to
verify the convergence is to validate that the imaginary part of
$\omega_l$ is zero without dissipation (the mode remains stable).

\begin{figure}[h!]
\includegraphics[width=0.49\textwidth]{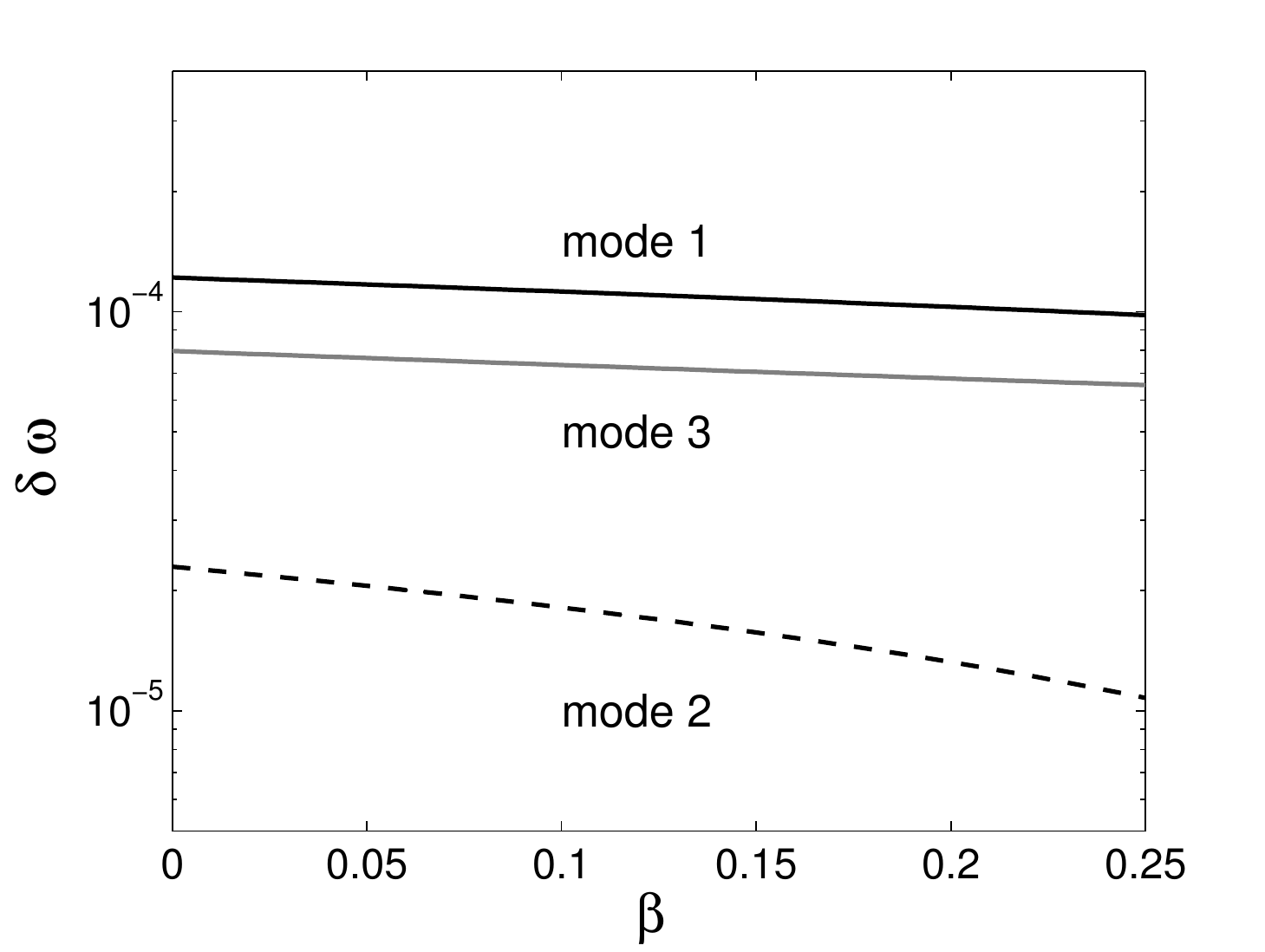} 
\caption{\label{fig_comp_cheb} 
  Difference
  between the eigenfrequencies calculated by a Chebyshev code  and
  our code. The differences are calculated with the three first retro-grade
  modes with $m=1$ and $n=1$.
}     
\end{figure}

We have verified our code by comparing our result with a code
developed by Antkowiak and Brancher \cite{antkowiak2007vortex} based
on the Chebyshev pseudo-spectral method
\cite{weideman2000matlab}. Unlike our code, the Chebyshev code
respects no-slip boundary conditions. We have used a sufficient number
of collocation points (typically $50$) to converge the results without
resolving the boundary layer in order to mimic a free-slip
condition. We fix a small Ekman number (${\rm{Ek}}<10^{-3}$) in order to
neglect the viscous dissipation in the bulk flow.  The difference
between the eigenfrequency of our method and the Chebyshev code for
$m=1$, $n=1$ and $l=(1,2,3)$, with $\omega_{0l}=(0.99,0.51,0.34)$, is
shown in Fig.\ref{fig_comp_cheb}. We have checked that the relative
difference is small, with values of order $10^{-4}$.

We point out that our code allows to follow continuously all Kelvin
modes by varying $\beta$, whereas spurious modes arise in the
Chebyshev code when the resolution is increased.

\providecommand{\noopsort}[1]{}\providecommand{\singleletter}[1]{#1}%
\end{document}